\documentclass[twocolumn,showpacs,preprintnumbers,amsmath,amssymb,notitlepage,nofootinbib]{revtex4}

\usepackage{amsfonts}
\usepackage{amsmath}
\usepackage{amssymb}
\usepackage{graphicx}
\usepackage{dcolumn}
\usepackage{bm}
\usepackage{tabularx}
\usepackage{epstopdf}
\usepackage{xcolor}
\usepackage{mathrsfs}
\usepackage{mathtools}
\usepackage{float}
\usepackage{algpseudocode}
\usepackage{verbatim}
\usepackage{wasysym}
\usepackage{footnote}
\usepackage{chngcntr}
\usepackage{natbib}
\usepackage{mathtools}
\usepackage{stmaryrd}
\usepackage{booktabs}

\def\multiset#1#2{\ensuremath{\left(\kern-.3em\left(\genfrac{}{}{0pt}{}{#1}{#2}\right)\kern-.3em\right)}}

\begin{document}

\title{Detecting Mesoscale Structures by Surprise}

\author{Emiliano Marchese}
\affiliation{IMT School for Advanced Studies, Piazza S.Francesco 19, 55100 Lucca (Italy)}
\author{Guido Caldarelli}
\affiliation{`Ca' Foscari' University of Venice, Dorsoduro 3246, 30123 Venice (Italy)}
\author{Tiziano Squartini}
\affiliation{IMT School for Advanced Studies, Piazza S.Francesco 19, 55100 Lucca (Italy)}
\affiliation{Institute for Advanced Study (IAS), University of Amsterdam, Oude Turfmarkt 145, 1012 GC Amsterdam (The Netherlands)}
\date{\today}

\begin{abstract}
The importance of identifying the presence of mesoscale structures in complex networks can be hardly overestimated. So far, much attention has been devoted to the detection of \emph{communities}, \emph{bipartite} and \emph{core-periphery} structures on \emph{binary} networks: such an effort has led to the definition of a unified framework based upon the score function called \emph{surprise}, i.e. a p-value that can be assigned to any given partition of nodes, on both \emph{undirected} and \emph{directed} networks. Here, we aim at making a step further, by extending the entire framework to the \emph{weighted} case: after reviewing the application of the surprise-based formalism to the detection of binary mesoscale structures, we present a suitable generalization of it for detecting weighted mesoscale structures, a topic that \textcolor{black}{has received much less attention.} To this aim, we analyze four variants of the surprise; from a technical point of view, this amounts at employing four variants of the hypergeometric distribution: the \emph{binomial} one for the detection of binary communities, the \emph{multinomial} one for the detection of binary `bimodular' structures and their \emph{negative} counterparts for the detection of communities and `bimodular' structures on weighted networks. On top of that, we define two `enhanced' variants of surprise, able to encode both binary and weighted constraints and whose definition rests upon two suitable generalizations of the hypergeometric distribution itself. As a result, we present a general, statistically-grounded approach to detect mesoscale structures on networks via a unified, suprise-based framework. To illustrate the performance of our methods, we\textcolor{black}{, first, test them on a variety of well-established, synthetic benchmarks and, then, apply them} to several real-world networks, including social, economic, financial and ecological ones. Moreover, we attach to the paper a Python code implementing all variants of the surprise considered in the present work.
\end{abstract}
\keywords{complex networks \and mesoscale structures detection \and surprise minimization}
\pacs{89.75.Fb; 02.50.Tt; 89.65.Gh}

\maketitle

\section*{INTRODUCTION}

The importance of identifying the signature of some kind of mesoscopic organization in complex networks (be it due to the presence of communities or bipartite, core-periphery, bow-tie structures) can be hardly overestimated \cite{Fortunato2016,Bisma2017}, the best example of complex systems whose behavior is deeply affected by their mesoscopic structural organization (e.g. resilience to the propagation of shocks, to the failure of nodes, etc.) being provided by financial networks \cite{Borgatti2000,Craig2010,IntVeld2014,Luu2018}. 

\textcolor{black}{So far, much attention has been devoted to the detection of \emph{binary mesoscale structures}, i.e. \emph{communities} and, to a far less extent, \emph{core-periphery structures}: the efforts to solve these problems have led to a number of approaches that are briefly sketched below (for a detailed review of them, see \cite{Fortunato2010,Cimini2021})}.

\textcolor{black}{Community detection has been initially approached by attempting a definition of `communities' based on the concepts of \emph{clustering}, \emph{cliques}, \emph{k-core}; core-periphery structures have, instead, been defined in a purely top-down fashion, by imagining a fully connected subgraph (i.e. the core) surrounded by (peripherical) vertices exclusively linked to the first ones \cite{Borgatti2000}. As stressed in \cite{IntVeld2014}, the deterministic character of these definitions makes their \emph{tout court} application to real-world systems extremely difficult. This is the reason why the intuitive requirements that `the number of internal edges is larger than the number of external edges' and that `the core portion of a network is densely connected, while its periphery is loosely connected' \cite{Cimini2021} are, now, interpreted in a purely probabilistic way: a community has, thus, become a subgraph whose vertices have a larger probability to be inter-connected than to be connected to any other vertex in the graph - and analogously for the core-periphery structure. In other words, the top-down approach defining a golden standard and looking for (deviations from) it has left the place to a bottom-up approach where structures are supposed to emerge as the result of non-trivial (i.e. non-casual) interactions between the nodes.}

\textcolor{black}{This change of perspective leads to a number of problems. The first one concerns the definition of models stating how edges are formed and has been solved by adopting the rich formalism defining the Exponential Random Graphs framework \cite{Fronczak2014}; the second, and most important, one concerns the definition of a (statistically-sound) procedure for selecting the best model among the ones providing competing descriptions of the data. This has led to the identification of} \textcolor{black}{three broad classes of algorithms\footnote{For a different classification of the algorithms proposed for the detection of mesoscale structures see \cite{Pesciotto2021}.}: according to our intuition, all of them implement some kind of statistical inference, the major difference lying in the way the corresponding test of hypothesis is implemented; from a practical point of view, instead, all these methods are designed for \emph{optimization}, the functional form of the specific score function determining the class to which a given algorithm belongs.}

\textcolor{black}{The most representative algorithms among those belonging to the first class are the ones based on modularity. Although these methods compare the empirical network with a benchmark, they do not provide any indication of the \emph{statistical significance} of the recovered partition, the reason being that none of them is designed as a proper statistical test. For instance, let us consider the definition of modularity, whose generic addendum is proportional to the term $(a_{ij}-p_{ij})$: while it embodies a comparison between the (empirical) adjacency matrix $\mathbf{A}$ and the matrix of probability coefficients $\mathbf{P}$ defining the benchmark, it does not implement any proper test of hypothesis. The algorithms prescribing to maximize a plain likelihood function belong to this group as well. Although popular, this way of proceeding is known to be affected by \emph{overfitting} issues: an example is provided by the straight maximization of the likelihood defining the Stochastic Block Model (SBM), or its degree-corrected version, over the whole set of possible partitions, outputting the trivial one where each vertex is a cluster on its own \cite{Karrer2011}.}

\textcolor{black}{The aforementioned, major limitation is overcome by the algorithms belonging to the second class. They implement tests of hypothesis either \emph{`a la Fisher} or \emph{\`a la Neyman-Pearson}, i.e. either defining a single benchmark (the \emph{null hypothesis} of the first scenario) or two, alternative ones (the \emph{null} and the \emph{alternative hypothesis} of the second scenario): from a practical point of view, such a result is achieved by identifying the aforementioned benchmarks with proper probability distributions and the best partition of nodes with the one minimizing the corresponding p-value. Surprise-based algorithms belong to this second group: as it has been shown in \cite{DeJeude2019} - for a particular case; such a result will be generalized in what follows - optimizing (asymptotic) surprise amounts at carrying out a (sort of) Likelihood Ratio Test aimed at choosing between two alternative models.}

\textcolor{black}{Hypothesis testing can be further refined by allowing for more than two hypotheses to be tested at a time: results of the kind are particularly useful for \emph{model selection} and, in fact, have produced a plethora of criteria (e.g. the Akaike Information Criterion, the Bayesian Information Criterion and the Minimum Description Length) for singling out the best statistical model out of a basket of competing ones. Generally speaking, optimization, here, seeks for the maximum of a `corrected' likelihood function embodying the trade-off between accuracy and parsimony of a description. An example of the algorithms belonging to this third class is represented by Infomap; other examples are represented by the recipes employing the SBM within a Bayesian framework (see \cite{Pesciotto2017} and the references therein).}\\

\textcolor{black}{With this paper, we pose ourselves within the \textcolor{black}{second} research line and adopt a bottom-up approach that prescribes to compare any empirical network structure with the outcome of a properly-defined benchmark model. To this aim, we devise a unified framework for mesoscale structures detection} based upon the score function called \emph{surprise}\footnote{Notice that our function is named \emph{surprise} since it generalizes the function proposed in \cite{Aldecoa2013} for community detection. However, in our case it indicates a proper probability and not its logarithm, as in \cite{Aldecoa2013}. The same holds true for its asymptotic expression.}, i.e. a p-value that can be assigned to any given partition of nodes, on both \emph{undirected} and \emph{directed} networks: while for \emph{binary community detection} this is achieved by employing the \emph{binomial hypergeometric distribution} \cite{Aldecoa2013,Nicolini2016}, with `bimodular' structures like the bipartite and the core-periphery ones, one needs to consider its \emph{multinomial} variant \cite{DeJeude2019}. Here, however, we aim at making a step further, by extending the entire framework to the \emph{weighted} case. As a result, we present a general, statistically-grounded approach to the problem of detecting mesoscale structures on networks via a unified, suprise-based framework.

\textcolor{black}{Before moving forward, we would like to stress that the use of the hypergeometric distribution to carry out tests of hypothesis on networks is not novel: examples are provided by the papers \cite{Tumminello2011} (where the authors introduce a method to provide a statistically-validated, monopartite projection of a bipartite network - the considered null hypothesis encoding the heterogeneity of the system), \cite{Bongiorno2017} (where the authors employ the same validation procedure to detect cores of communities within each set of nodes of a bipartite system) and \cite{Musciotto2021} (where the authors extend the framework proposed in the aforementioned references to carry out a statistical validation of motifs observed in hypergraphs). For a review on the use of the hypergeometric distribution for network analyses see \cite{Micciche2019} and the references therein.}

\section*{MESOSCALE STRUCTURES DETECTION\\VIA EXACT TESTS}

\textcolor{black}{Surprise has recently received a lot of attention: the advantages of employing such a score function have been extensively discussed in \cite{Aldecoa2013,Nicolini2016,DeJeude2019,Jiang2014,Delser2016,Tang2019,Gamermann2020} where researchers have tested and compared its performance from a purely numerical perspective. However, a characterization of the statistical properties of surprise is still missing: for this reason, we will, first, make an effort to `translate' the problem of detecting a given mesoscale network structure into a proper \emph{exact significance test} (to the best of our knowledge, the only other attempt of the kind - specifically, to detect communities - is the one in \cite{Sadamori2018}) and, then, show how the rich - yet, still underexplored - surprise-based formalism can properly answer such a question.}

\textcolor{black}{The basic equation underlying exact tests reads}

\begin{eqnarray}
\text{Pr}(x\geq x^*)=\sum_{x\geq x^*}f(x)
\end{eqnarray}
\textcolor{black}{and returns the probability of observing an outcome of the random variable $X$ which is `more extreme' than the realized one, i.e. $x^*$. In the setting above, $f$ represents the distribution encoding the \emph{null hypothesis} and $\text{Pr}(x\geq x^*)$ - commonly known with the name of \emph{p-value} - answers the question \emph{is the realized value $X=x^*$ compatible with the (null) hypothesis that $X$ is distributed according to $f$?}}

\textcolor{black}{The p-value constitutes the basic quantity for carrying out any \emph{significance test}: hence, in what follows we will tackle the problem of detecting the signature of \emph{statistically-significant} mesoscale organizations by individuating specific tests (or, equivalently, suitable functional forms for $f$).}

\section*{DETECTION OF MESOSCALE STRUCTURES\\IN BINARY NETWORKS}

\noindent{\bf{Community detection.}} The topic of nodes partitioning into densely connected groups has traditionally received a lot of attention, with applications ranging from the analysis of social networks to the definition of novel recommendation systems \cite{Fortunato2016,Bisma2017}. \textcolor{black}{Within the surprise-based framework}, binary community detection is carried out \textcolor{black}{via the identification}

\begin{eqnarray}
f(l_\bullet)&\equiv&\text{H}(l_\bullet|V,V_\bullet,L)\nonumber\\
&=&\frac{\prod_{i=\bullet,\circ}\binom{V_i}{l_i}}{\binom{V}{L}}=\frac{\binom{V_\bullet}{l_\bullet}\binom{V_\circ}{l_\circ}}{\binom{V}{L}}=\frac{\binom{V_\bullet}{l_\bullet}\binom{V-V_\bullet}{L-l_\bullet}}{\binom{V}{L}}
\end{eqnarray}
\textcolor{black}{i.e. by calculating the p-value}

\begin{equation}
\mathscr{S}\equiv\sum_{l_\bullet\geq l_\bullet^*}f(l_\bullet)
\label{eq1}
\end{equation}
\textcolor{black}{of a \emph{binomial hypergeometric distribution} whose parameters read as above. In this formalism, the $\bullet$ subscript will be meant to indicate quantities that are \emph{internal} to communities, while the $\circ$ subscript will be meant to indicate quantities that are \emph{external} to communities. More precisely,} the binomial coefficient $\binom{V_\bullet}{l_\bullet}$ enumerates the number of ways $l_\bullet$ links can be redistributed \emph{within} communities, i.e. over the available $V_\bullet$ node pairs, while the binomial coefficient $\binom{V_\circ}{l_\circ}$ enumerates the number of ways the remaining $l_\circ=L-l_\bullet$ links can be redistributed \emph{between} communities, i.e. over the remaining $V_\circ=V-V_\bullet$ node pairs. Notice that, although $l_\bullet$ is `naturally' bounded by the value $V_\bullet$, it cannot exceed $L$ - whence the usual requirement $l_\bullet\in[l_\bullet^*,\min\{L,V_\bullet\}]$.

\textcolor{black}{From a merely statistical point of view, surprise `considers' a network as a population of $V$ node pairs, $L$ of which have been drawn; out of the $L$ extracted ones, $l_\bullet$ node pairs have the desired feature of being `internal' to communities, since they connect some of the node pairs belonging to the set $V_\bullet$. Hence, for a given partition of nodes into communities, $\mathscr{S}$ quantifies the probability of observing at least $l_\bullet^*$ `successes' (i.e. intra-cluster edges) out of $L$ draws: the lower this probability, the `more surprising' the observation of the corresponding partition, hence the `better' the partition itself.}\\

\noindent{\bf{Asymptotic results.}} We can gain more insight into the surprise-based formalism above upon deriving an asymptotic expression for $\mathscr{S}$ \cite{Traag2015}. To this aim, let us consider that it can be simplified upon Stirling-approximating the binomial coefficients that appear within it. By exploiting the recipe $n!\simeq\sqrt{2\pi n}\left(\frac{n}{e}\right)^n$, $\mathscr{S}$ can be rewritten as

\begin{equation}
\mathscr{S}\simeq\sum_{l_\bullet\geq l^*_\bullet}A(l_\bullet)\left[\frac{\text{Ber}(V,L,p)}{\prod_{i=\bullet,\circ}\text{Ber}(V_i,l_i,p_i)}\right]
\label{eq2}
\end{equation}
where the expression

\begin{equation}
\text{Ber}(x,y,z)=z^y(1-z)^{x-y}
\label{eq3}
\end{equation}
\textcolor{black}{defines a \emph{Bernoulli} probability mass function, the parameters appearing in eq. (\ref{eq2}) read $p=\frac{L}{V}$ and $p_i=\frac{l_i}{V_i}$ and the coefficient in front of the sum is}

\begin{equation}
A(l_\bullet)=\sqrt{\frac{\sigma^2}{2\pi\prod_{i=\bullet,\circ}\sigma^2_i}}
\end{equation}
with $\sigma^2=Vp(1-p)$ and $\sigma^2_i=V_ip_i(1-p_i)$. Equation (\ref{eq2}) makes it explicit that employing $\mathscr{S}$ for binary community detection ultimately amounts at comparing the description of a networked configuration provided by the Random Graph Model (RGM), and encoded into the expression 

\begin{equation}
\text{Ber}(V,L,p)=p^L(1-p)^{V-L},
\end{equation}
with the description of the same configuration provided by the Stochastic Block Model (SBM) \cite{Karrer2010}, and encoded into the expression

\begin{equation}
\prod_{i=\bullet,\circ}\text{Ber}(V_i,l_i,p_i)=p_\bullet^{l_\bullet}(1-p_\bullet)^{V_\bullet-l_\bullet}\cdot p_\circ^{l_\circ}(1-p_\circ)^{V_\circ-l_\circ}
\end{equation}
\textcolor{black}{where $p_\bullet=\frac{l_\bullet}{V_\bullet}$ and $p_\circ=\frac{l_\circ}{V_\circ}$. Naturally, the SBM takes as input the two sets indexed by $\bullet$ and $\circ$ and distinguish the connections found `within the clusters' - contributing to the probability of the whole configuration with the term $\text{Ber}(V_\bullet,l_\bullet,p_\bullet)=p_\bullet^{l_\bullet}(1-p_\bullet)^{V_\bullet-l_\bullet}$ - from the ones found `between the clusters' - contributing to the probability of the whole configuration with the term $\text{Ber}(V_\circ,l_\circ,p_\circ)=p_\circ^{l_\circ}(1-p_\circ)^{V_\circ-l_\circ}$.}

\textcolor{black}{Notice also that the asymptotic expression of the surprise guarantees that the parameters of the null models defining it are tuned according to the maximum-of-the-likelihood principle. To see this explicitly, let us consider $\text{Ber}(x,y,z)$ whose log-likelihood reads}

\begin{equation}
\mathscr{L}=y\ln z+(x-y)\ln(1-z);
\end{equation}
\textcolor{black}{upon maximizing it with respect to $z$, one finds $z=\frac{y}{x}$.}

\textcolor{black}{The way $\mathscr{S}$ works, i.e. by comparing two different null models, is reminiscent of more traditional likelihood ratio tests, where a null hypothesis $H_0$ (in our case, a given partition is compatible with the RGM) is tested against an alternative hypothesis $H_1$ (in our case, the given partition is compatible with the SBM): as the asymptotic expression of the surprise clarifies, minimizing it amounts at finding the partition least likely to occour under the RGM than under the SBM.}\\

\noindent{\bf{`Bimodular' structures detection.}} The surprise-based framework can be easily extended to detect what can be called `bimodular structures', a term that will be used to compactly indicate core-periphery \cite{Zhang2014,Barucca2015,Kojaku2018} and bipartite structures \cite{Holme2003,Estrada2005}. The reason for adopting such a terminology lies in the evidence that both kinds of structures are defined by bimodular partitions, i.e. partitions of nodes into two different groups.

As shown elsewhere \cite{DeJeude2019}, the issue of detecting binary `bimodular' structures can be addressed by considering \textcolor{black}{a \emph{multivariate} (or \emph{multinomial}) \emph{hypergeometric distribution}, i.e. by identifying}

\begin{eqnarray}
f(l_\bullet,l_\circ)&\equiv&\text{MH}(l_\bullet,l_\circ|V,V_\bullet,V_\circ,L)\nonumber\\
&=&\frac{\prod_{i=\bullet,\circ,\top}\binom{V_i}{l_i}}{\binom{V}{L}}=\frac{\binom{V_\bullet}{l_\bullet}\binom{V_\circ}{l_\circ}\binom{V_\top}{l_\top}}{\binom{V}{L}}\nonumber\\
&=&\frac{\binom{V_\bullet}{l_\bullet}\binom{V_\circ}{l_\circ}\binom{V-(V_\bullet+V_\circ)}{L-(l_\bullet+l_\circ)}}{\binom{V}{L}}
\end{eqnarray}
\textcolor{black}{where $V_\top\equiv V-(V_\bullet+V_\circ)$ indicates the number of node pairs between the modules $\bullet$ and $\circ$ and $l_\top\equiv L-(l_\bullet+l_\circ)$ indicates the number of links that must be assigned therein. While the binomial coefficient $\binom{V_\bullet}{l_\bullet}$ enumerates the number of ways $l_\bullet$ links can redistributed \emph{within} the first module (e.g. the core portion) and the binomial coefficient $\binom{V_\circ}{l_\circ}$ enumerates the number of ways $l_\circ$ links can redistributed \emph{within} the second module (e.g. the periphery portion), the third binomial coefficient $\binom{V-(V_\bullet+V_\circ)}{L-(l_\bullet+l_\circ)}$ enumerates the number of ways the remaining $L-(l_\bullet+l_\circ)$ links can be redistributed \emph{between} the first and the second module, i.e. over the remaining $V-(V_\bullet+V_\circ)$ node pairs.}

\textcolor{black}{This choice induces the definition of the \emph{binary bimodular surprise}}

\begin{equation}
\mathscr{S}_\sslash\equiv\sum_{l_\bullet\geq l_\bullet^*}\sum_{l_\circ\geq l_\circ^*}f(l_\bullet,l_\circ);
\end{equation}
\textcolor{black}{analogously to the univariate case, $l_\bullet$ and $l_\circ$ are `naturally' bounded by the values $V_\bullet$ and $V_\circ$ - notice, however, that the sum $l_\bullet+l_\circ$ cannot exceed $L$ (although it may not reach such a value, e.g. in case $V_\bullet+V_\circ<L$).}\\

\noindent{\bf{Asymptotic results.}} \textcolor{black}{Analogously to the univariate case, the asymptotic expression for $\mathscr{S}_\sslash$ can be derived upon Stirling-approximating the binomial coefficients appearing within it. This time, the recipe $n!\simeq\sqrt{2\pi n}\left(\frac{n}{e}\right)^n$ leads to}

\begin{equation}
\mathscr{S}_\sslash\simeq\sum_{l_\bullet\geq l^*_\bullet}\sum_{l_\circ\geq l^*_\circ}B(l_\bullet,l_\circ)\left[\frac{\text{Ber}(V,L,p)}{\prod_{i=\bullet,\circ,\top}\text{Ber}(V_i,l_i,p_i)}\right]
\label{eq11}
\end{equation}
where, as before, $\text{Ber}(x,y,z)=z^y(1-z)^{x-y}$ defines a Bernoulli probability mass function and the parameters read $p=\frac{L}{V}$ and $p_i=\frac{l_i}{V_i}$; the numerical coefficient appearing in front of the whole expression, now, reads

\begin{equation}
B(l_\bullet,l_\circ)=\frac{1}{2\pi}\sqrt{\frac{\sigma^2}{\prod_{i=\bullet,\circ,\top}\sigma^2_i}}
\end{equation}
with $\sigma^2=Vp(1-p)$ and $\sigma^2_i=V_ip_i(1-p_i)$. \textcolor{black}{The quantity $\mathscr{S}_\sslash$ compares the description of a networked configuration provided by the RGM, and encoded into the expression $\text{Ber}(V,L,p)=p^L(1-p)^{V-L}$, with the description of the same configuration provided by the SBM (now, defined by three - instead of two - different blocks), `represented' by the denominator of the expression defined in eq. (\ref{eq11}), i.e.}

\begin{widetext}
\begin{equation}
\prod_{i=\bullet,\circ,\top}\text{Ber}(V_i,l_i,p_i)=p_\bullet^{l_\bullet}(1-p_\bullet)^{V_\bullet-l_\bullet}\cdot p_\circ^{l_\circ}(1-p_\circ)^{V_\circ-l_\circ}\cdot p_\top^{l_\top}(1-p_\top)^{V_\top-l_\top}.
\end{equation}
\end{widetext}

\section*{DETECTION OF MESOSCALE STRUCTURES\\IN WEIGHTED NETWORKS}

\noindent\textcolor{black}{{\bf Community detection.} Within the surprise-based framework, the problem of detecting binary communities has been rephrased as an aleatory experiment whose random variable is the number of links within communities. Interestingly enough, such an experiment can be easily mapped into a counting problem, allowing us to interpret $\mathscr{S}$ as indicating the number of configurations whose number of `internal' links (i.e. within communities) is larger than the observed one.}

\textcolor{black}{When dealing with weighted networks, we would like to proceed along similar guidelines and consider the total, `internal' weight as our new random variable, to be redistributed across the available node pairs. Adopting this approach has four major consequences: 1) weights must be considered as composed by an integer number of binary links, 2) each node pair must be allowed to be occupied by more than one link and 3) the total weight must be allowed to vary even beyond the network size (when handling real-world networks, the case $W\gg V$ is often encountered).}

\textcolor{black}{In this case, the proper setting to define an aleatory experiment satisfying the requests above is provided by the so-called \emph{stars and bars} model, a combinatorial technique that has been introduced to handle the counting of configurations with multiple occupancies. Basically, the problem of counting in how many ways $w_\bullet$ particles (our links) can be redistributed among $V_\bullet$ boxes (our node pairs), while allowing more than one particle to occupy each box, can be tackled by allowing \emph{both} the particles \emph{and} the bars `delimiting' the boxes to be permuted \cite{Feller1950}. Since $V_\bullet$ boxes are delimited by $V_\bullet-1$ bars, a term like $\binom{V_\bullet+w_\bullet-1}{w_\bullet}$ is needed.}

\textcolor{black}{In order to better grasp the meaning of such a term, let us make a simple example. Let us imagine to observe a network with three nodes and two links, carrying a weight of 1 and 2, respectively. Now, were we interested in a purely binary analysis, we may ask ourselves in how many ways we could place the two links among the $\frac{N(N-1)}{2}=\frac{3(3-1)}{2}=3$ available pairs: the answer is provided by the `binary' binomial coefficient $\binom{V_\bullet}{l_\bullet}=\binom{3}{2}=3$. The implicit assumption we make is that the three links must not occupy the same node pairs - otherwise the total number of connections wouldn't be preserved.}

\textcolor{black}{This perspective changes from the purely weighted point of view. Since we are now interested in preserving just the total weight of our network, irrespectively of the number of connections it is placed upon, the number of admissible configurations amounts precisely at $\binom{V_\bullet+w_\bullet-1}{w_\bullet}=\binom{2+3}{3}=10$. Such a number is larger than before since, now, weights are `disaggregated' into binary links and multiple occupations of the latter ones are allowed.}

\textcolor{black}{The considerations above lead us to generalize the community detection problem to the weighted case by identifying}

\begin{eqnarray}
f(w_\bullet)&\equiv&\text{NH}(w_\bullet|V+W,W,V_\bullet)\nonumber\\
&=&\frac{\prod_{i=\bullet,\circ}\binom{V_i+w_i-1}{w_i}}{\binom{V+W-1}{W}}=\frac{\binom{V_\bullet+w_\bullet-1}{w_\bullet}\binom{V_\circ+w_\circ-1}{w_\circ}}{\binom{V+W-1}{W}}\nonumber\\
&=&\frac{\binom{V_\bullet+w_\bullet-1}{w_\bullet}\binom{(V-V_\bullet)+(W-w_\bullet)-1}{W-w_\bullet}}{\binom{V+W-1}{W}}
\end{eqnarray}
\textcolor{black}{i.e. by replacing the binomial hypergeometric distribution considered in the purely binary case with a \emph{negative hypergeometric distribution}, a choice inducing the definition of the \emph{weighted surprise}}

\begin{equation}
\mathscr{W}\equiv\sum_{w_\bullet\geq w_\bullet^*}f(w_\bullet)
\end{equation}
\textcolor{black}{where the binomial coefficient $\binom{V_\bullet+w_\bullet-1}{w_\bullet}=\binom{V_\bullet+w_\bullet-1}{V_\bullet-1}$ enumerates the number of ways $w_\bullet$ links can be redistributed \emph{within} communities, i.e. over the available $V_\bullet$ node pairs, and the binomial coefficient $\binom{V_\circ+w_\circ-1}{w_\circ}=\binom{V_\circ+w_\circ-1}{V_\circ-1}$ enumerates the number of ways the remaining $w_\circ=W-w_\bullet$ links can be redistributed \emph{between} communities, i.e. over the remaining $V_\circ=V-V_\bullet$ node pairs.} Differently from the binary case, the sum ranges up to the maximum empirical weight of the network, i.e. $w_\bullet\in[w_\bullet^*,W]$.\\

\noindent\textcolor{black}{{\bf Asymptotic results.} The asymptotic expression for $\mathscr{W}$ can be deduced by following the same reasoning that has allowed us to derive the asymptotic expression for $\mathscr{S}$. Stirling-approximating the binomial coefficients entering into the definition of $\mathscr{W}$ leads to the writing}

\begin{equation}
\mathscr{W}\simeq\sum_{w_\bullet\geq w^*_\bullet}C(w_\bullet)\left[\frac{\text{Geo}(V,W,q)}{\prod_{i=\bullet,\circ}\text{Geo}(V_i,w_i,q_i)}\right]
\label{eq8}
\end{equation}
where the expression

\begin{equation}
\text{Geo}(x,y,z)=z^y(1-z)^x
\end{equation}
\textcolor{black}{defines a \emph{geometric} probability mass function and the parameters appearing in eq. (\ref{eq8}) read $q=\frac{W}{V+W-1}$ and $q_i=\frac{w_i}{V_i+w_i-1}$.} In the weighted case, the Bernoulli probability mass function appearing in the asymptotic expression of $\mathscr{S}$ is replaced by a geometric probability mass function: this implies that (asymptotically) the comparison is, now, carried out between the description of a networked configuration provided by the Weighted Random Graph Model (WRGM), and encoded into the expression

\begin{equation}
\text{Geo}(V,W,q)=q^W(1-q)^V,
\end{equation}
with the description of the same configuration provided by the Weighted Stochastic Block Model (WSBM) and encoded into the expression

\begin{equation}
\prod_{i=\bullet,\circ}\text{Geo}(V_i,w_i,q_i)=q_\bullet^{w_\bullet}(1-q_\bullet)^{V_\bullet}\cdot q_\circ^{w_\circ}(1-q_\circ)^{V_\circ}
\end{equation}
\textcolor{black}{where $q_\bullet=\frac{w_\bullet}{V_\bullet+w_\bullet}$ and $q_\circ=\frac{w_\circ}{V_\circ+w_\circ}$. As for the binary case, the asymptotic expression of the weighted surprise clarifies that minimizing it amounts at finding the partition least likely to occour under the WRGM than under the WSBM.}

\textcolor{black}{As in the binary case, the parameters characterizing the geometric probability mass functions defining the asymptotic weighted surprise can be estimated via the maximum-of-the-likelihood principle, according to which the log-likelihood of the expression $\text{Geo}(x,y,z)$ reads}

\begin{equation}
\mathscr{L}=y\ln z+x\ln (1-z);
\end{equation}
\textcolor{black}{upon maximizing it with respect to $z$, one finds $z=\frac{y}{x+y}$: hence, one can pose $q\simeq\frac{W}{V+W}$ and $q_i\simeq\frac{w_i}{V_i+w_i}$.}

\textcolor{black}{The numerical coefficient appearing in front of the whole expression, instead, reads}

\begin{equation}
C(w_\bullet)=\sqrt{\frac{\mu}{2\pi\prod_{i=\bullet,\circ}\mu_i}}
\end{equation}
with $\mu\simeq Vq$ and $\mu_i\simeq V_iq_i$.\\

\noindent\textcolor{black}{{\bf `Bimodular' structures detection.} Let us now introduce the third generalization of the suprise-based formalism: following the same line of reasoning that led us to approach the detection of binary `bimodular' structures by considering the multinomial analogoue of the distribution introduced for binary community detection, we focus on the \emph{multinomial} (or \emph{multivariate}) \emph{negative hypergeometric distribution}, i.e.}

\begin{eqnarray}
f(w_\bullet,w_\circ)&\equiv&\text{MNH}(w_\bullet,w_\circ|V+W,W,V_\bullet,V_\circ)\nonumber\\
&=&\frac{\prod_{i=\bullet,\circ,\top}\binom{V_i+w_i-1}{w_i}}{\binom{V+W-1}{W}}\nonumber\\
&=&\frac{\binom{V_\bullet+w_\bullet-1}{w_\bullet}\binom{V_\circ+w_\circ-1}{w_\circ}\binom{V_\top+w_\top-1}{w_\top}}{\binom{V+W-1}{W}}\nonumber\\
&=&\frac{\binom{V_\bullet+w_\bullet-1}{w_\bullet}\binom{V_\circ+w_\circ-1}{w_\circ}\binom{V-(V_\bullet+V_\circ)+W-(w_\bullet+w_\circ)-1}{W-(w_\bullet+w_\circ)}}{\binom{V+W-1}{W}};\nonumber\\
\end{eqnarray}
while the binomial coefficient $\binom{V_\bullet+w_\bullet-1}{w_\bullet}$ enumerates the number of ways $w_\bullet$ links can redistributed \emph{within} the first module (e.g. the core portion), the binomial coefficient $\binom{V_\circ+w_\circ-1}{w_\circ}$ enumerates the number of ways $w_\circ$ links can be redistributed \emph{within} the second module (e.g. the periphery portion) and the binomial coefficient $\binom{V-(V_\bullet+V_\circ)+W-(w_\bullet+w_\circ)-1}{W-(w_\bullet+w_\circ)}$ enumerates the number of ways the remaining $w_\top\equiv W-(w_\bullet+w_\circ)$ links can be redistributed \emph{between} the first and the second module, i.e. over the remaining $V_\top\equiv V-(V_\bullet+V_\circ)$ node pairs. \textcolor{black}{Such a position induces the definition of the \emph{weighted bimodular surprise}}

\begin{equation}
\mathscr{W}_\sslash\equiv\sum_{w_\bullet\geq w_\bullet^*}\sum_{w_\circ\geq w_\circ^*}f(w_\bullet,w_\circ);
\end{equation}
as for the (weighted) community detection, weights are understood as integer numbers - equivalently, as composed by an integer number of binary links. \textcolor{black}{For what concerns the limits of the summations, $w_\bullet$ and $w_\circ$ are `naturally' bounded by $W$; notice, however, that the sum $w_\bullet+w_\circ$ itself cannot exceed such a value.}\\

\noindent{\textcolor{black}{{\bf Asymptotic results.} Let us now derive the asymptotic expression for $\mathscr{W}_\sslash$. As usual, let us Stirling-approximate the binomial coefficients entering into the definition of $\mathscr{W}_\sslash$; such a simplification leads us to the expression}}

\begin{equation}
\mathscr{W}_\sslash\simeq\sum_{w_\bullet\geq w^*_\bullet}\sum_{w_\circ\geq w^*_\circ}D(w_\bullet,w_\circ)\left[\frac{\text{Geo}(V,W,q)}{\prod_{i=\bullet,\circ,\top}\text{Geo}(V_i,w_i,q_i)}\right]
\label{eq39}
\end{equation}
\textcolor{black}{where, as before, $\text{Geo}(x,y,z)=z^y(1-z)^x$ defines a geometric probability mass function and the parameters read $q=\frac{W}{V+W-1}$ and $q_i=\frac{w_i}{V_i+w_i-1}$ but can be approximated, according to the maximum-of-the-likelihood principle, as $q\simeq\frac{W}{V+W}$ and $q_i\simeq\frac{w_i}{V_i+w_i}$; the numerical coefficient multiplying the whole expression reads}

\begin{equation}
D(w_\bullet,w_\circ)=\frac{1}{2\pi}\sqrt{\frac{\mu}{\prod_{i=\bullet,\circ,\top}\mu_i}}
\end{equation}
with $\mu\simeq Vq$ and $\mu_i\simeq V_iq_i$. \textcolor{black}{The analysis of $\mathscr{W}_\sslash$ in the asymptotic regime reveals that it compares the description of a networked configuration provided by the WRGM, and encoded into the expression $\text{Geo}(V,W,q)=q^W(1-q)^V$, with the description of the same configuration provided by the WSBM (now, defined by three - instead of two - different blocks), `represented' by the denominator of the expression defined in eq. (\ref{eq39}), i.e.}

\begin{widetext}
\begin{equation}
\prod_{i=\bullet,\circ,\top}\text{Geo}(V_i,w_i,q_i)=q_\bullet^{w_\bullet}(1-q_\bullet)^{V_\bullet}\cdot q_\circ^{w_\circ}(1-q_\circ)^{V_\circ}\cdot q_\top^{w_\top}(1-q_\top)^{V_\top}.
\end{equation}
\end{widetext}

\begin{figure*}[t!]
\includegraphics[width=0.5\textwidth]{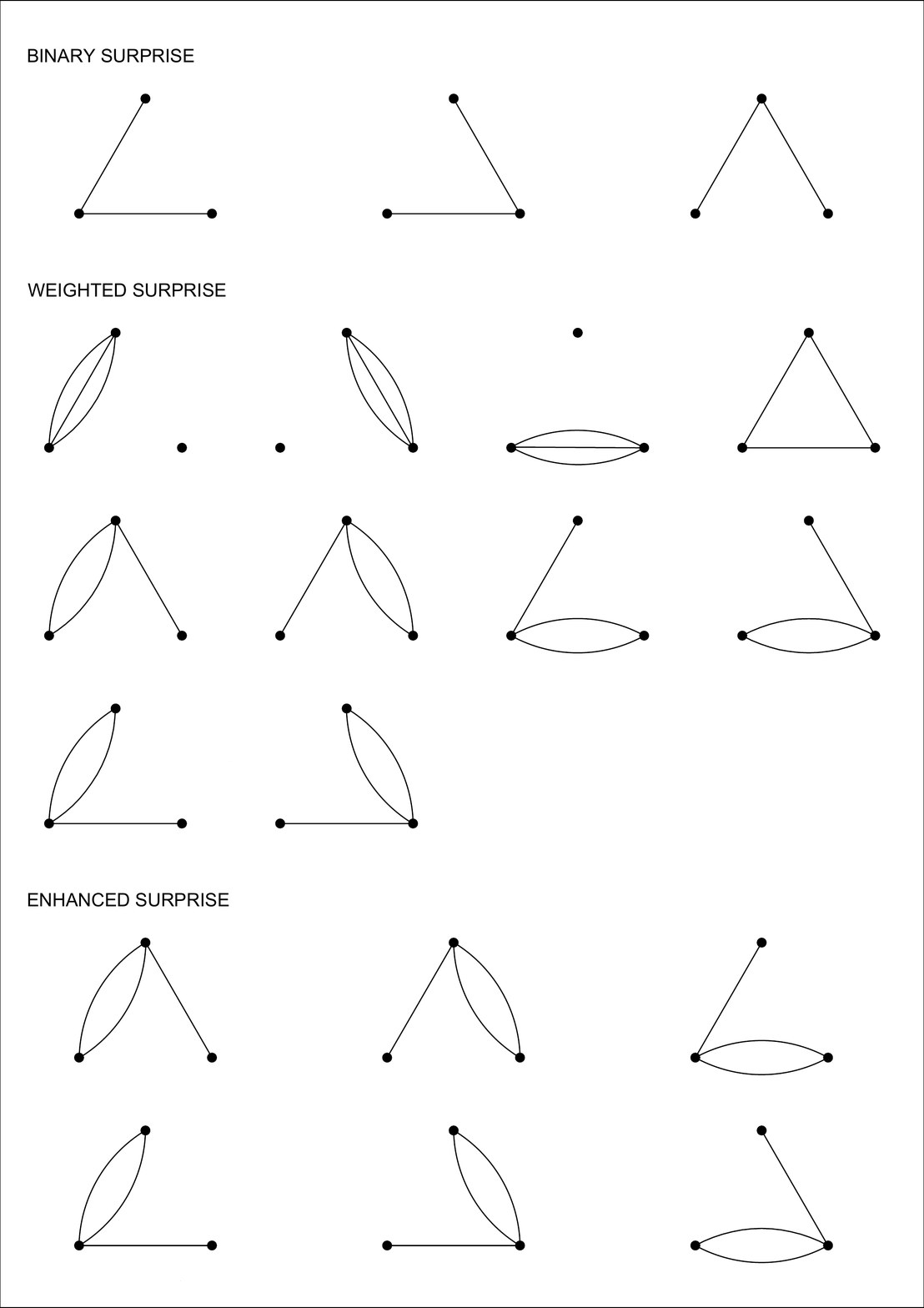}
\caption{Graphical comparison of the three different ways of counting the admissible configurations when dealing with the purely binary, the purely \textcolor{black}{weighted} and the enhanced surprise. Let us imagine to observe a network with three nodes and two links, carrying a weight of 1 and 2, respectively. Were we interested in a purely binary analysis, we may ask ourselves in how many ways we could place the two links among the $3$ available pairs: the answer is provided by the `binary' binomial coefficient $\binom{V_\bullet}{l_\bullet}=\binom{3}{2}=3$. Were we interested in a purely weighted analysis, we may ask ourselves in how many ways we could place the two links among the $3$ available pairs while preserving the total weight of our network, irrespectively of the number of connections it is placed upon; the number of admissible configurations becomes $\binom{V_\bullet+w_\bullet-1}{w_\bullet}=\binom{2+3}{3}=10$. Such a number is larger than before since, now, weights are `disaggregated' into binary links and multiple occupations of the latter ones are allowed. Were we interested in the enhanced analysis, we may ask ourselves in how many ways we could place the two links among the $3$ available pairs while preserving both the total number of links and the total weight of the network; the number of admissible configurations becomes $\binom{V_\bullet}{l_\bullet}\binom{w_\bullet-1}{w_\bullet-l_\bullet}=\binom{3}{2}\binom{3-1}{3-2}=3\cdot2=6$, as can be easily verified upon explicitly listing them.}
\label{fig1}
\end{figure*}

\section*{ENHANCED DETECTION OF MESOSCALE\\STRUCTURES IN NETWORKS}

\noindent{\textcolor{black}{{\bf Community detection.} The recipe to detect communities on weighted networks can be further refined to account for the information encoded into the total number of links, beside the one provided by the total weight. Generally speaking, this can be realized by `combining' two of the distributions introduced above. To this aim, let us proceed in a two-step fashion: first, let us recall that the number of ways $L$ links can be placed among $V$ node pairs, in such a way that $l_\bullet$ connections are `internal' to the clusters while the remaining $L-l_\bullet$ ones are, instead, `external' is precisely}}

\begin{equation}
\text{H}(l_\bullet|V,V_\bullet,L)=\frac{\binom{V_\bullet}{l_\bullet}\binom{V-V_\bullet}{L-l_\bullet}}{\binom{V}{L}};
\end{equation}
\textcolor{black}{now, for each of the binary configurations listed above, $W-L$ links remain to be assigned: while $w_\bullet-l_\bullet$ of them must be placed within the clusters, on top of the $l_\bullet$ available `internal' links, the remaining $(W-L)-(w_\bullet-l_\bullet)$ ones must be placed between the clusters, on top of the $L-l_\bullet$ available inter-cluster connections. Hence, the `conditional' negative hypergeometric distribution reading}

\begin{widetext}
\begin{equation}
\text{NH}(w_\bullet|W,W-L,l_\bullet)=\frac{\binom{l_\bullet+(w_\bullet-l_\bullet)-1}{w_\bullet-l_\bullet}\binom{(L-l_\bullet)+(W-L)-(w_\bullet-l_\bullet)-1}{(W-L)-(w_\bullet-l_\bullet)}}{\binom{L+(W-L)-1}{W-L}}
\label{eq14}
\end{equation}
\end{widetext}
\textcolor{black}{remains naturally defined; now, `combining' the two distributions above, simplifying and re-arranging, the generic term of the \emph{enhanced hypergeometric distribution} can be rewritten as}

\begin{widetext}
\begin{eqnarray}
\text{EH}(l_\bullet,w_\bullet|V,V_\bullet,L,W)&=&\text{H}(l_\bullet|V,V_\bullet,L)\cdot\text{NH}(w_\bullet|W,W-L,l_\bullet)\nonumber\\
&=&\frac{\binom{V_\bullet}{l_\bullet}\binom{V_\circ}{l_\circ}}{\binom{V}{L}}\cdot\frac{\binom{w_\bullet-1}{w_\bullet-l_\bullet}\binom{w_\circ-1}{w_\circ-l_\circ}}{\binom{W-1}{W-L}}
\end{eqnarray}
\end{widetext}
\textcolor{black}{with a clear meaning of the symbols. An analytical characterization of it is provided into the Appendices: for the moment, let us simply notice that the definition provided above works for the values $0<l_\bullet<L$. By posing $f(l_\bullet,w_\bullet)\equiv\text{EH}(l_\bullet,w_\bullet|V,V_\bullet,L,W)$, our novel distribution induces the definition of the \emph{enhanced surprise}, i.e.}

\begin{equation}
\mathscr{E}\equiv\sum_{l_\bullet\geq l_\bullet^*}\sum_{w_\bullet\geq w_\bullet^*}f(l_\bullet,w_\bullet);
\end{equation}
\textcolor{black}{although $l_\bullet$ and $w_\bullet-l_\bullet$ are `naturally' bounded by $V_\bullet$ and $W-L$, respectively, the former one cannot exceed $L$.}\\

\textcolor{black}{In order to better understand how the enhanced surprise works, let us consider again the aforementioned example: given a network with three nodes and two links, carrying a weight of 1 and 2, respectively, we observe $\binom{V_\bullet}{l_\bullet}=\binom{3}{2}=3$ (purely binary) configurations with exactly the same number of links and $\binom{V_\bullet+w_\bullet-1}{w_\bullet}=\binom{2+3}{3}=10$ (purely weighted) configurations with exactly the same total weight. If we, now, constrain both the total number of links and the total weight of the network, the number of admissible configurations becomes $\binom{V_\bullet}{l_\bullet}\binom{w_\bullet-1}{w_\bullet-l_\bullet}=\binom{3}{2}\binom{3-1}{3-2}=3\cdot2=6$, as it can be easily verified upon explicitly listing them. Naturally, the configurations `admissible' by the enhanced surprise are a subset of the configurations `admissible' by the weighted surprise, i.e. precisely the ones with the desired number of links (see also fig. \ref{fig1}).}\\

\noindent\textcolor{black}{{\bf Asymptotic results.} Analogously to the other functionals, the asymptotic expression of $\mathscr{E}$ can be derived by Stirling-approximating the binomial coefficients entering into its definition as well:}

\begin{equation}
\mathscr{E}\simeq\sum_{l_\bullet\geq l_\bullet^*}\sum_{w_\bullet\geq w_\bullet^*}E(l_\bullet,w_\bullet)\left[\frac{\text{BF}(V,L,W,p,r)}{\prod_{i=\bullet,\circ}\text{BF}(V_i,l_i,w_i,p_i,r_i)}\right];
\label{eq17}
\end{equation}
\textcolor{black}{in the formula above, the expression}

\begin{eqnarray}
\text{BF}(x,y,u,z,t)&=&z^y(1-z)^{x-y}\cdot t^{u-y}(1-t)^y\nonumber\\
&=&\text{Ber}(x,y,z)\cdot\text{Geo}(y,u-y,t)
\end{eqnarray}
\textcolor{black}{defines a \emph{Bose-Fermi} probability mass function \cite{Garlaschelli2009}. While a Bernoulli probability mass function characterizes the asymptotic behavior of the purely binary surprise and a geometric probability mass function characterizes the asymptotic behavior of the purely weighted surprise, the enhanced surprise is asymptotically characterized by a distribution whose functional form is halfway between the two previous ones: while its Bernoulli-like portion controls for the `presence' of the links, its `conditional' geometric-like portion controls for the magnitude of the `remaining' weights. To stress its `mixed' character, such a distribution has been named `Bose-Fermi': remarkably, it can be retrieved within the Exponential Random Graphs framework as a consequence of Shannon entropy maximization, constrained to simultaneously reproduce both the total number of links and the total weight of a network \cite{Garlaschelli2009,Mastrandrea2014}.}

\textcolor{black}{The parameters appearing in eq. (\ref{eq17}) read $p=\frac{L}{V}$, $p_i=\frac{l_i}{V_i}$ and $r=\frac{W-L}{W-1}$, $r_i=\frac{w_i-l_i}{w_i-1}$; while the first class of parameters can be tuned according to the maximum-of-the-likelihood principle, the ones belonging to the second class can be approximated according to the same recipe. To see this explicitly, let us consider $\text{BF}(x,y,u,z,t)$, whose log-likelihood reads}

\begin{equation}
\mathscr{L}=y\ln z+(x-y)\ln(1-z)+(u-y)\ln t+y\ln(1-t);
\end{equation}
\textcolor{black}{upon maximizing it with respect to $z$, one finds $z=\frac{y}{x}$; upon maximizing it with respect to $t$ one finds $t=\frac{u-y}{u}$ - hence, one can pose $r\simeq\frac{W-L}{W}$, $r_i\simeq\frac{w_i-l_i}{w_i}$.}

\textcolor{black}{The numerical coefficient multiplying the whole expression is defined as}

\begin{equation}
E(l_\bullet,w_\bullet)=\frac{1}{2\pi}\sqrt{\frac{\sigma^2\mu}{\prod_{i=\bullet,\circ}\sigma^2_i\mu_i}}
\end{equation}
\textcolor{black}{where $\sigma^2=Vp(1-p)$, $\sigma^2_i=V_ip_i(1-p_i)$, $\mu\simeq Lr$ and $\mu_i\simeq l_ir_i$.}

\textcolor{black}{Similarly to the other cases, the asymptotic expression of the enhanced suprise compares the description of a networked configuration provided by the Enhanced Random Graph Model (ERGM), and encoded into the expression $\text{BF}(V,L,W,p,r)=p^L(1-p)^{V-L}\cdot r^{W-L}(1-r)^L$, with the description of the same configuration provided by its block-wise counterpart, i.e. the Enhanced Stochastic Block Model (ESBM) and encoded into the expression $\prod_{i=\bullet,\circ}\text{BF}(V_i,l_i,w_i,p_i,r_i)$.}\\

\noindent{\textcolor{black}{{\bf `Bimodular' structures detection.} The last generalization of surprise concerns its use for the detection of `bimodular' structures within the enhanced framework. This amounts at considering the following `multinomial variant' of the enhanced hypergeometric distribution}}

\begin{widetext}
\begin{eqnarray}
\text{MEH}(l_\bullet,l_\circ,w_\bullet,w_\circ|V,V_\bullet,V_\circ,L,W)
&=&\frac{\binom{V_\bullet}{l_\bullet}\binom{V_\circ}{l_\circ}\binom{V_\top}{l_\top}}{\binom{V}{L}}\cdot\frac{\binom{w_\bullet-1}{w_\bullet-l_\bullet}\binom{w_\circ-1}{w_\circ-l_\circ}\binom{w_\top-1}{w_\top-l_\top}}{\binom{W-1}{W-L}}\nonumber\\
&=&\frac{\binom{V_\bullet}{l_\bullet}\binom{V_\circ}{l_\circ}\binom{V-(V_\bullet+V_\circ)}{L-(l_\bullet+l_\circ)}}{\binom{V}{L}}\cdot\frac{\binom{w_\bullet-1}{w_\bullet-l_\bullet}\binom{w_\circ-1}{w_\circ-l_\circ}\binom{W-(w_\bullet+w_\circ)-1}{(W-L)-((w_\bullet+w_\circ)-(l_\bullet+l_\circ))}}{\binom{W-1}{L-1}}
\end{eqnarray}
\end{widetext}
\textcolor{black}{where $V_\top\equiv V-(V_\bullet+V_\circ)$ indicates the number of node pairs between the modules $\bullet$ and $\circ$ and $l_\top\equiv L-(l_\bullet+l_\circ)$ indicates the number of links that must be assigned therein. An analytical characterization of it is provided into the Appendices: for the moment, let us simply notice that the definition provided above works for the values $0<l_\bullet, l_\circ<L$. The position $f(l_\bullet,l_\circ,w_\bullet,w_\circ)\equiv\text{MEH}(l_\bullet,l_\circ,w_\bullet,w_\circ|V,V_\bullet,V_\circ,L,W)$ induces the definition of the \emph{enhanced bimodular surprise}}

\begin{equation}
\mathscr{E}_\sslash=\sum_{l_\bullet\geq l_\bullet^*}\sum_{l_\circ\geq l_\circ^*}\sum_{w_\bullet\geq w_\bullet^*}\sum_{w_\circ\geq w_\circ^*}f(l_\bullet,l_\circ,w_\bullet,w_\circ).
\end{equation}
\textcolor{black}{Notice that $l_\bullet$ and $l_\circ$ are `naturally' bounded by $V_\bullet$ and $V_\circ$: still, their sum cannot exceed $L$; analogously, $w_\bullet-l_\bullet$ and $w_\circ-l_\circ$ are `naturally' bounded by $W-L$: still, their sum itself cannot exceed $W-L$.}

\textcolor{black}{As for its binomial counterpart, the expression of the MEH can be rearranged in a term-by-term fashion, in such a way that the module-specific binomial coefficients can be grouped together. Upon doing so, it becomes clearer that the MEH counts the number of ways $w_\bullet-l_\bullet$ links can be placed on top of the $l_\bullet$ binary links characterizing the connectance of the $\bullet$ module, times the number of ways $w_\circ-l_\circ$ links can be placed on top of the $l_\circ$ binary links characterizing the connectance of the $\circ$ module, times the number of ways the remaining $W-(w_\bullet+w_\circ)-(L-(l_\bullet+l_\circ))$ links can be placed on top of the $L-(l_\bullet+l_\circ)$ binary links characterizing the connectance of the third module.}\\

\begin{center}
\begin{table*}[t!]
\begin{tabular}{c|c|c}
\hline
\hline
\text{Full expression} & \text{Community detection} & \text{`Bimodular' structures detection} \\
\hline
\text{Binary case} & $f=\frac{\prod_{i=\bullet,\circ}\binom{V_i}{l_i}}{\binom{V}{L}}$ & $f=\frac{\prod_{i=\bullet,\circ,\top}\binom{V_i}{l_i}}{\binom{V}{L}}$ \\
\hline
\text{Weighted case} & $f=\frac{\prod_{i=\bullet,\circ}\binom{V_i+w_i-1}{w_i}}{\binom{V+W-1}{W}}=\frac{\prod_{i=\bullet,\circ}\multiset{V_i}{w_i}}{\multiset{V}{W}}$ & $f=\frac{\prod_{i=\bullet,\circ,\top}\binom{V_i+w_i-1}{w_i}}{\binom{V+W-1}{W}}=\frac{\prod_{i=\bullet,\circ,\top}\multiset{V_i}{w_i}}{\multiset{V}{W}}$
\\
\hline
\text{Enhanced case} & $f=\frac{\prod_{i=\bullet,\circ}\binom{V_i}{l_i}\binom{w_i-1}{l_i-1}}{\binom{V}{L}\binom{W-1}{W-L}}$ & $f=\frac{\prod_{i=\bullet,\circ,\top}\binom{V_i}{l_i}\binom{w_i-1}{l_i-1}}{\binom{V}{L}\binom{W-1}{W-L}}$ \\
\hline
\hline
\text{Asymptotic expression} & \text{Community detection} & \text{`Bimodular' structures detection} \\
\hline
\text{Binary case} & $f\propto\frac{\text{Ber}(V,L,p)}{\prod_{i=\bullet,\circ}\text{Ber}(V_i,l_i,p_i)}$ & $f\propto\frac{\text{Ber}(V,L,p)}{\prod_{i=\bullet,\circ,\top}\text{Ber}(V_i,l_i,p_i)}$ \\
\hline
\text{Weighted case} & $f\propto\frac{\text{Geo}(V,W,q)}{\prod_{i=\bullet,\circ}\text{Geo}(V_i,w_i,q_i)}$ & $f\propto\frac{\text{Geo}(V,W,q)}{\prod_{i=\bullet,\circ,\top}\text{Geo}(V_i,w_i,q_i)}$ \\
\hline
\text{Enhanced case} & $f\propto\frac{\text{BF}(V,L,W,p,r)}{\prod_{i=\bullet,\circ}\text{BF}(V_i,l_i,w_i,p_i,r_i)}$ & $f\propto\frac{\text{BF}(V,L,W,p,r)}{\prod_{i=\bullet,\circ,\top}\text{BF}(V_i,l_i,w_i,p_i,r_i)}$ \\
\hline
\end{tabular}
\caption{Table illustrating all the generalizations of the surprise-based formalism proposed in the present paper and showing both the full and the asymptotic expression of each probability mass function - with the only exception of the numerical coefficient characterizing each asymptotic expression. To sum up, detecting a weighted mesoscale structure implies considering the \emph{negative} version of the probability mass function working in the corresponding binary case (e.g. moving from the hypergeometric to the negative hypergeometric one); detecting a `bimodular' structure, instead, implies considering the \emph{multinomial} version of the probability mass function working in the corresponding binary case (e.g. moving from the binomial hypergeometric to the multinomial one). Upon employing the multiset notation, a nice formal symmetry can be recovered between the purely binary and the purely weighted cases.}
\label{tab1}
\end{table*}
\end{center}

\noindent\textcolor{black}{{\bf Asymptotic results.} The enhanced bimodular surprise admits an asymptotic expression as well, i.e.}

\begin{widetext}
\begin{equation}
\label{eq64}
\mathscr{E}_\sslash\simeq\sum_{l_\bullet\geq l_\bullet^*}\sum_{l_\circ\geq l_\circ^*}\sum_{w_\bullet\geq w_\bullet^*}\sum_{w_\circ\geq w_\circ^*}F(l_\bullet,l_\circ,w_\bullet,w_\circ)\left[\frac{\text{BF}(V,L,W,p,r)}{\prod_{i=\bullet,\circ,\top}\text{BF}(V_i,l_i,w_i,p_i,r_i)}\right]
\end{equation}
\end{widetext}
\textcolor{black}{that can be recovered by Stirling-approximating the binomial coefficients entering into the definition of $\mathscr{E}_\sslash$. While the expression \text{BF} denotes a Bose-Fermi probability mass function \cite{Garlaschelli2009}, the parameters appearing in eq. (\ref{eq64}) read $p=\frac{L}{V}$, $p_i=\frac{l_i}{V_i}$ and $r=\frac{W-L}{W-1}$, $r_i=\frac{w_i-l_i}{w_i-1}$; as for $\mathscr{E}$, the maximum-of-the-likelihood principle determines (approximates) the first (second) class of parameters.}

\textcolor{black}{The numerical coefficient multiplying the whole expression is defined as}

\begin{equation}
F(l_\bullet,l_\circ,w_\bullet,w_\circ)=\frac{1}{(2\pi)^2}\sqrt{\frac{\sigma^2\mu}{\prod_{i=\bullet,\circ,\top}\sigma^2_i\mu_i}}
\end{equation}
\textcolor{black}{where $\sigma^2=Vp(1-p)$, $\sigma^2_i=V_ip_i(1-p_i)$, $\mu\simeq Lr$ and $\mu_i\simeq l_ir_i$. As observed for the other functionals, the asymptotic expression of the enhanced bimodular suprise compares the description of a networked configuration provided by the ERGM, and encoded into the expression $\text{BF}(V,L,W,p,r)=p^L(1-p)^{V-L}\cdot r^{W-L}(1-r)^L$, with the description of the same configuration provided by its block-wise counterpart, i.e. the ESBM (now, defined by three - instead of two - different blocks), `represented' by the denominator of the expression defined in eq. (\ref{eq64}), i.e. $\prod_{i=\bullet,\circ,\top}\text{BF}(V_i,l_i,w_i,p_i,r_i)$}\\

\textcolor{black}{Table \ref{tab1} gathers all the variants of the surprise-based formalism, illustrating both the full and the asymptotic expression for each of them. To sum up, detecting a weighted mesoscale structure implies considering the \emph{negative} version of the probability mass function working in the corresponding binary case (e.g. moving from the hypergeometric to the negative hypergeometric one); detecting a `bimodular' structure, instead, implies considering the \emph{multinomial} version of the probability mass function working in the corresponding binary case (e.g. moving from the binomial hypergeometric to the multinomial one).}

\begin{figure*}[t!]
\includegraphics[width=0.3\textwidth]{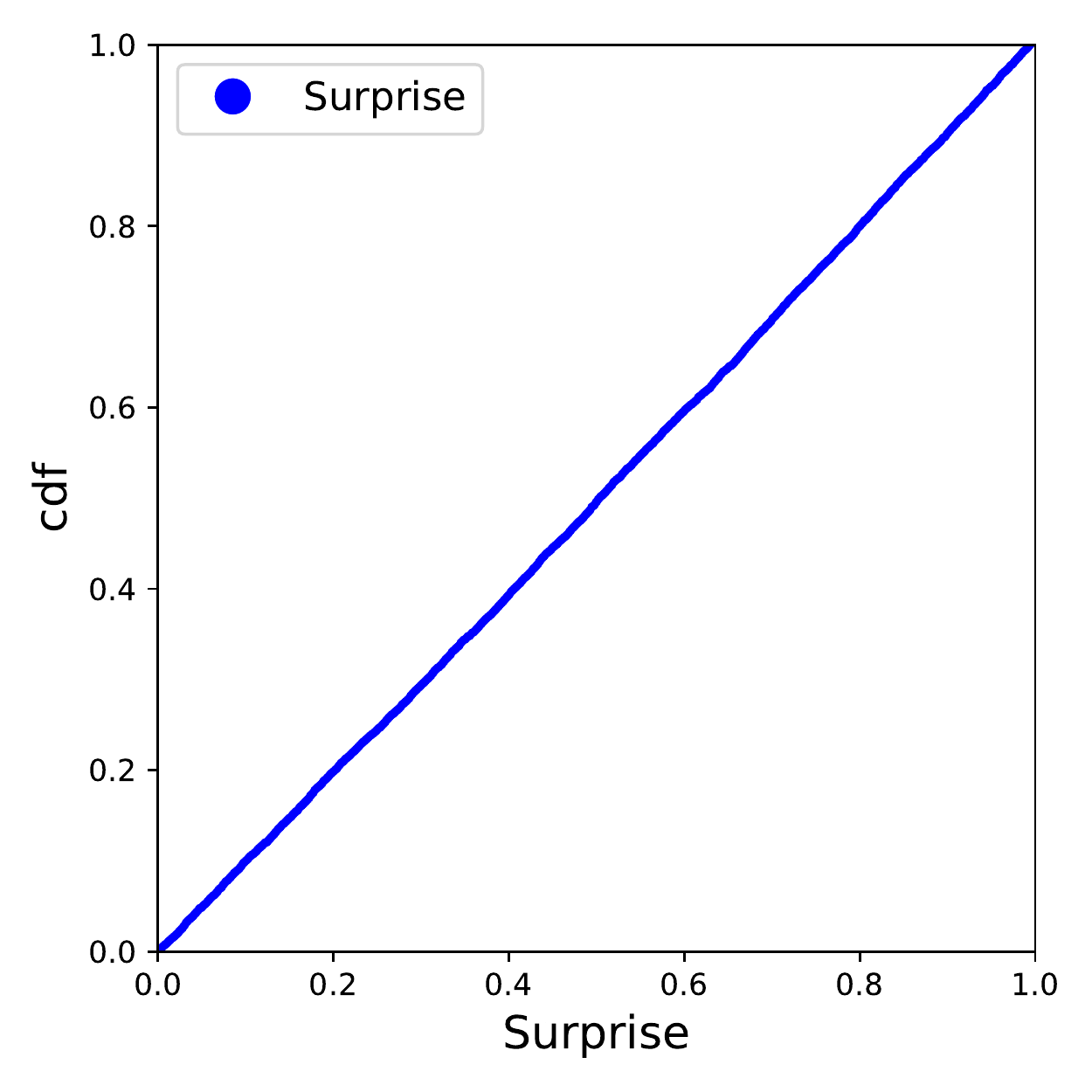}
\hspace{5mm}
\includegraphics[width=0.3\textwidth]{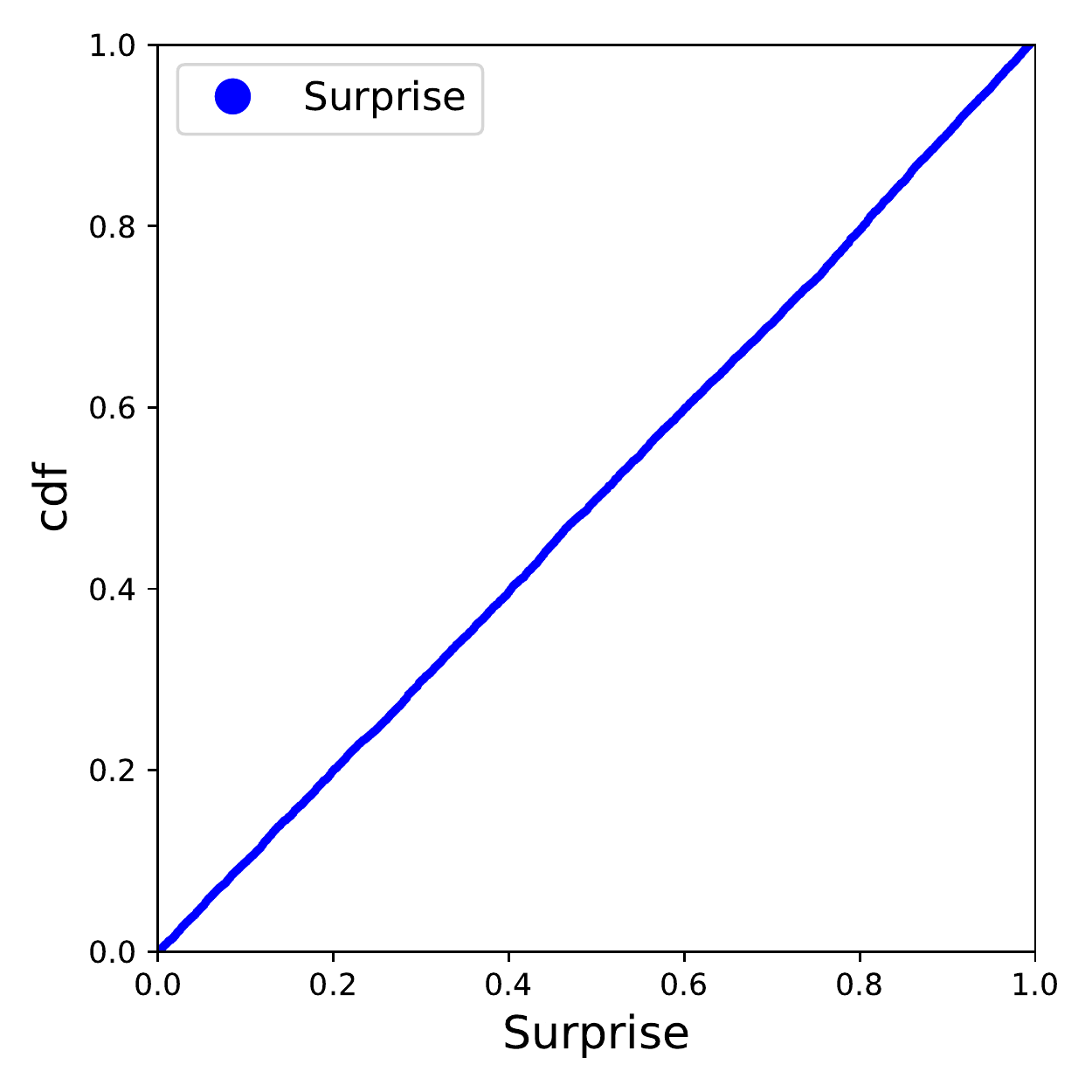}
\hspace{5mm}
\includegraphics[width=0.3\textwidth]{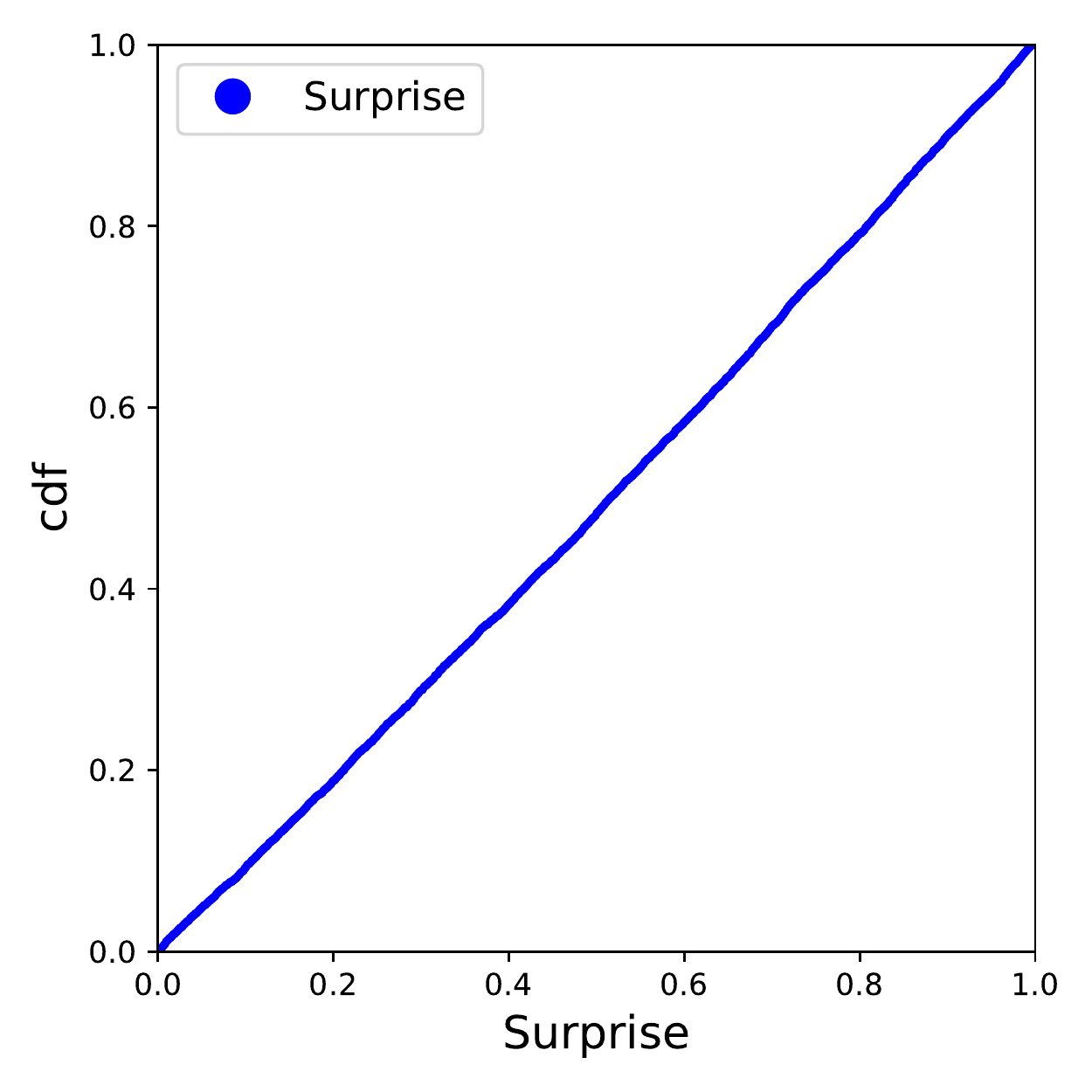}\\
\includegraphics[width=0.3\textwidth]{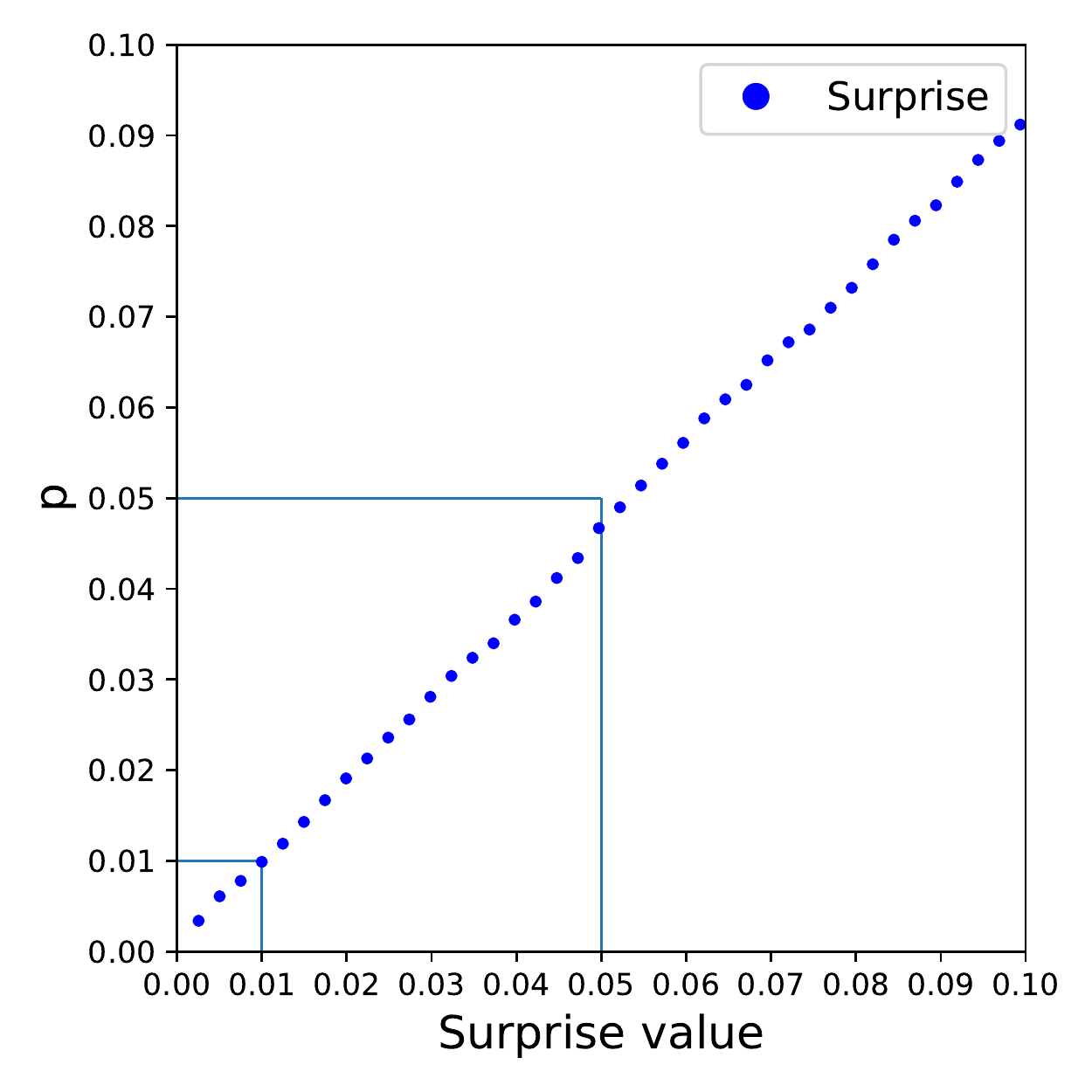}
\hspace{5mm}
\includegraphics[width=0.3\textwidth]{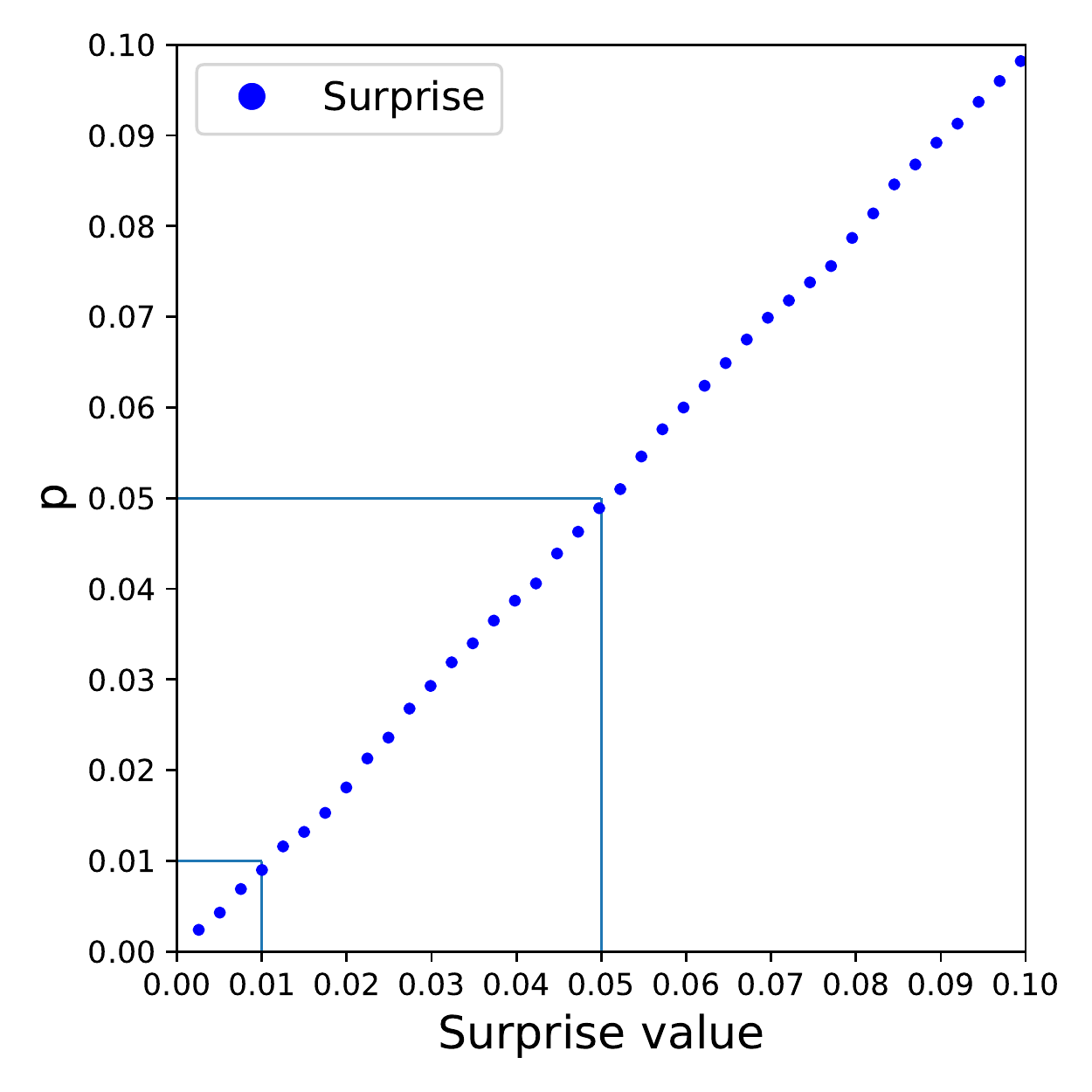}
\hspace{5mm}
\includegraphics[width=0.3\textwidth]{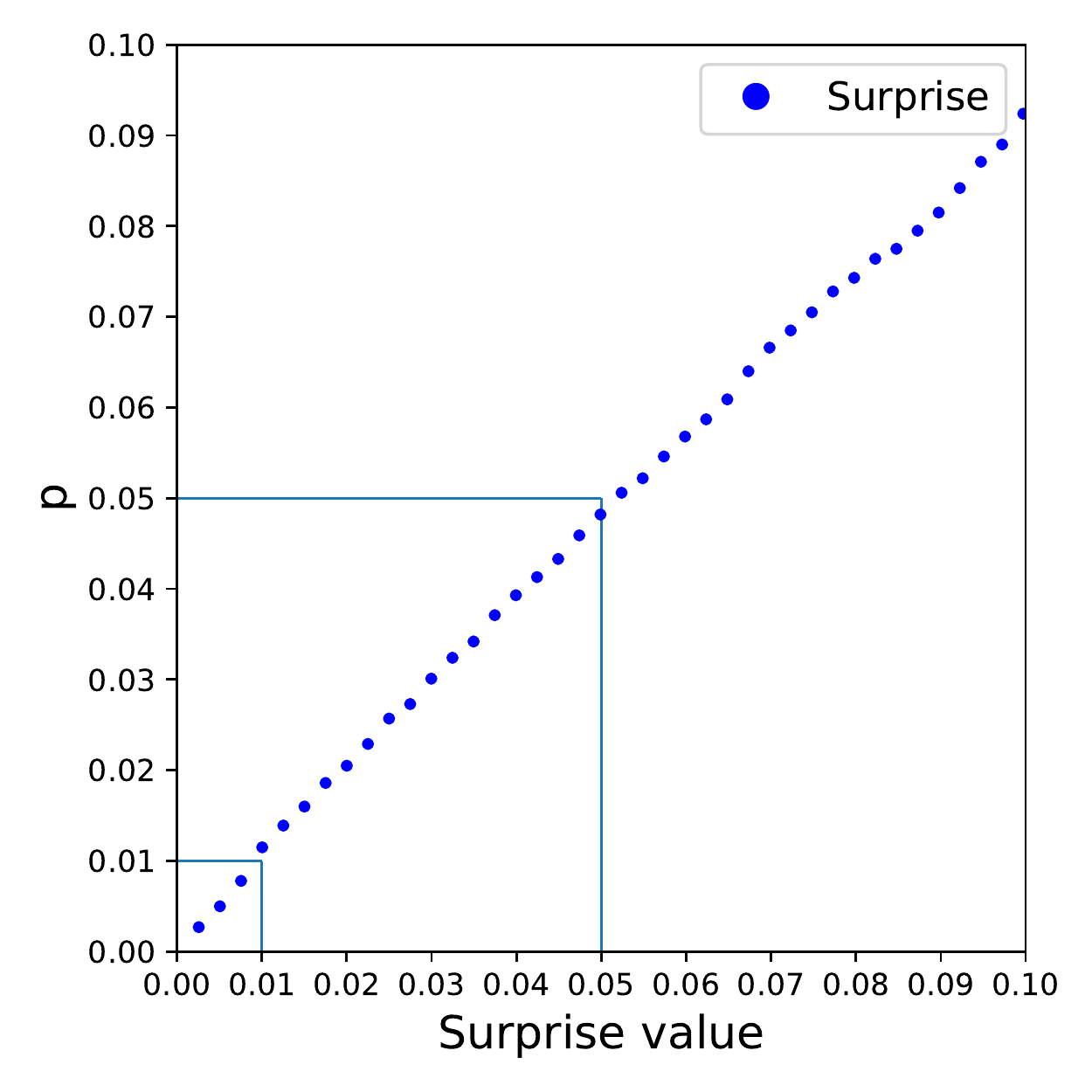}
\caption{Consistency check of our surprise-based formalism in the simplest case of community detection on binary networks. Any exact statistical test is characterized by a parameter, usually denoted with $\alpha$ and known as the \emph{type I error rate}, that quantifies the percentage of times the test provides a \emph{false positive}: it can be kept below a given threshold by adjusting the threshold of the p-value of the corresponding test, e.g. by deeming its response as significant in case it is less than $0.05$. Over 10.000 networks sampled by the RGM with $p=0.2$, a given planted partition - from left to right: two, three and four clusters with different dimensions - is indeed recovered as significant $5\%$ of the times. Notably, since the three upper panels show the CDF of the empirical values of $\mathscr{S}$, they also provide an information about its distribution, which is uniform over the unit interval.}
\label{fig2}
\end{figure*}

\section*{RESULTS}

\textcolor{black}{The previous sections have been devoted to the description of the surprise-based formalism for detecting a number of mesoscale structures; let us now test it on a bunch of synthetic and real-world configurations.}\\

\noindent{\textcolor{black}{{\bf Consistency checks.} Let us start by checking the consistency of our surprise-based formalism. Since we have rephrased the problem of detecting any mesoscale structure into an exact significance test, the limitations of our formalism are the same ones affecting the tests of the kind. More precisely, any significance test is characterized by a parameter known as \emph{type I error rate} and usually denoted with $\alpha$: it quantifies the percentage of times the considered test provides a \emph{false positive}. In other words, any statistical test is known to `fail'; in our case, this amounts at recognizing that a significant mesoscale structure can be detected, \emph{in case there is none}, a percentage $\alpha$ of the times: this number can be kept `small' by adjusting the threshold of the p-value of the corresponding test - e.g. by deeming its response as significant in case it is less than $0.05$.}}

\textcolor{black}{Remarkably, the aforementioned behavior can be explicitly tested for each of the cases considered above. Whenever exercises like these are carried out, it is of utmost importance to be as clear as possible about the null hypothesis tested: here, we aim at testing whether \emph{a given partition} is significant or not when the null hypothesis is true. For the sake of illustration, let us consider the problem of detecting communities on binary networks: as we learnt from the asymptotic expression of $\mathscr{S}$, it compares the description of a network provided by the RGM with that of the same network provided by the SBM, thus suggesting the RGM as the model playing the role of $H_0$. Hence, let us generate many networks from the RGM and, for each of them, let us impose a partition - the same for each sampled configuration, over which our SBM is tuned; finally, let us calculate $\mathscr{S}$ on each of them.}

\textcolor{black}{The results of such an experiment are shown in fig. \ref{fig2}: over 10.000 networks sampled from the RGM, whose only parameter has been set to $p=0.2$, the (given) planted partition - i.e. two, three and four clusters with different dimensions - is, indeed, recovered as significant $5\%$ of the times; notice that such a result holds true irrespectively of the details of the partition imposed on the sampled configurations.}

\begin{figure*}[t!]
\includegraphics[width=0.3\textwidth]{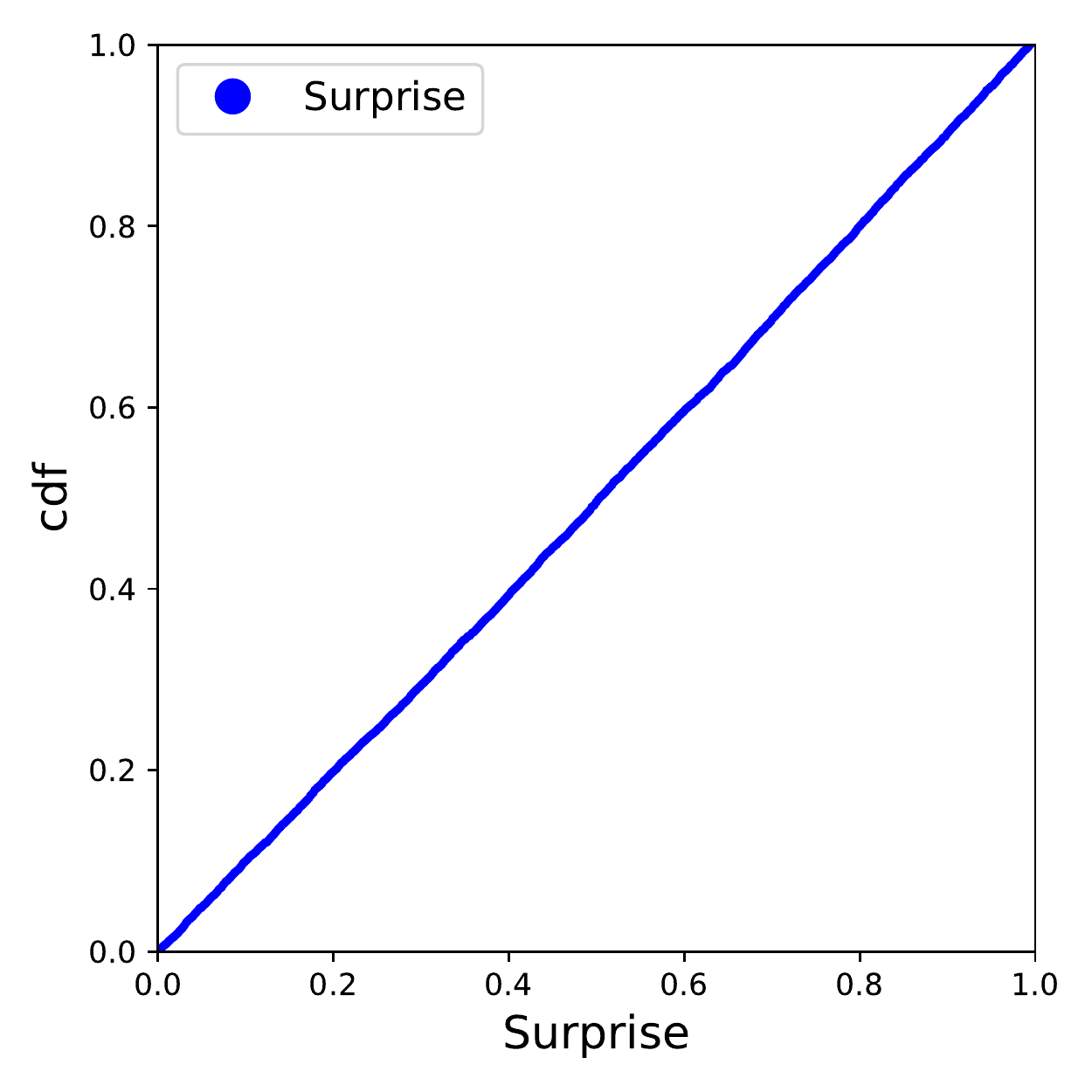}
\hspace{5mm}
\includegraphics[width=0.3\textwidth]{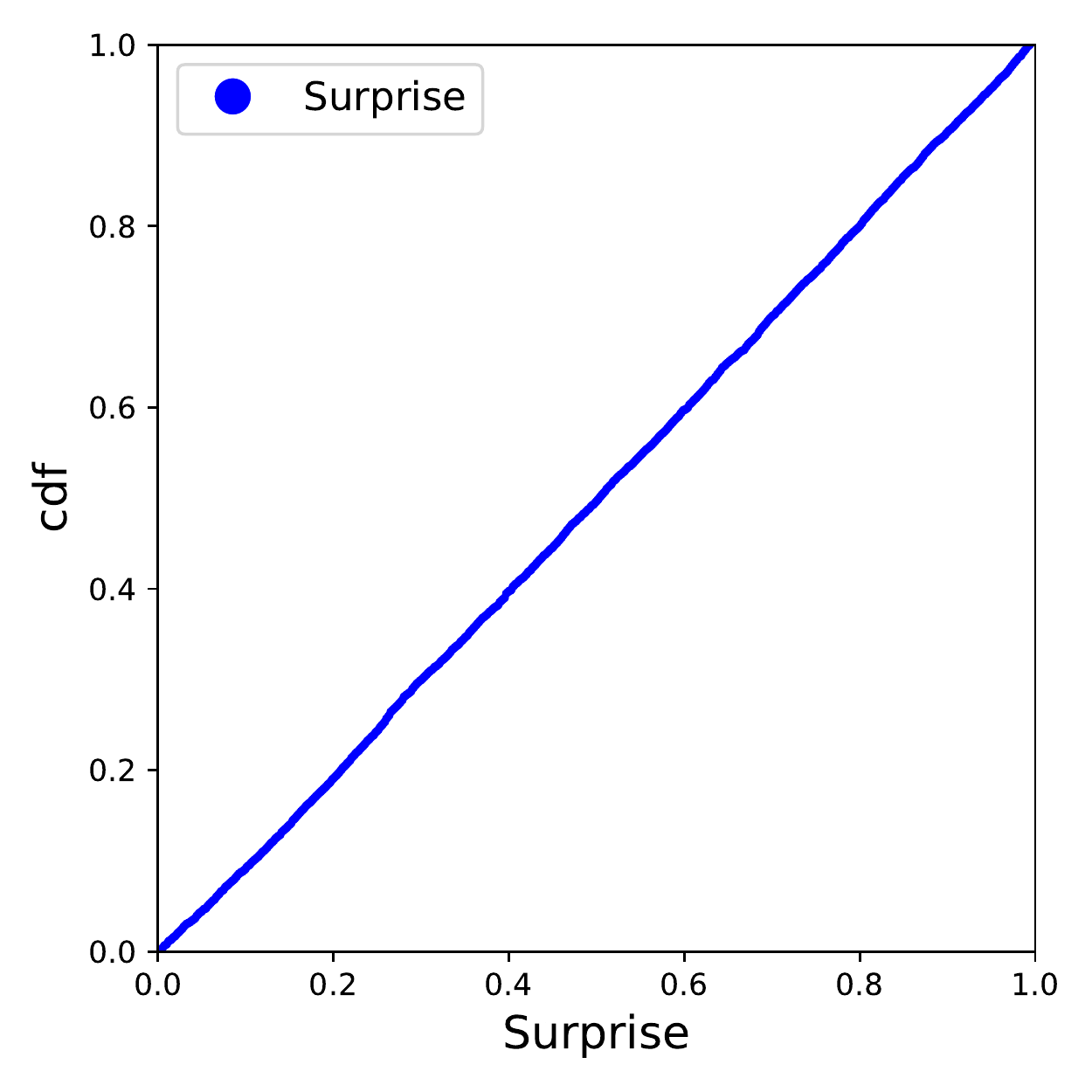}
\hspace{5mm}
\includegraphics[width=0.3\textwidth]{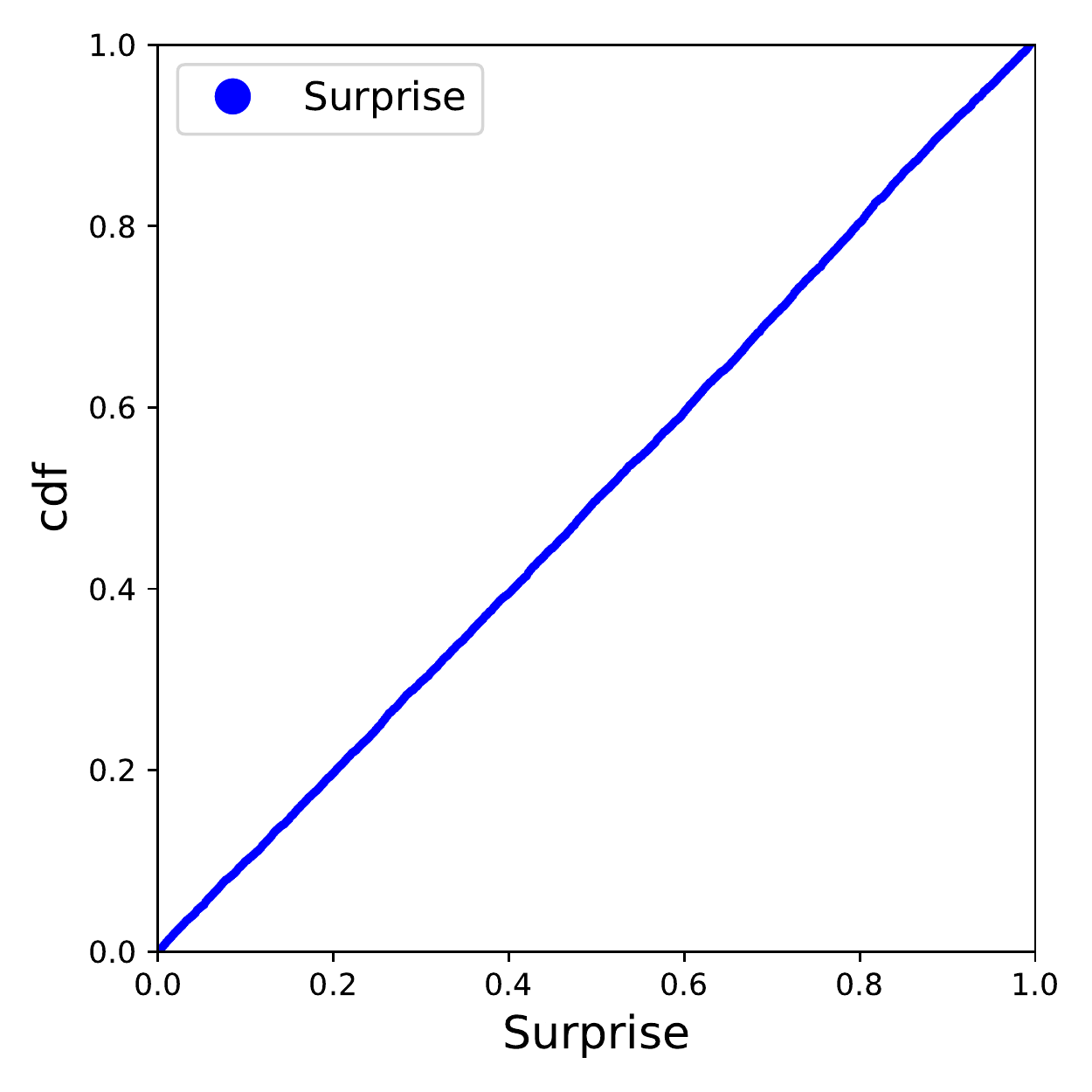}\\
\includegraphics[width=0.3\textwidth]{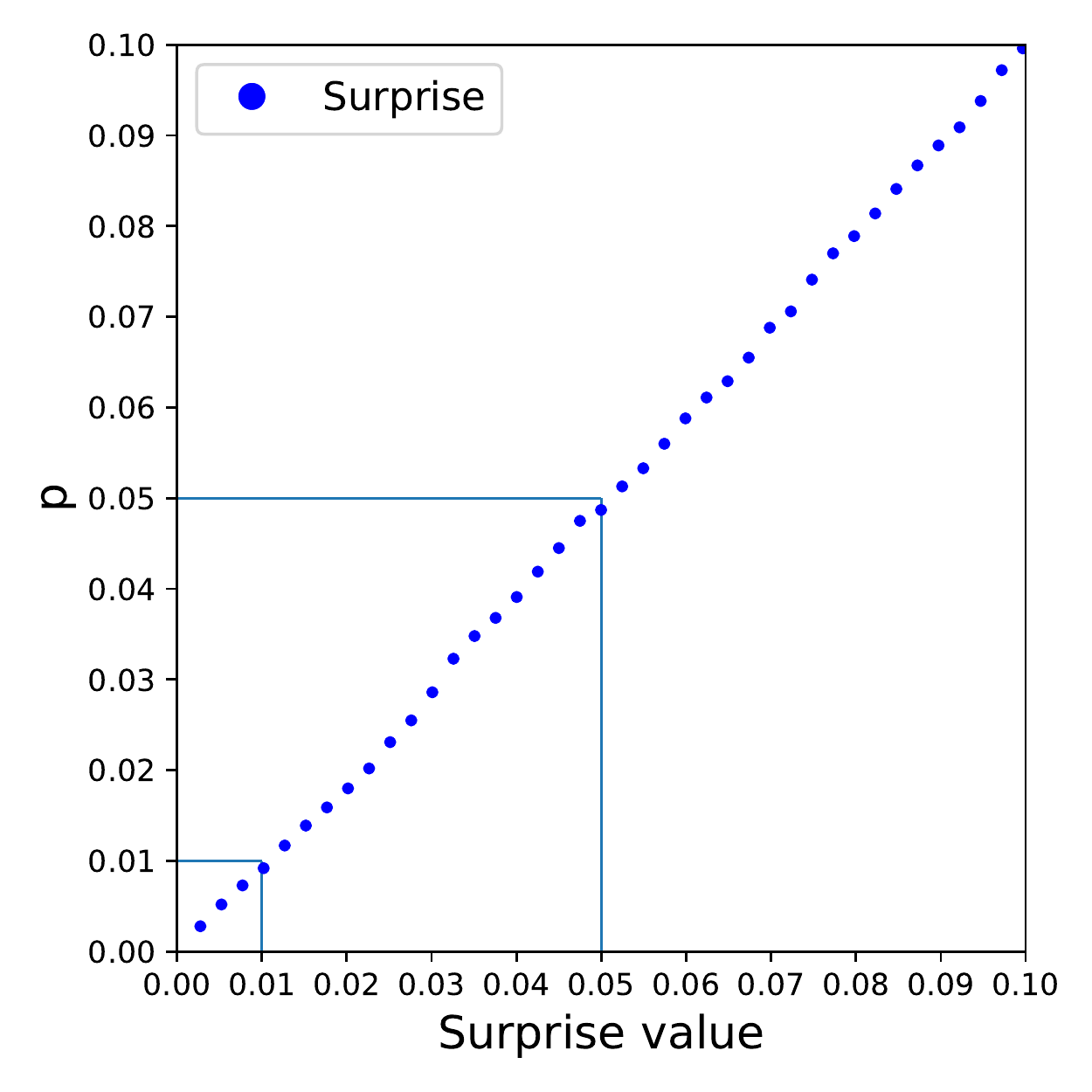}
\hspace{5mm}
\includegraphics[width=0.3\textwidth]{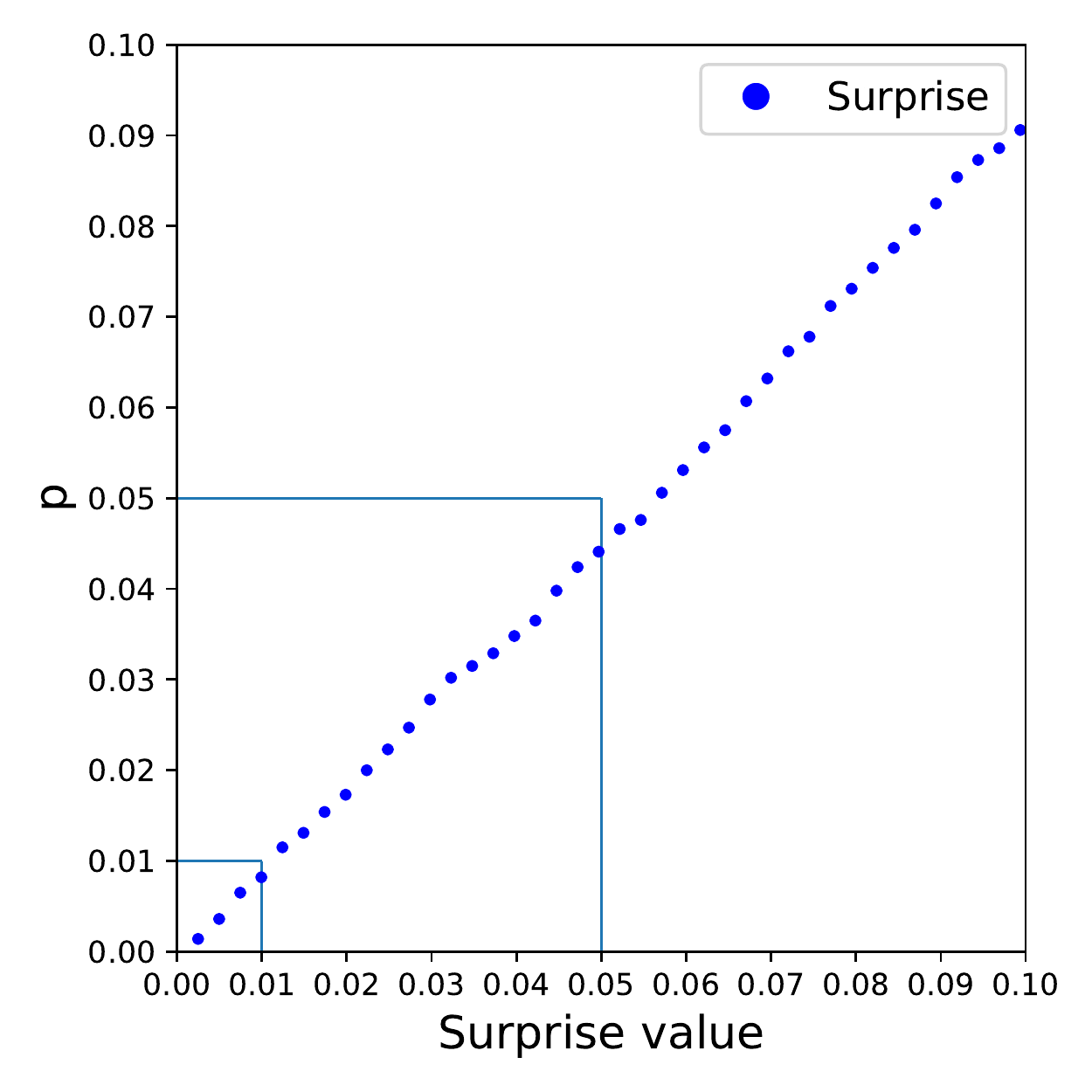}
\hspace{5mm}
\includegraphics[width=0.3\textwidth]{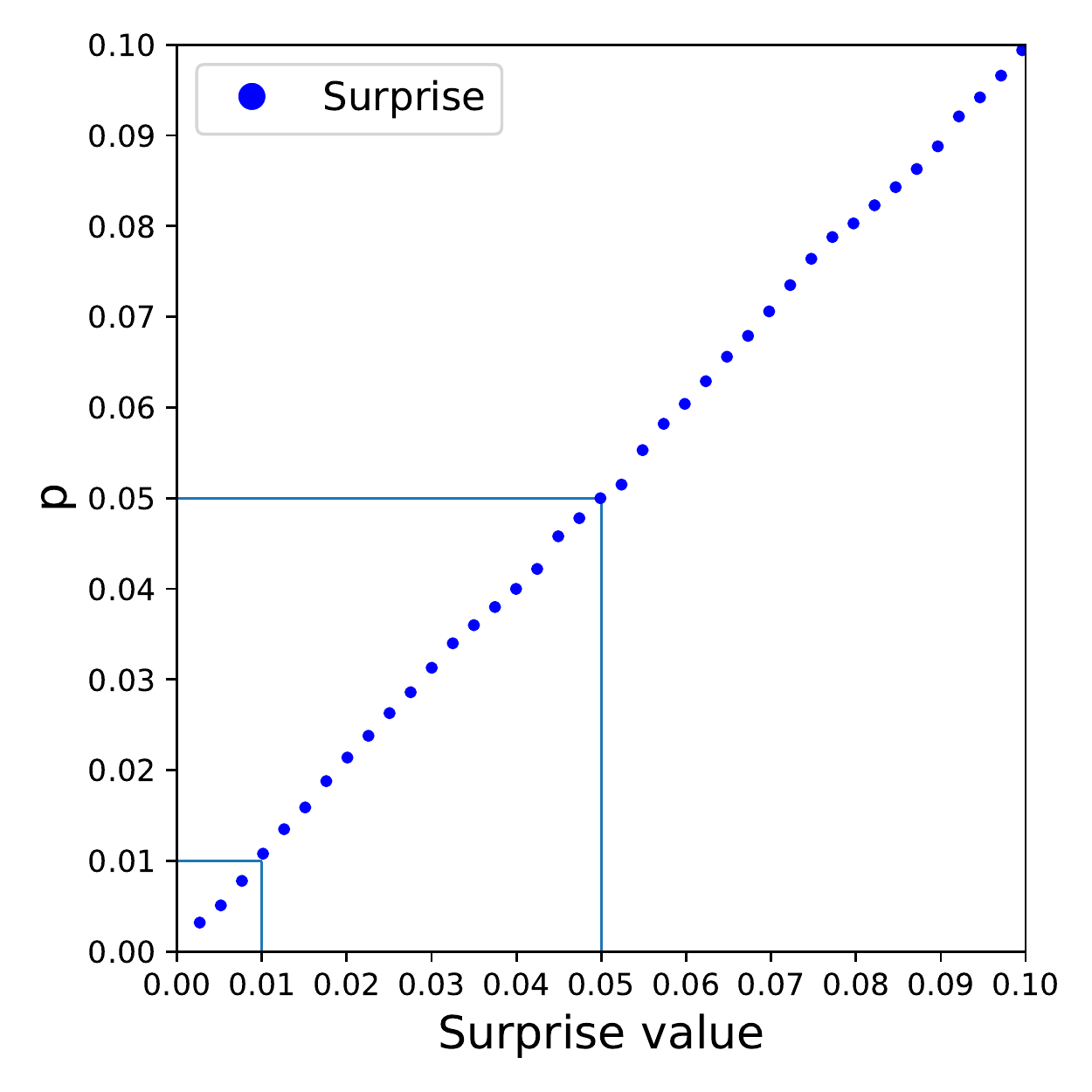}
\caption{Consistency check of our surprise-based formalism in the case of community detection on weighted networks. As in the binary case, the type I error rate can be kept below a given threshold by adjusting the threshold of the p-value of the corresponding test, e.g. by deeming its response as significant in case it is less than $0.05$. Over 10.000 networks sampled by the WRGM with $q=0.2$, a given planted partition - from left to right: two, three and four clusters with different dimensions - is indeed recovered as significant $5\%$ of the times. Notably, since the three upper panels show the CDF of the empirical values of $\mathscr{W}$, they also provide an information about its distribution, which is uniform over the unit interval.}
\label{fig3}
\end{figure*}

\textcolor{black}{We have also repeated such an experiment for the problem of detecting communities on weighted networks: as fig. \ref{fig3} shows, the same results are recovered, i.e. over 10.000 networks sampled from the WRGM, whose only parameter has been set to $q=0.2$, a (given) planted partition of two, three and four clusters with different dimensions is recognized as significant $5\%$ of the times; again, such a result holds true irrespectively of the details of the partition imposed on our sampled configurations.}

\textcolor{black}{Our experiments tell us something deeper about the behavior of $\mathscr{S}$ and $\mathscr{W}$: since each of the three upper panels in figs. \ref{fig2} and \ref{fig3} shows the cumulative distribution functions (CDF) of the empirical values of $\mathscr{S}$ and $\mathscr{W}$, we learn that the latter ones are distributed \emph{uniformly} over the unit interval, i.e. $\mathscr{S}\sim\text{U}[0,1]$ and $\mathscr{W}\sim\text{U}[0,1]$ - an evidence further confirming the behavior of the p-value of any exact significance test.}\\

\begin{figure*}[t!]
\includegraphics[width=0.47\textwidth]{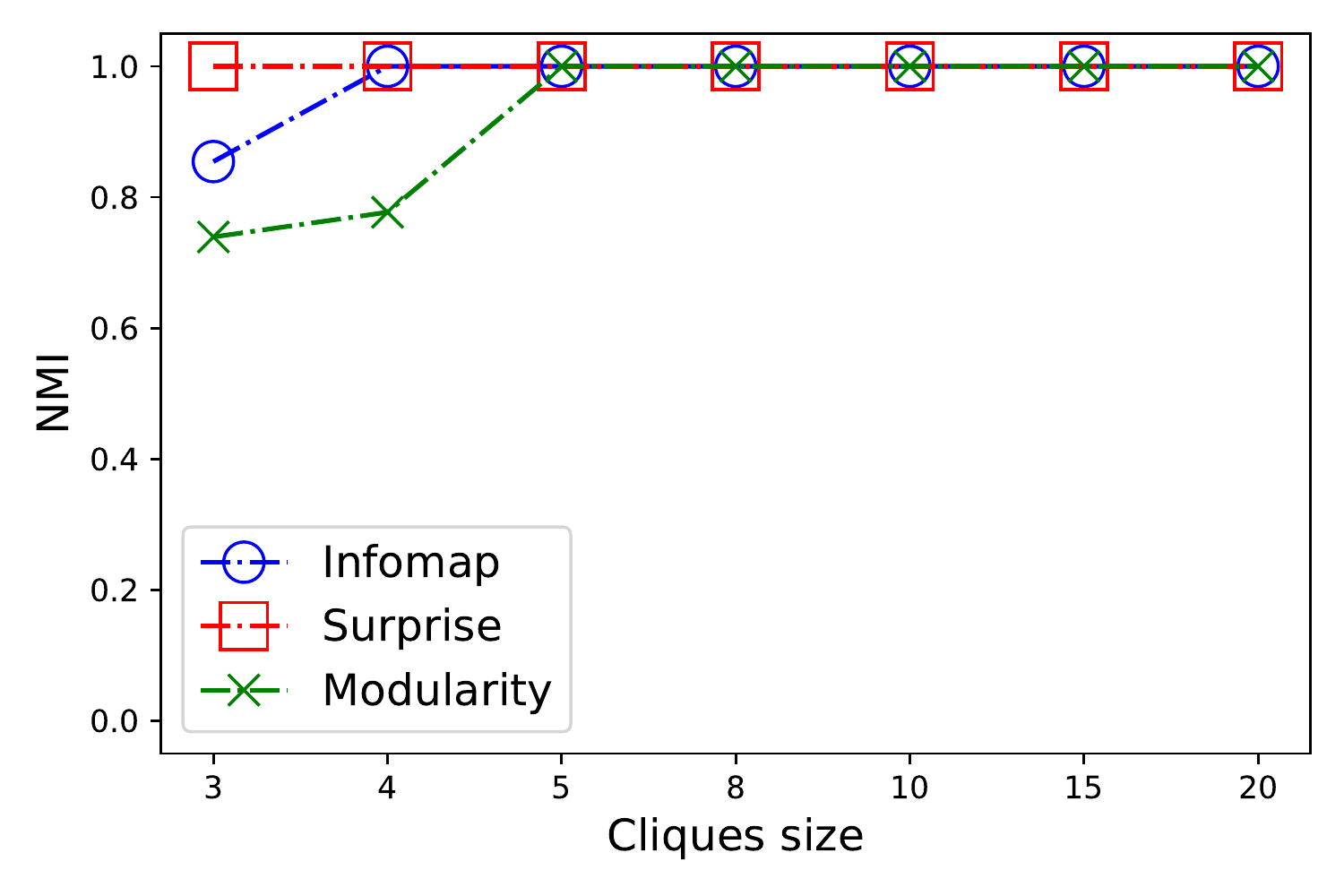}
\hspace{5mm}
\includegraphics[width=0.47\textwidth]{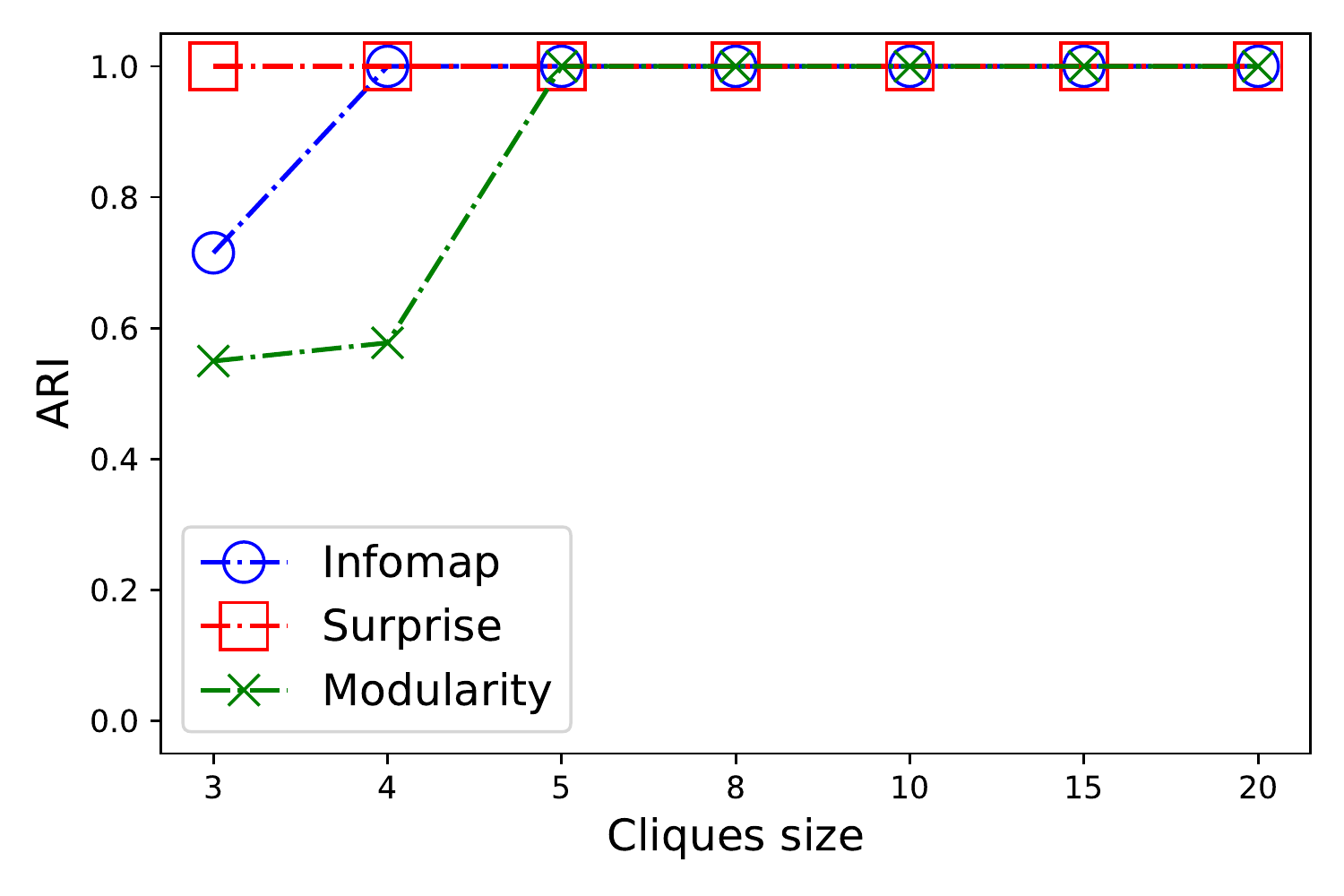}\\
\includegraphics[width=0.47\textwidth]{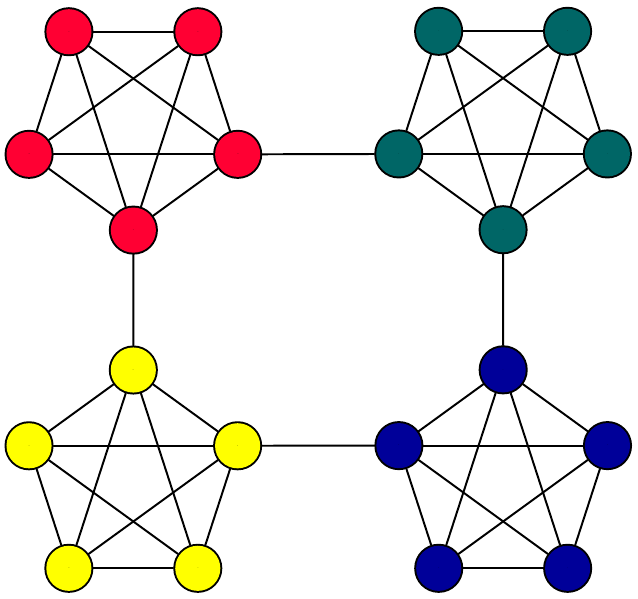}
\hspace{5mm}
\includegraphics[width=0.47\textwidth]{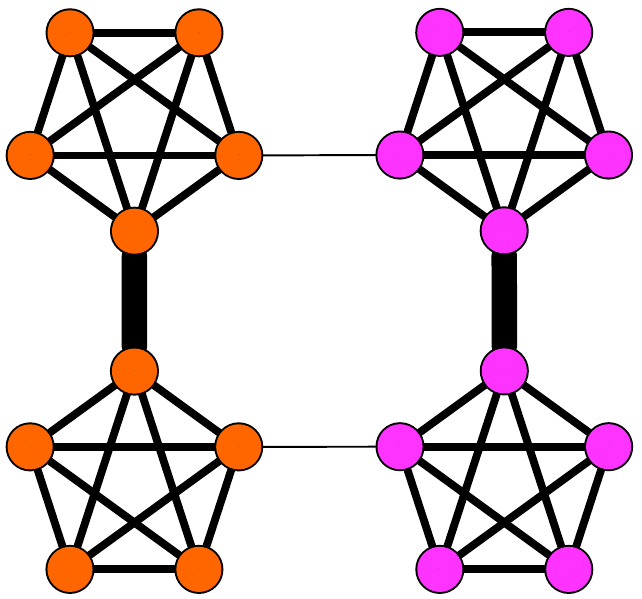}
\caption{\textcolor{black}{Upper row: performance of modularity, surprise and Infomap in correctly identifying the partitions induced by 7, different, `homogeneous' rings of cliques. For each configuration, 20 cliques have been considered; the size of the latter ones is left to vary as follows: $K_3$, $K_4$, $K_5$, $K_8$, $K_{10}$, $K_{15}$, $K_{20}$. While surprise always recovers the planted partition, modularity misses the partitions with $K_3$ and $K_4$ and Infomap misses the partition with $K_3$ - a result that may be a consequence of the resolution limit which is known to affect the last two algorithms.} Bottom row: comparison between the `ring of binary cliques' and the `ring of weighted cliques' cases. The result according to which surprise minimization is able to discriminate the cliques linked in a ring-like fashion changes once the weights come into play: in fact, as the weight of the links connecting any two cliques is risen, the algorithm reveals as `communities' pairs of tightly-connected cliques.}
\label{fig4}
\end{figure*}

\textcolor{black}{Checking the behavior of the multivariate versions of our hypergeometric distributions is, instead, much more difficult, the reason lying in the evidence that the null hypothesis is, now, a multivariate one and few results about the behavior of multivariate p-values are known. Still, we can say something about the behavior of the \emph{marginal p-values}; for the sake of illustration, let us consider the ones induced by $\mathscr{S}_\sslash$ and $\mathscr{W}_\sslash$ and reading}

\begin{eqnarray}
\mathscr{S}_\sslash^{(\bullet)}&=&\sum_{l_\bullet\geq l_\bullet^*}\sum_{l_\circ}f(l_\bullet,l_\circ)=\sum_{l_\bullet\geq l_\bullet^*}f(l_\bullet),\\
\mathscr{S}_\sslash^{(\circ)}&=&\sum_{l_\circ\geq l_\circ^*}\sum_{l_\bullet}f(l_\bullet,l_\circ)=\sum_{l_\circ\geq l_\circ^*}f(l_\circ)
\end{eqnarray}
and 

\begin{eqnarray}
\mathscr{W}_\sslash^{(\bullet)}&=&\sum_{w_\bullet\geq w_\bullet^*}\sum_{w_\circ}f(w_\bullet,w_\circ)=\sum_{w_\bullet\geq w_\bullet^*}f(w_\bullet),\\
\mathscr{W}_\sslash^{(\circ)}&=&\sum_{w_\circ\geq w_\circ^*}\sum_{w_\bullet}f(w_\bullet,w_\circ)=\sum_{w_\circ\geq w_\circ^*}f(w_\circ)
\end{eqnarray}
\textcolor{black}{respectively. As evident from the formulas above, marginal distributions of multivariate hypergeometric distributions are hypergeometric themselves: hence, the behavior of our marginal p-values is expected to match the one observed in the univariate cases, leading to recover $\mathscr{S}_\sslash^{(\bullet)}\sim\text{U}[0,1]$, $\mathscr{S}_\sslash^{(\circ)}\sim\text{U}[0,1]$, $\mathscr{W}_\sslash^{(\bullet)}\sim\text{U}[0,1]$ and $\mathscr{W}_\sslash^{(\circ)}\sim\text{U}[0,1]$. Analogously, for the enhanced surprise.}\\

\begin{figure*}[t!]
\includegraphics[width=0.47\textwidth]{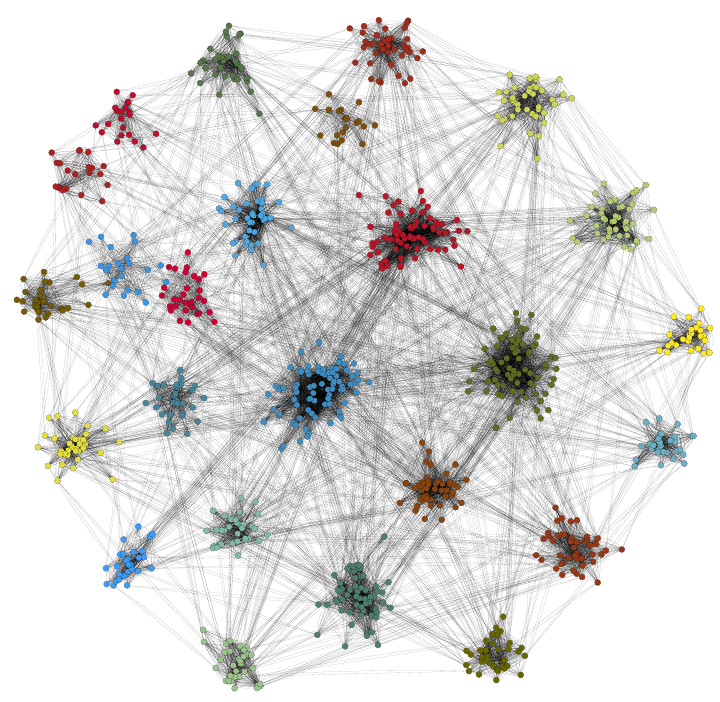}
\hspace{5mm}
\includegraphics[width=0.47\textwidth]{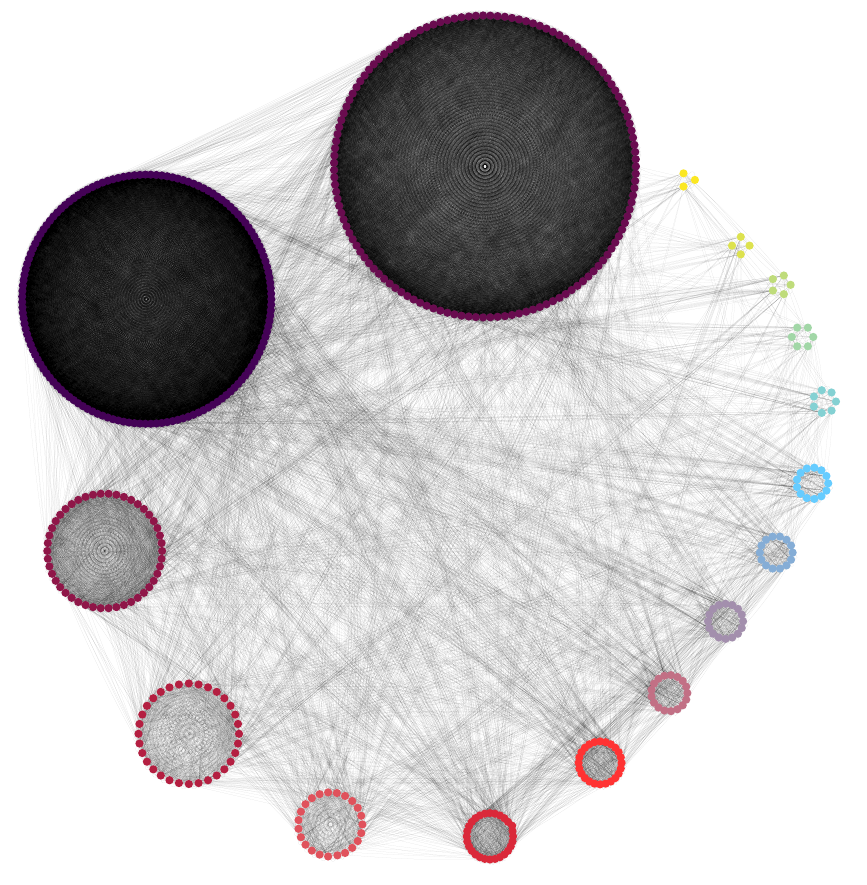}
\caption{\textcolor{black}{Examples of the benchmarks used in the present paper to compare modularity, surprise and Infomap: Lancichinetti-Fortunato-Radicchi benchmark (LFR, left panel), constituted by 1000 nodes arranged in `big' communities and with mixing coefficient $\mu_t=0.1$; Aldecoa's `relaxed-caveman' benchmark (RC, right panel), constituted by 16 communities whose size is distributed according to a power-law and with `degradation' coefficient $p=0.1$.}}
\label{fig5}
\end{figure*}

\textcolor{black}{Our findings can be also described from a different perspective, i.e. by answering the question \emph{what is the expected value of the surprise on a network sampled from the RGM ensemble?} To answer this question let us focus on the simplest case of detecting communities on binary networks and consider that, irrespectively from the partition that is planted on the sampled configuration, the expected number of links will be $\langle L\rangle=pV$ while the expected number of `internal' links will be $\langle l_\bullet^*\rangle=pV_\bullet$, where $p$ is the parameter defining the RGM; hence, the surprise becomes}

\begin{equation}
\mathscr{S}=\sum_{l_\bullet\geq\langle l_\bullet^*\rangle}\frac{\binom{V_\bullet}{l_\bullet}\binom{V-V_\bullet}{\langle L\rangle-l_\bullet}}{\binom{V}{\langle L\rangle}}=\sum_{l_\bullet\geq pV_\bullet}\frac{\binom{V_\bullet}{l_\bullet}\binom{V-V_\bullet}{pV-l_\bullet}}{\binom{V}{pV}}
\end{equation}
\textcolor{black}{(to be fully consistent we should have rounded both the value of $\langle L\rangle$ and that of $\langle l_\bullet^*\rangle$ to the nearest integer but our conclusions are still valid). In order to evaluate the expression above, let us consider that the expected value of the hypergeometric distribution defining the surprise reads}

\begin{equation}
\langle l_\bullet\rangle=\langle L\rangle\frac{V_\bullet}{V}=pV_\bullet
\end{equation}
\textcolor{black}{the validity of the first passage resting upon the evidence that the network is generated via the RGM. Hence, the surprise becomes a sum over the values $l_\bullet\geq\langle l_\bullet\rangle$, i.e. those that are more extreme that the average. Since the hypergeometric distribution is peaked around its average, the result above suggests that summing over larger values encodes half of the probability mass, i.e. $\simeq\frac{1}{2}$: as a consequence, the expected value of surprise, on a network generated via the RGM, amounts at $\simeq\frac{1}{2}$, irrespectively from the partition that is planted on the sampled configuration; its fluctuations are, instead, compatible with those of a uniform distribution whose support coincides with the unit interval.}\\

\noindent\textcolor{black}{{\bf Comparison with the binary modularity.} For the sake of comparison, let us consider the modularity function for the problem of community detection on binary, undirected networks. It is defined as}

\begin{equation}
\textcolor{black}{Q=\frac{1}{2L}\sum_{i\neq j}(a_{ij}-\langle a_{ij}\rangle)\delta_{c_i,c_j}}
\end{equation}
\textcolor{black}{where, for consistency, we employ the RGM as a benchmark, i.e. $\langle a_{ij}\rangle=p=\frac{L}{V}=\frac{2L}{N(N-1)}$, $\forall\:i<j$. From the definition of modularity, it follows that}

\begin{align}
Q&=\frac{1}{2L}\sum_{i\neq j}a_{ij}\delta_{c_i,c_j}-\frac{1}{2L}\sum_{i\neq j}p\delta_{c_i,c_j}\\
&=\frac{1}{2L}\sum_c[2l_c^*-pN_c(N_c-1)]\\
&=\frac{1}{2L}\sum_c\left[2l_c^*-2L\frac{N_c(N_c-1)}{N(N-1)}\right]\\
&=\left(\frac{l_\bullet^*}{L}-\frac{V_\bullet}{V}\right)
\end{align}
\textcolor{black}{where $N_c$ and $l_c^*$ indicate the size of community $c$ and the number of links \emph{within} it, respectively - naturally, $l_\bullet^*=\sum_cl_c^*$ and $V_\bullet=\sum_c N_c(N_c-1)$. This calculation clarifies that a positive value of $Q$ implies that the `internal' probability of connection $p_\bullet$ is larger than the one `predicted' by the RGM, i.e.}

\begin{equation}
Q\geq 0\Longrightarrow p_\bullet=\frac{l_\bullet^*}{V_\bullet}\geq\frac{L}{V}=p
\end{equation}
\textcolor{black}{while a null modularity value implies that $p_\bullet$ matches the only parameter defining the RGM. The result above can be also restated as follows: a null modularity value points out that the mesoscale structure of the configuration at hand \emph{has not been generated} by a SBM (or, equivalently, \emph{does not need} a SBM to be explained). Notice, however, that the modularity provides no indication about the statistical significance of the recovered partition.}\\

\begin{figure*}[t!]
\includegraphics[width=0.28\textwidth]{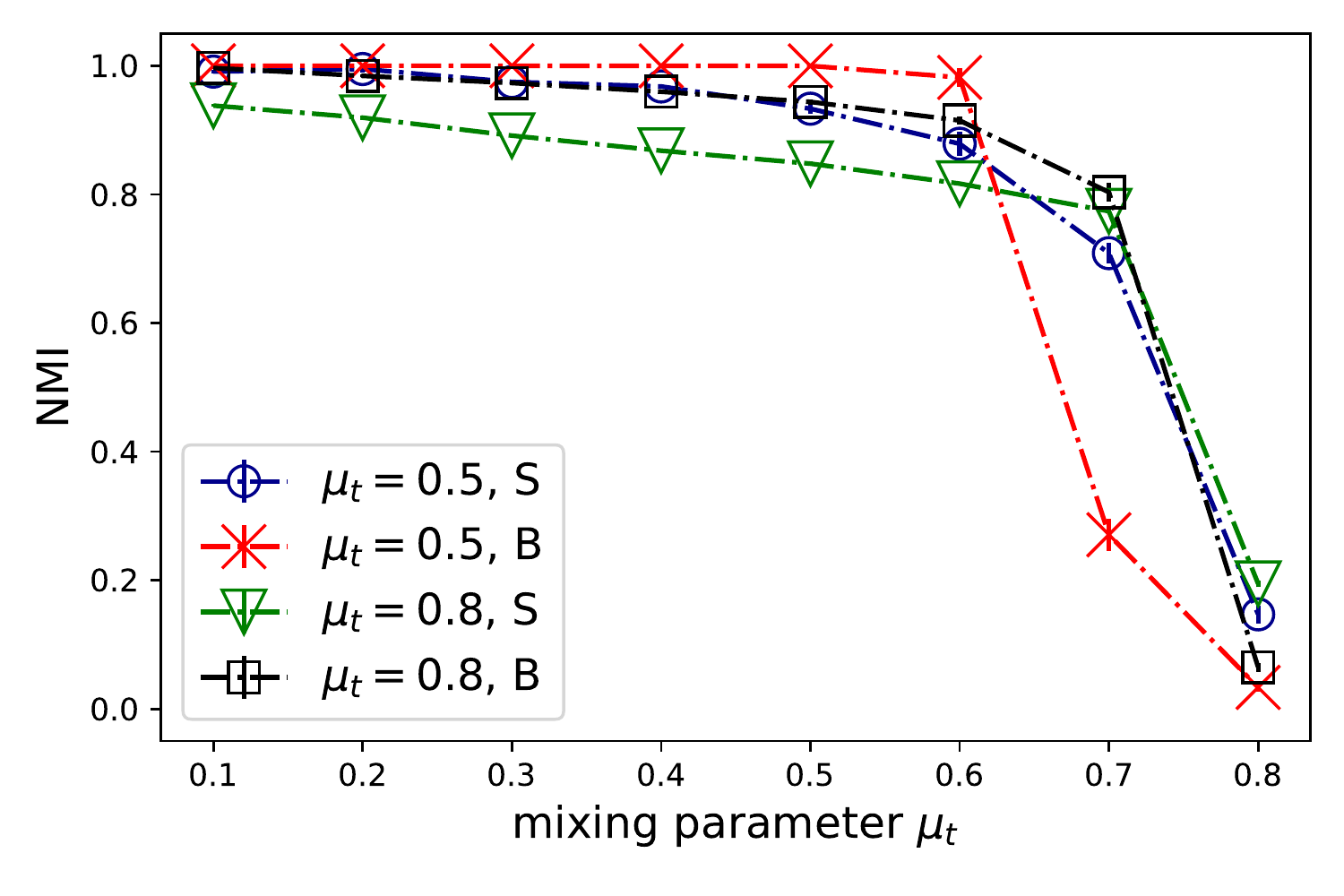}
\hspace{5mm}
\includegraphics[width=0.28\textwidth]{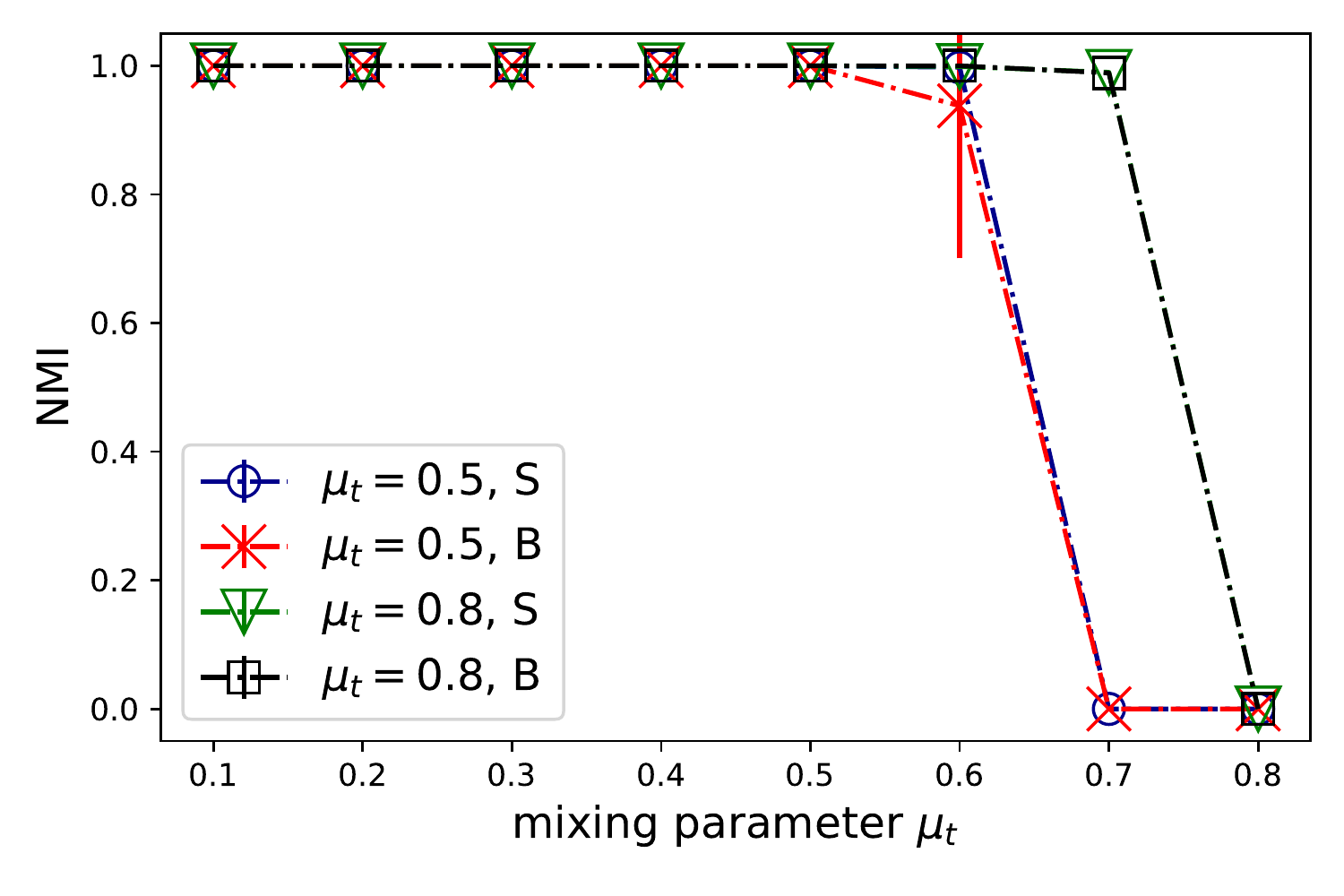}
\hspace{5mm}
\includegraphics[width=0.28\textwidth]{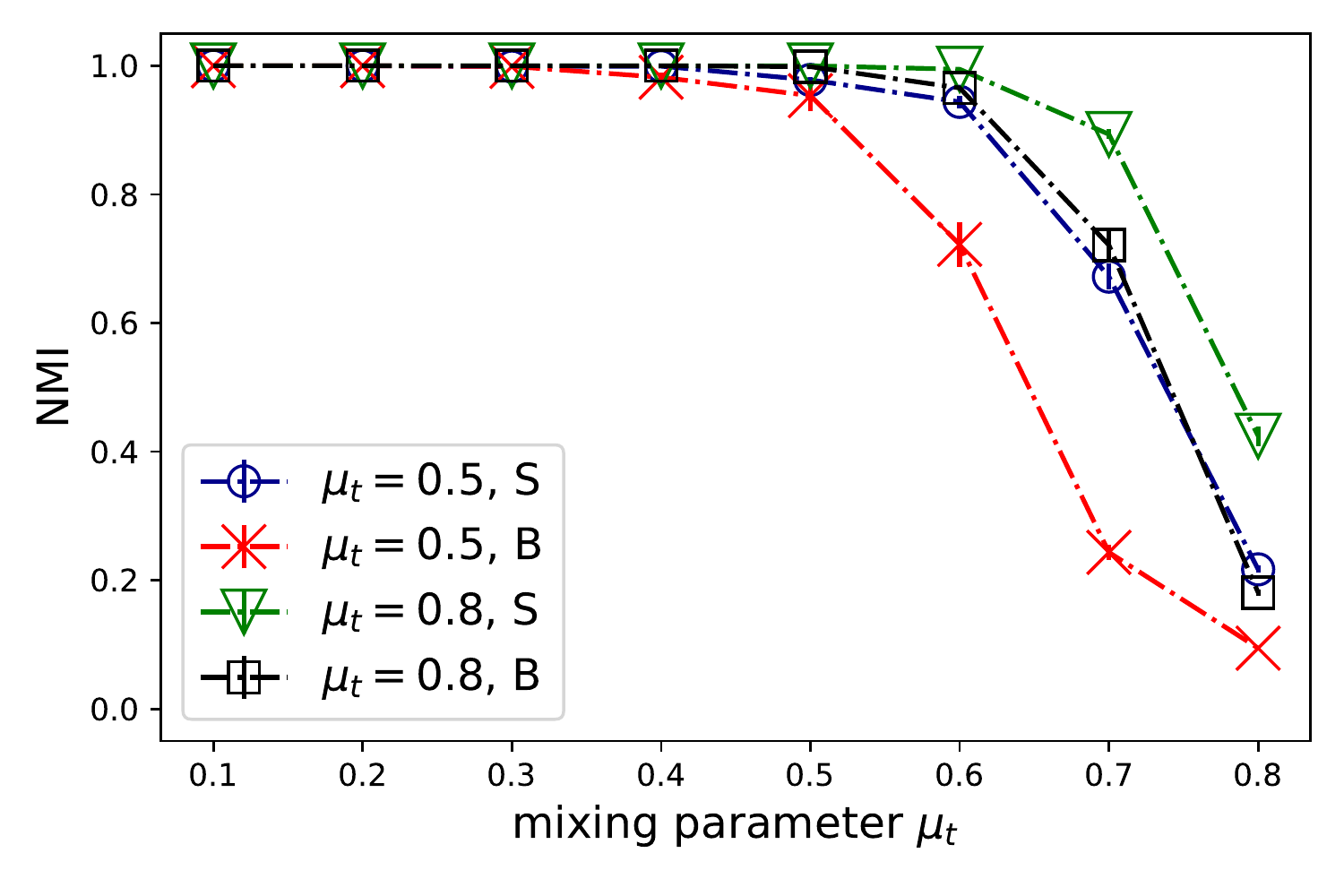}\\
\includegraphics[width=0.28\textwidth]{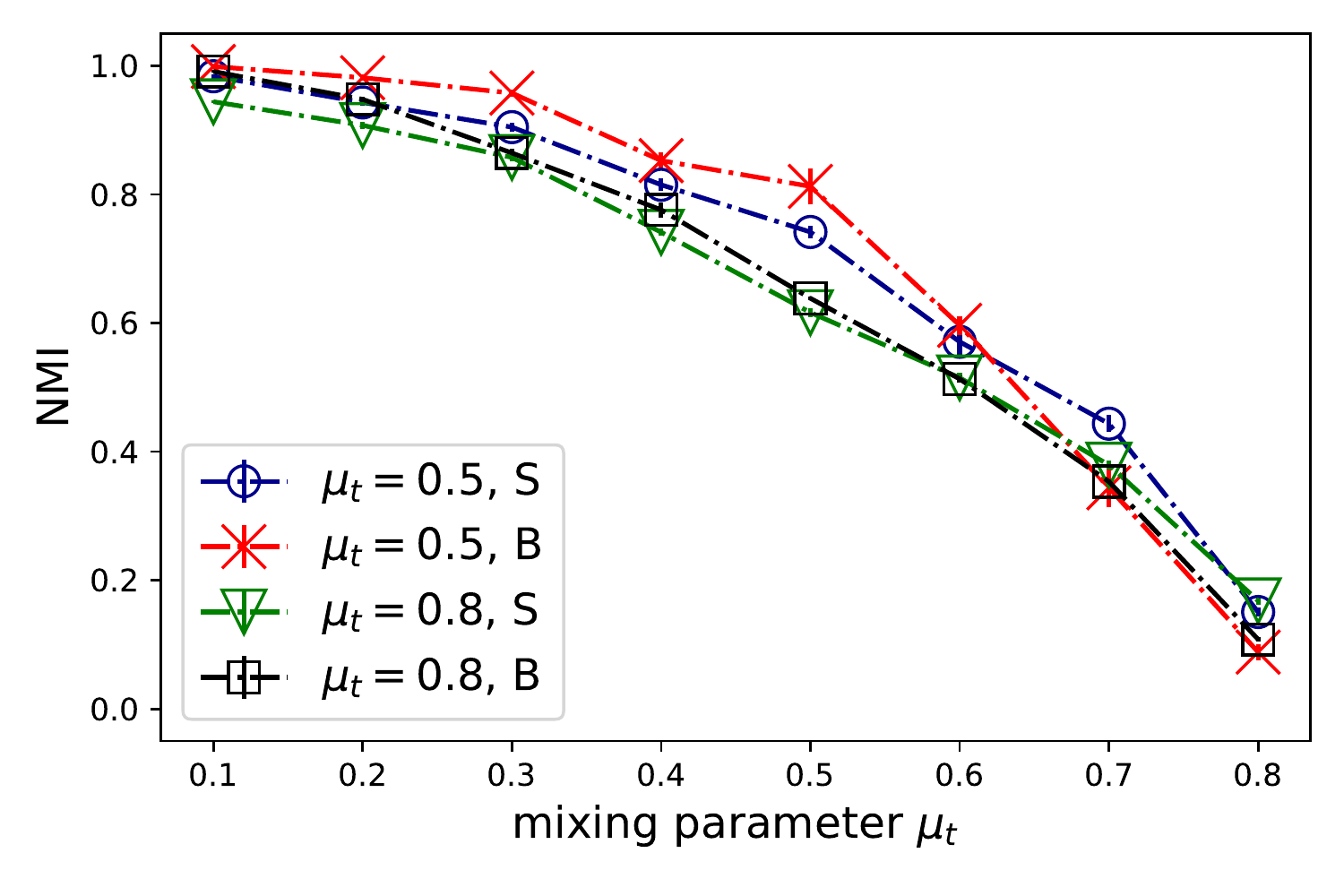}
\hspace{5mm}
\includegraphics[width=0.28\textwidth]{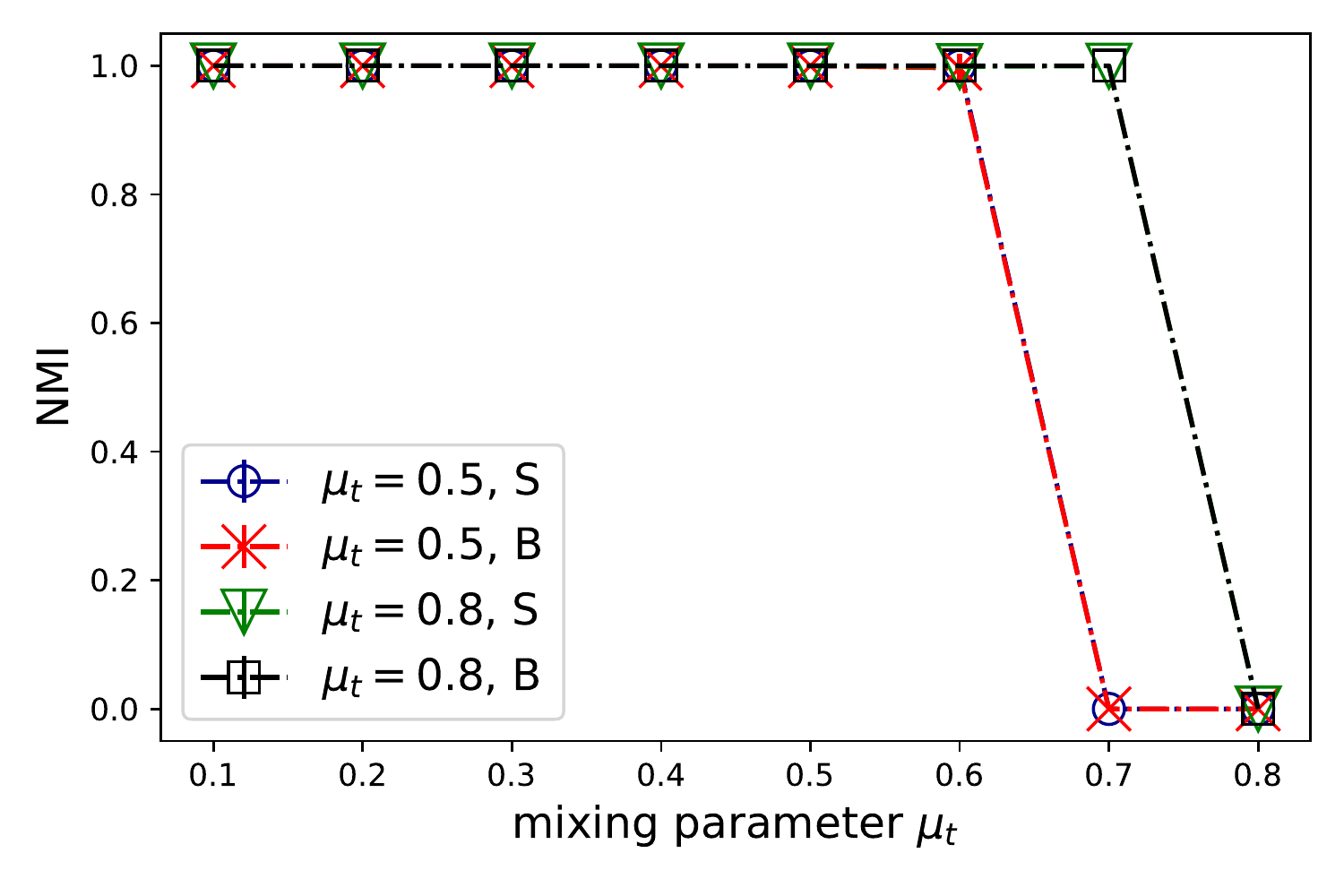}
\hspace{5mm}
\includegraphics[width=0.28\textwidth]{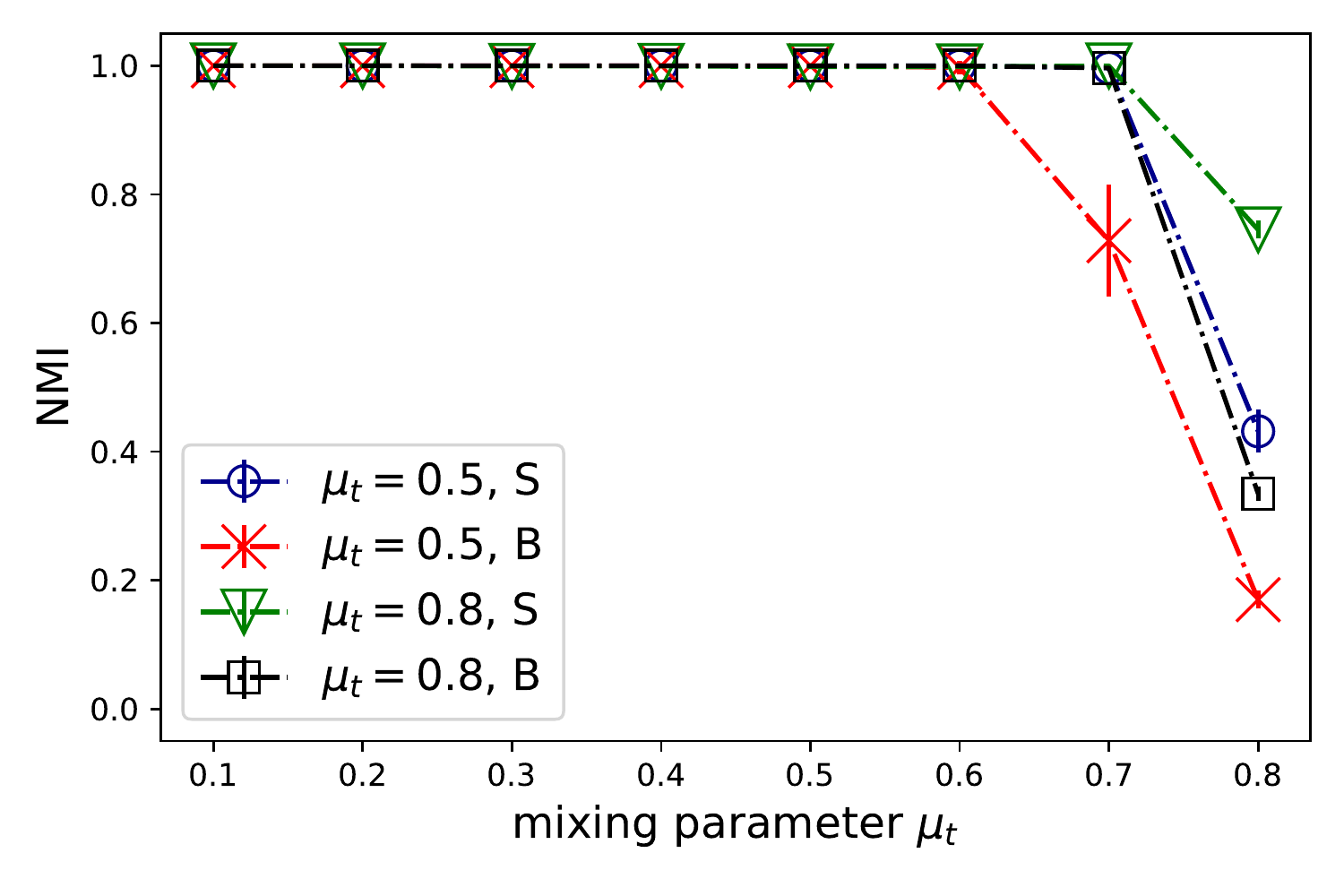}\\
\includegraphics[width=0.28\textwidth]{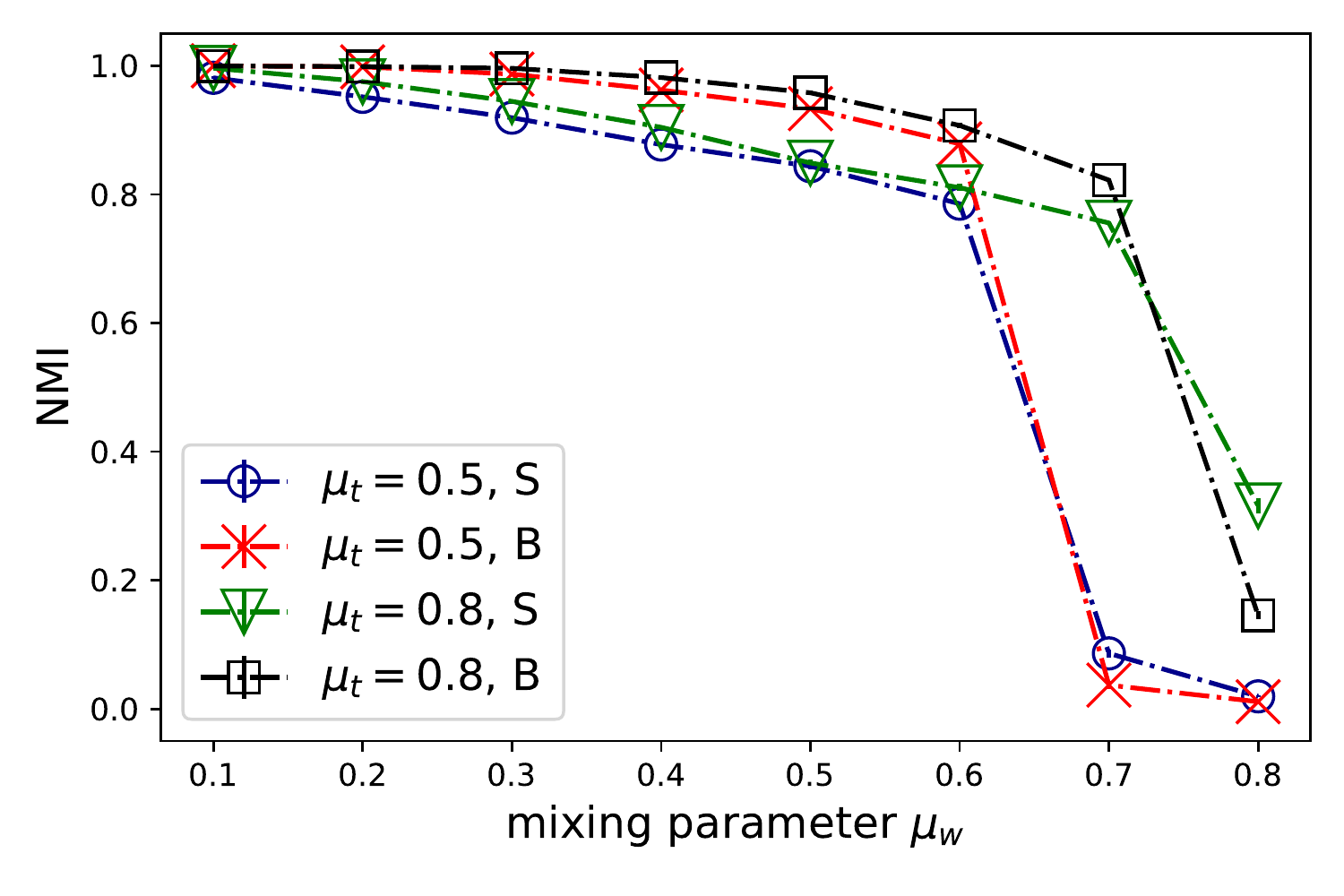}
\hspace{5mm}
\includegraphics[width=0.28\textwidth]{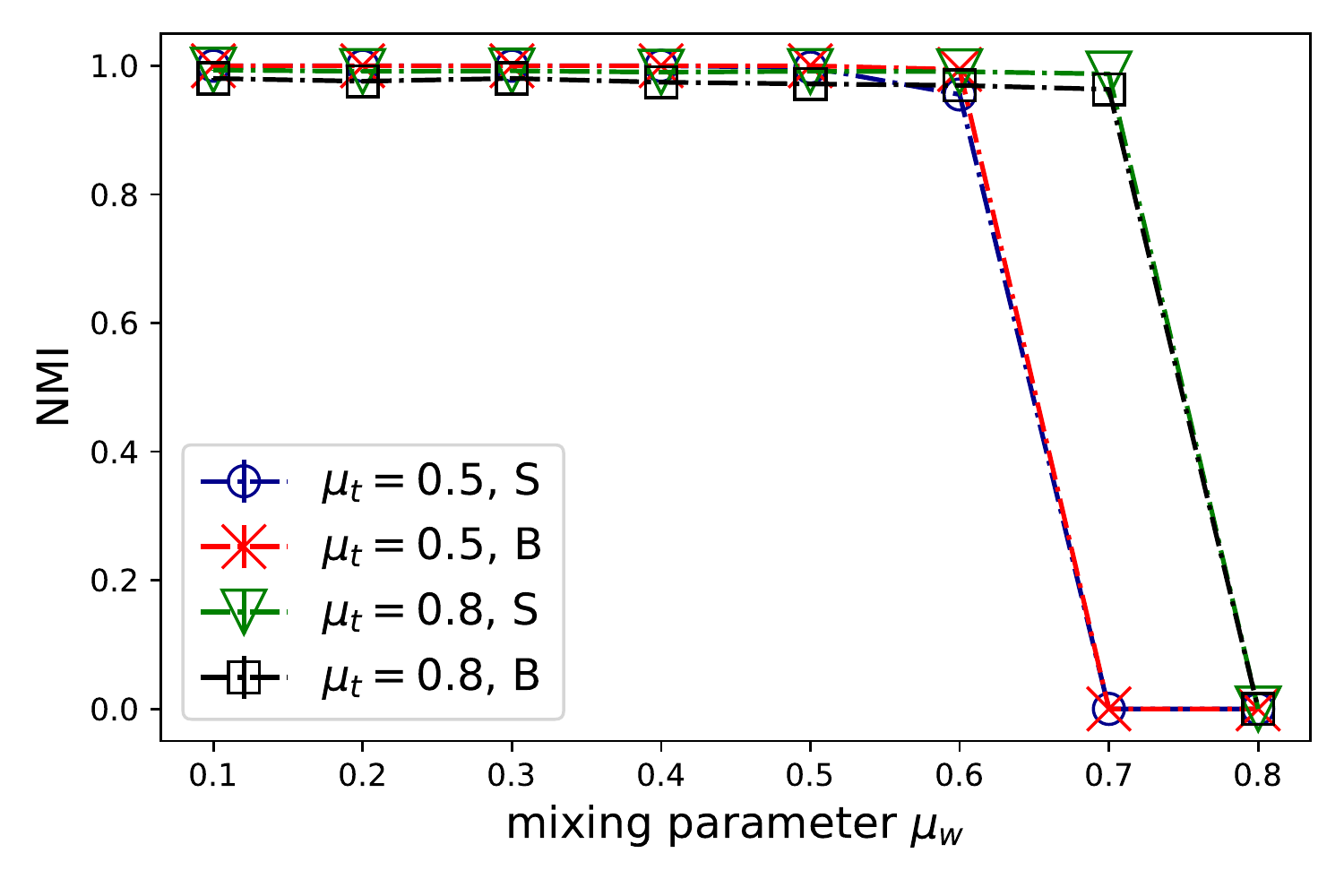}
\hspace{5mm}
\includegraphics[width=0.28\textwidth]{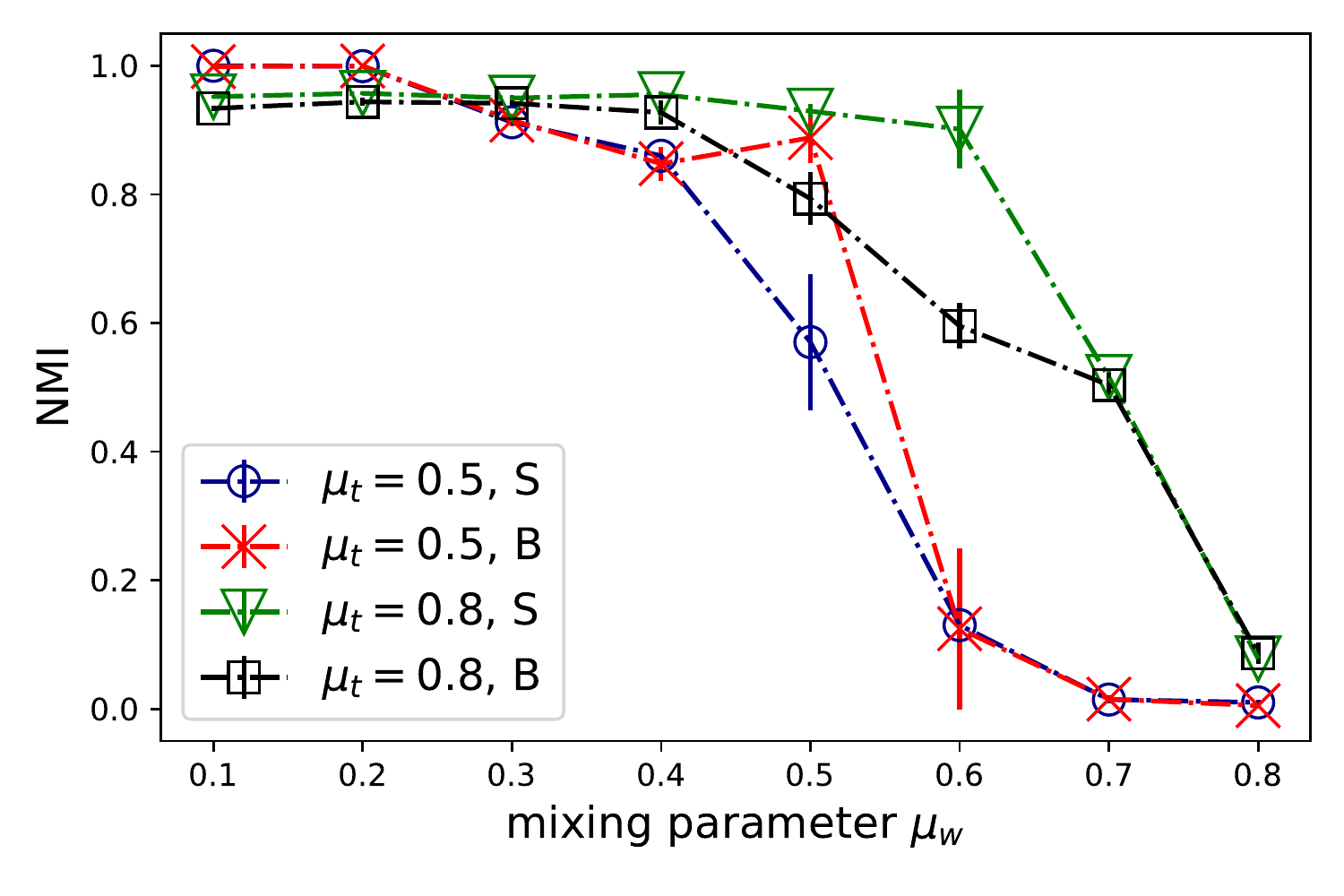}\\
\includegraphics[width=0.28\textwidth]{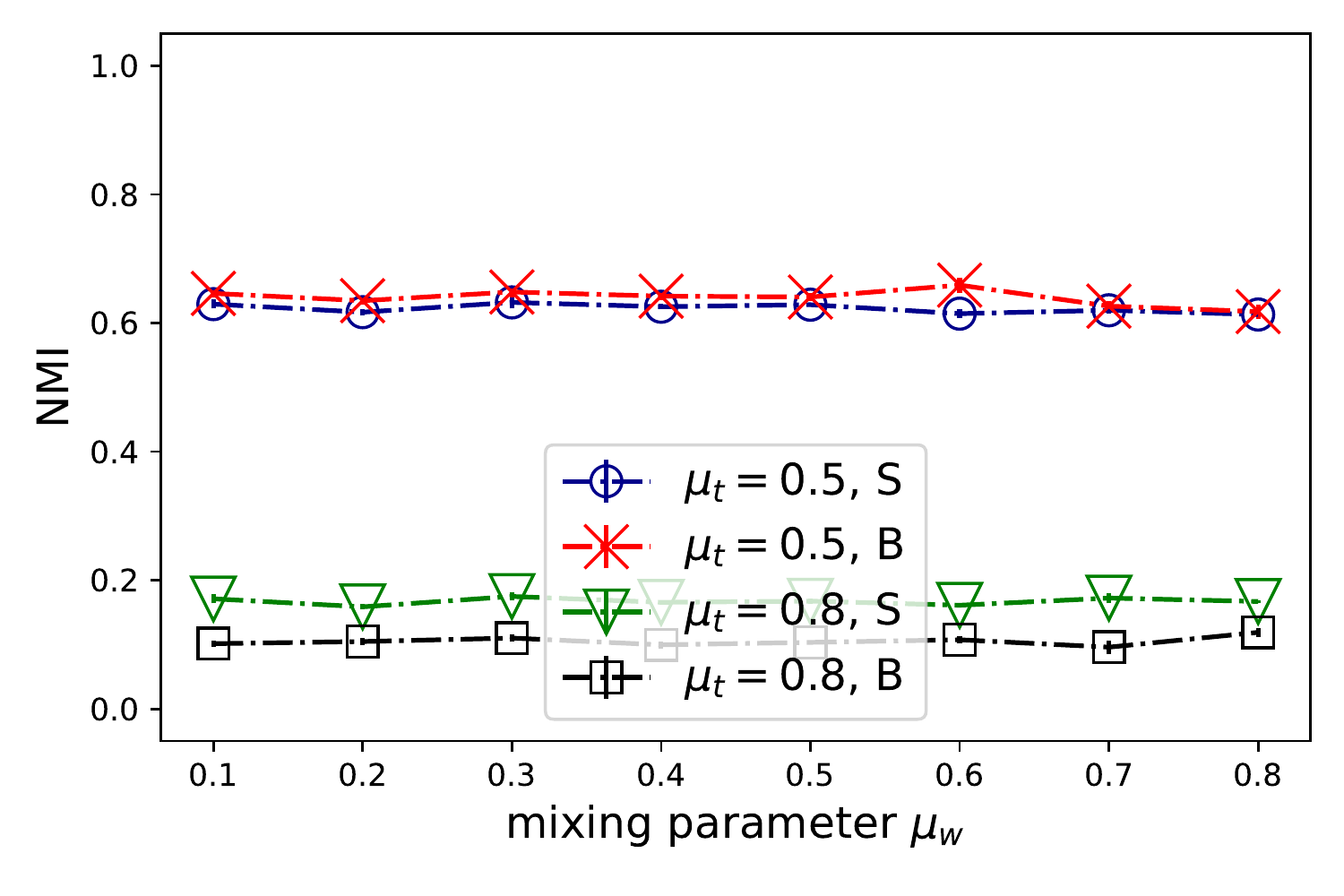}
\hspace{5mm}
\includegraphics[width=0.28\textwidth]{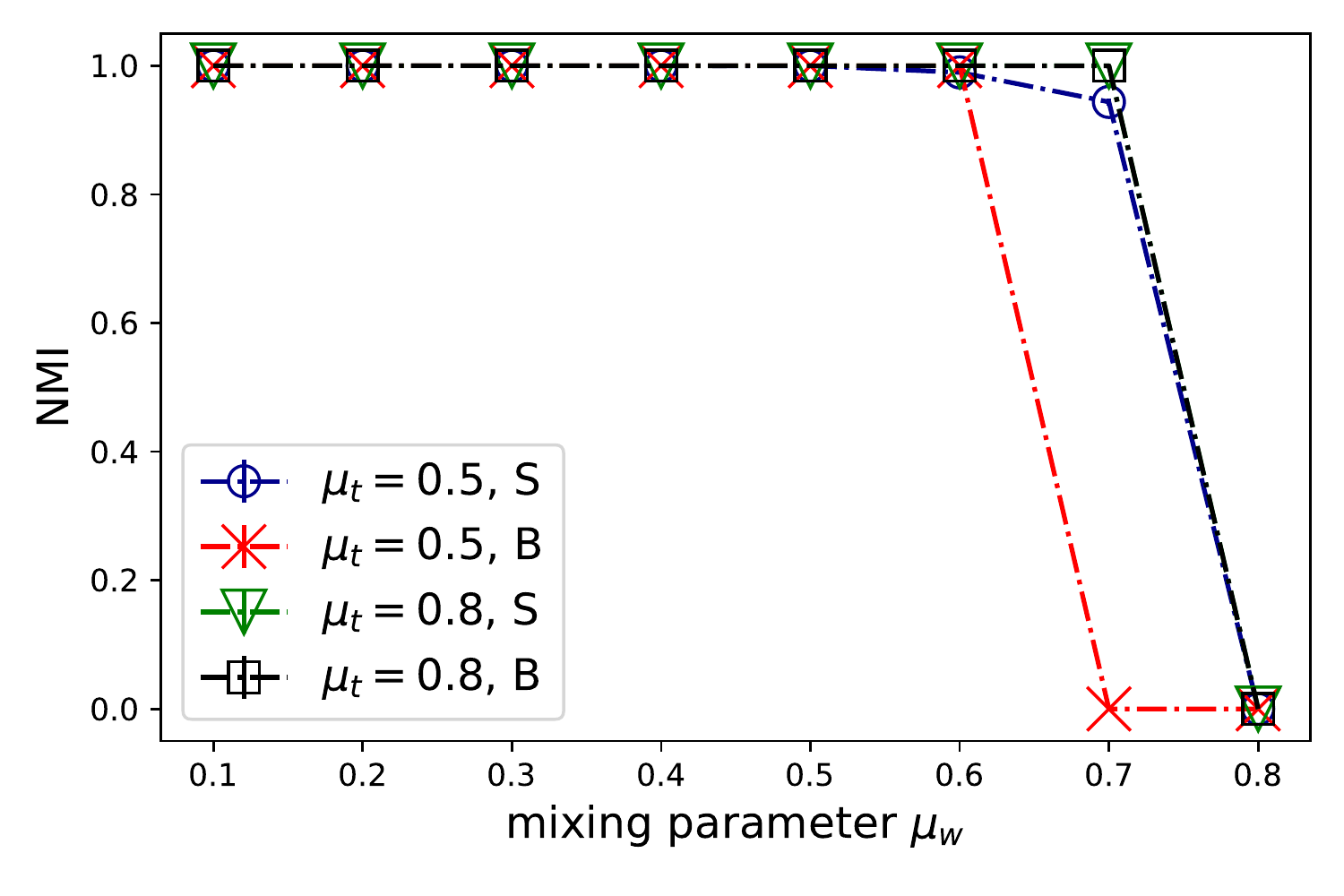}
\hspace{5mm}
\includegraphics[width=0.28\textwidth]{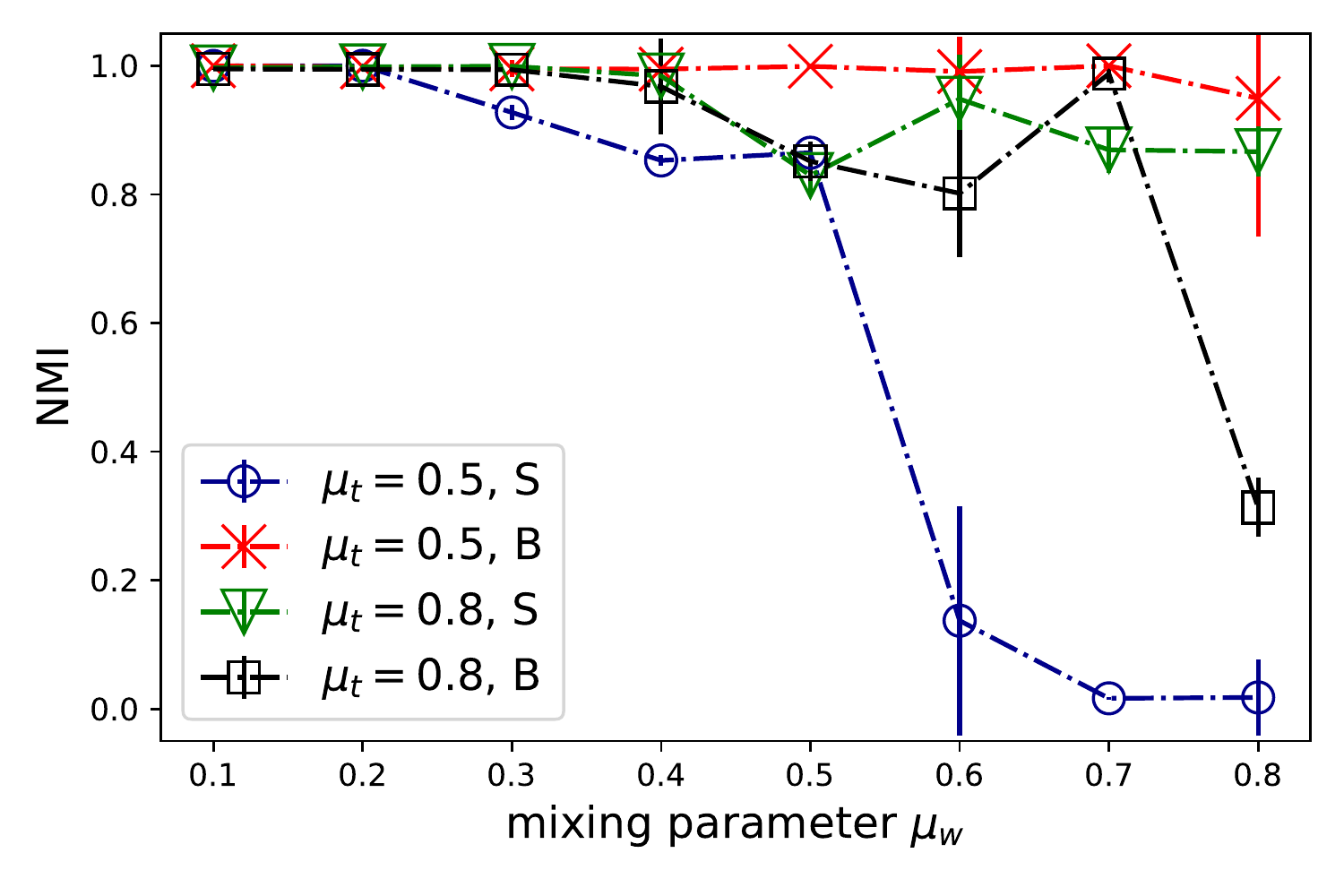}
\caption{\textcolor{black}{Comparison of three different algorithms for community detection on the LFR benchmark, i.e. modularity maximization (left column), Infomap (central column) and surprise minimization (right column): the chosen configurations are binary undirected networks (first row), binary directed networks (second row), weighted undirected networks (third row) and weighted directed networks (fourth row). Four curves are shown, corresponding to two different network sizes for binary configurations (blue and red: 1000 nodes; green and black: 5000 nodes) and, for a given size, to two different ranges for the size of communities, indicated by the letters S (`small', communities have between 10 and 50 nodes) and B (`big', communities have between 20 and 100 nodes). For weighted configurations, instead, all networks have the same size (5000 nodes): what differentiates the two pairs of curves is the value of the binary mixing parameter. When focusing on binary, undirected networks, we have considered $\tau_1=-2$ and $\tau_2=-1$ (the average degree is 20 and the maximum degree is 50). When binary, directed configurations are considered, $\mu_t$ refers to in-degrees, which are distributed according to a power-law while the out-degrees are kept constant for all nodes; the other input parameters, instead, are the same used for undirected configurations. When weighted, undirected configurations are considered, an additional mixing parameters is needed, i.e. $\mu_w$, accounting for the percentage of a node strength to be distributed on the links that connect it to the nodes outside its own community; the exponent of the strength distribution has been set to 1.5 for all realizations considered here. When weighted, directed configurations are considered, $\mu_w$ refers to in-strengths. The trend of the NMI, plotted as a function of the mixing parameters, reveals Infomap to be a strong performer even if its performance decreases abruptly as the mixing parameters exceed a threshold value that depends on the particular setting of the benchmark; the performance of modularity, instead, `degrades' less rapidly. The performance of surprise constitutes a good compromise between the robustness of modularity and the steadily high accuracy of Infomap.}}
\label{fig6}
\end{figure*}

\noindent\textcolor{black}{{\bf Comparison with the weighted modularity.} A similar conclusion can be reached upon considering the modularity function for community detection on weighted, undirected networks, defined as}

\begin{equation}
\textcolor{black}{Q=\frac{1}{2W}\sum_{i\neq j}(w_{ij}-\langle w_{ij}\rangle)\delta_{c_i,c_j}}
\end{equation}
\textcolor{black}{where we have employed the Weighted Random Graph Model (WRGM) as a benchmark, according to which $\langle w_{ij}\rangle_\text{WRGM}=\frac{W}{V}=\frac{2W}{N(N-1)}$, $\forall\:i<j$ \cite{Garlaschelli2009b}. From the definition of weighted modularity provided above, it follows that}

\begin{align}
Q&=\frac{1}{2W}\sum_{i\neq j}w_{ij}\delta_{c_i,c_j}-\frac{1}{2W}\sum_{i\neq j}\langle w_{ij}\rangle_\text{WRGM}\delta_{c_i,c_j}\\
&=\frac{1}{2W}\sum_c[2w_c^*-\langle w_{ij}\rangle_\text{WRGM} N_c(N_c-1)]\\
&=\frac{1}{2W}\sum_c\left[2w_c^*-2W\frac{N_c(N_c-1)}{N(N-1)}\right]\\
&=\left(\frac{w_\bullet^*}{W}-\frac{V_\bullet}{V}\right)
\end{align}
\textcolor{black}{where $N_c$ and $w_c^*$ indicate the number of nodes of community $c$ and the weight of links \emph{within} community $c$, respectively - naturally, $w_\bullet^*=\sum_cw_c^*$. As in the binary case, the calculation above clarifies that a positive value of $Q$ implies that the expected weight of any `internal' link is larger than the one predicted by the WRGM, i.e.}

\begin{equation}
Q\geq 0\Longrightarrow\langle w_{ij}\rangle_\text{WSBM}=\frac{w_\bullet^*}{V_\bullet}\geq\frac{W}{V}=\langle w_{ij}\rangle_\text{WRGM};
\end{equation}
\textcolor{black}{a null modularity value, on the other hand, implies that $\langle w_{ij}\rangle_\text{WSBM}$ coincides with the expectation coming from the WRGM. Analogously to the binary case, the result above can be also restated as follows: a null modularity value points out that the mesoscale structure of the configuration at hand \emph{has not been generated} by a WSBM (or, equivalently, \emph{does not need} a WSBM to be explained). Again, however, the modularity provides no indication about the statistical significance of the recovered partition.}\\

\begin{figure*}[t!]
\includegraphics[width=0.47\textwidth]{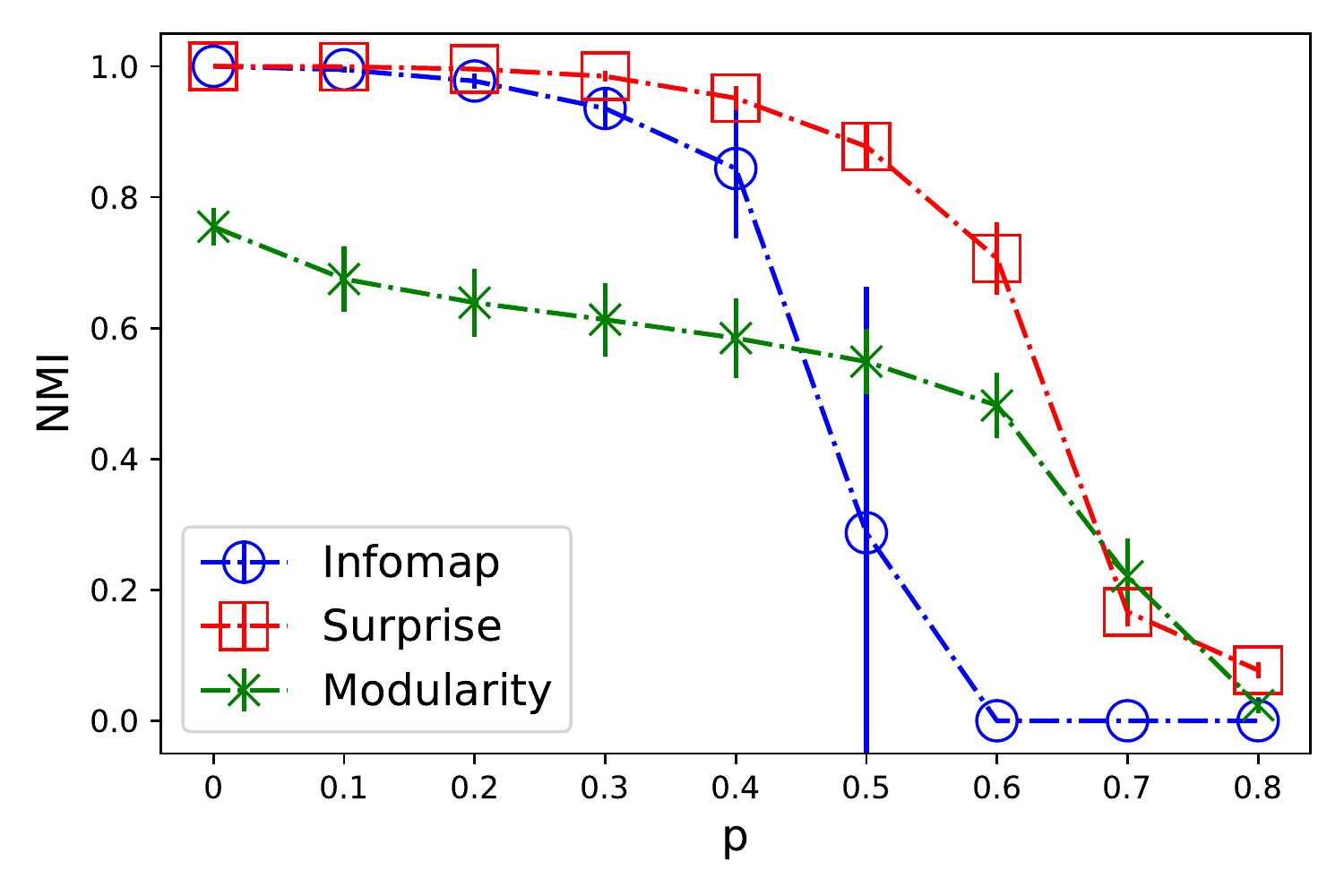}
\hspace{5mm}
\includegraphics[width=0.47\textwidth]{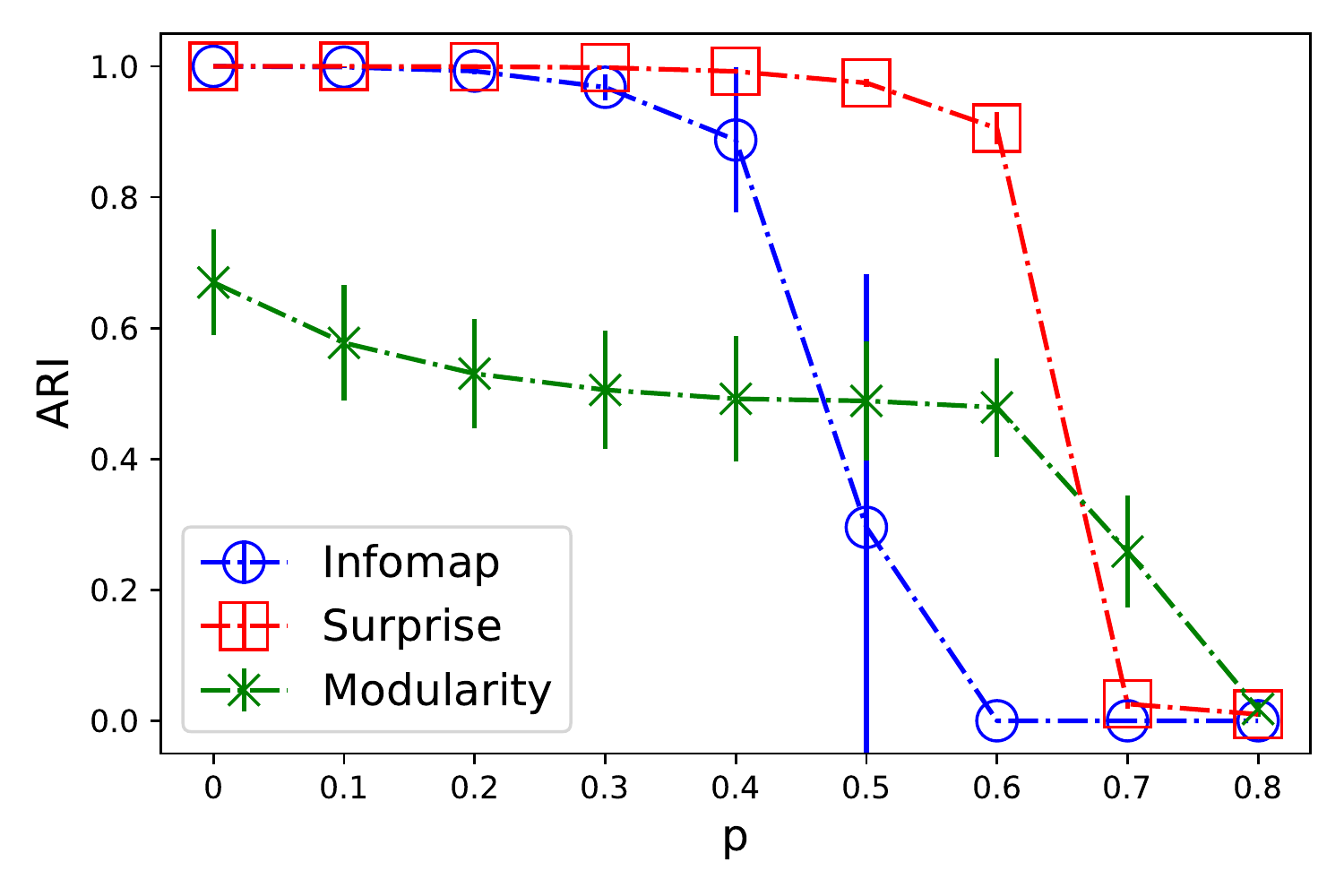}\\
\includegraphics[width=0.47\textwidth]{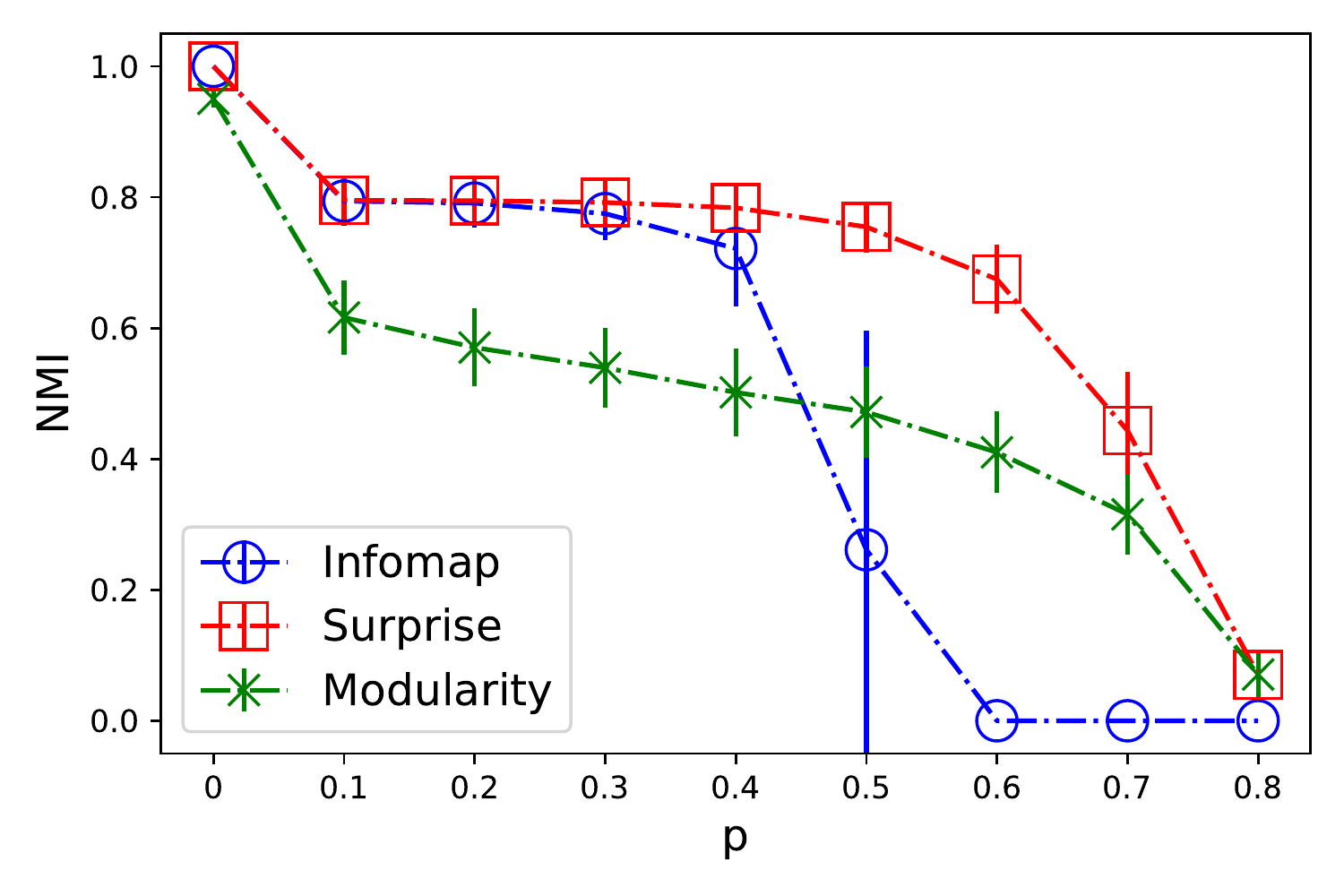}
\hspace{5mm}
\includegraphics[width=0.47\textwidth]{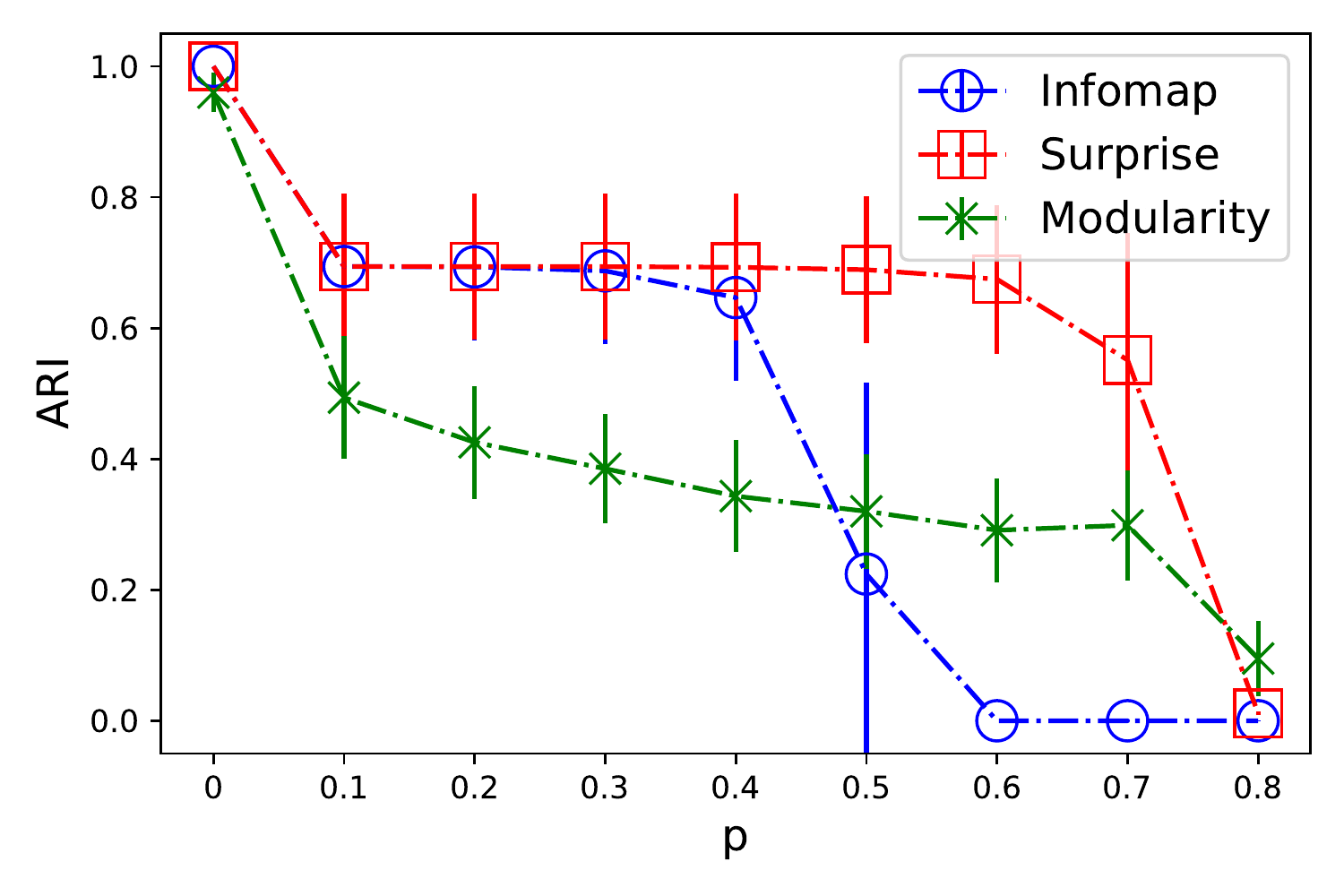}
\caption{\textcolor{black}{Comparison of three different algorithms for community detection on the RC benchmark, i.e. modularity maximization, Infomap and surprise minimization. The chosen configurations are binary, undirected (first row) and directed (second row) networks, consisting of 512 nodes, grouped in 16 communities arranged in a ring-like fashion and whose sizes obey a power-law whose exponent has been set to 1.8; the smallest community is composed by 3 nodes. Such initial configurations are progressively `degraded' according to the following mechanism: first, a percentage $p$ of links is selected randomly and removed; afterwards, a percentage $p$ of links is selected randomly and rewired. The trend of the NMI, plotted as a function of the (single) `degradation' parameter $p$, driving the evolution of the initial ring-of-cliques towards a progressively less-defined clustered configuration, reveals surprise to outperform both modularity and Infomap. From a more general perspective, these results confirm what has been already observed elsewhere, i.e. that the best-performing algorithms on the LFR benchmarks often perform poorly on the RC benchmarks and vice versa.}}
\label{fig7}
\end{figure*}

\begin{figure*}[t!]
\includegraphics[width=0.278\textwidth]{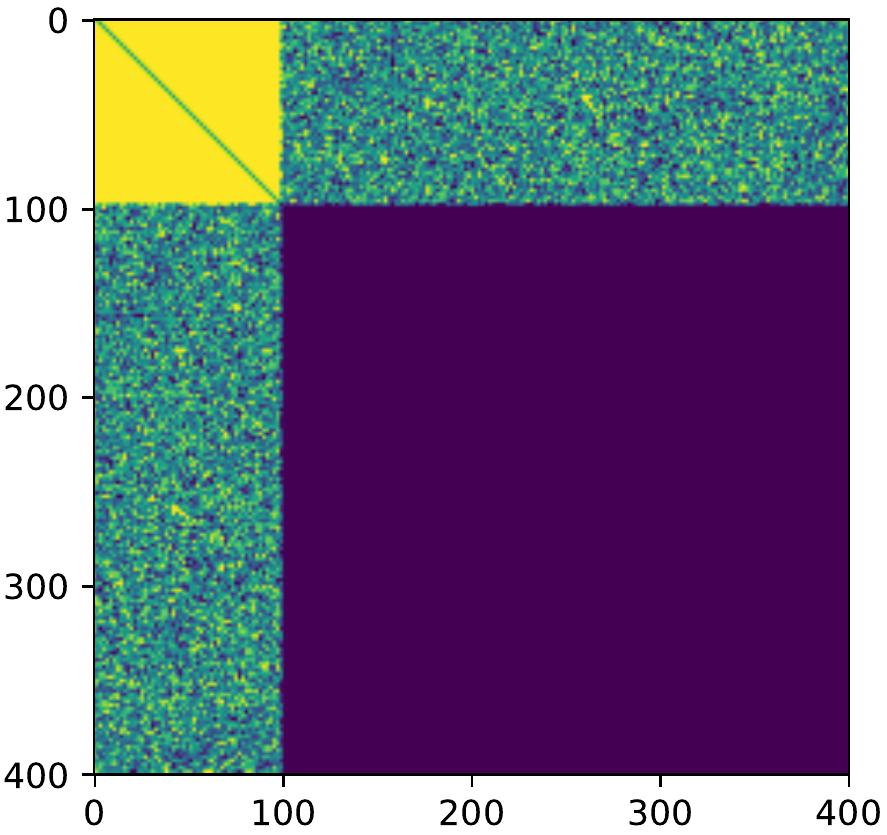}
\hspace{1mm}
\includegraphics[width=0.278\textwidth]{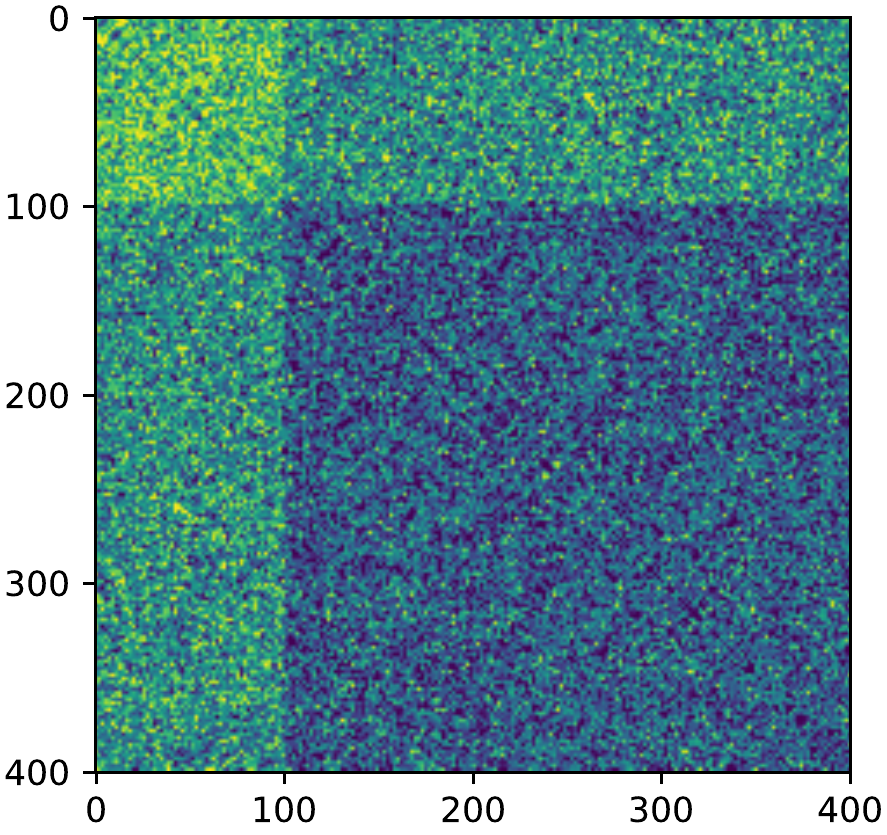}
\hspace{1mm}
\includegraphics[width=0.278\textwidth]{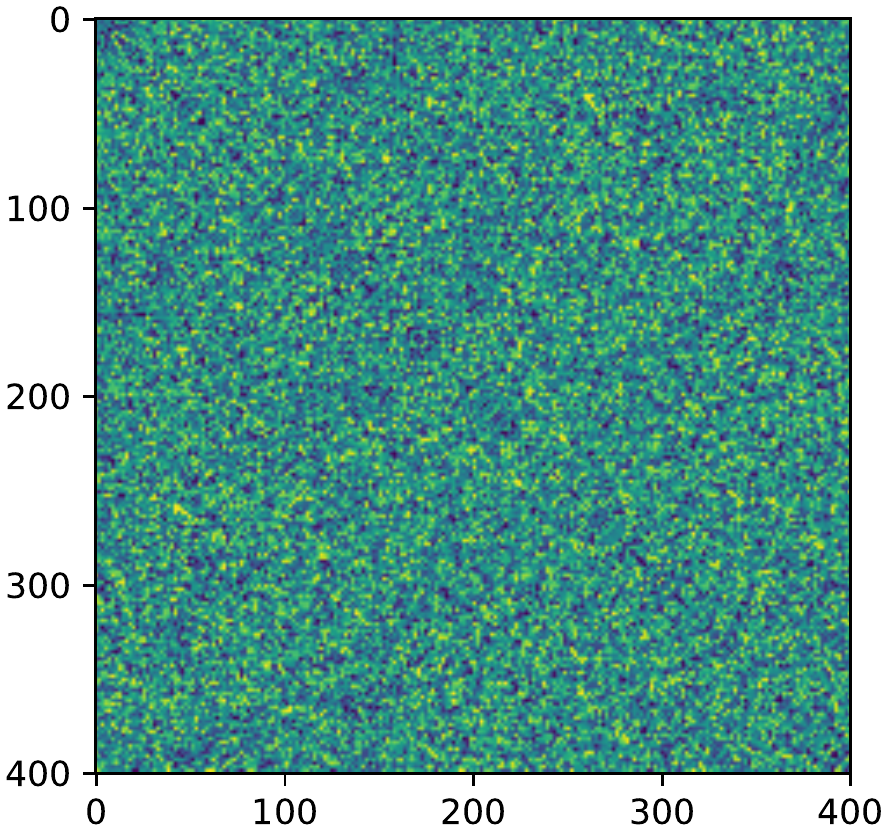}\\
\includegraphics[width=0.42\textwidth]{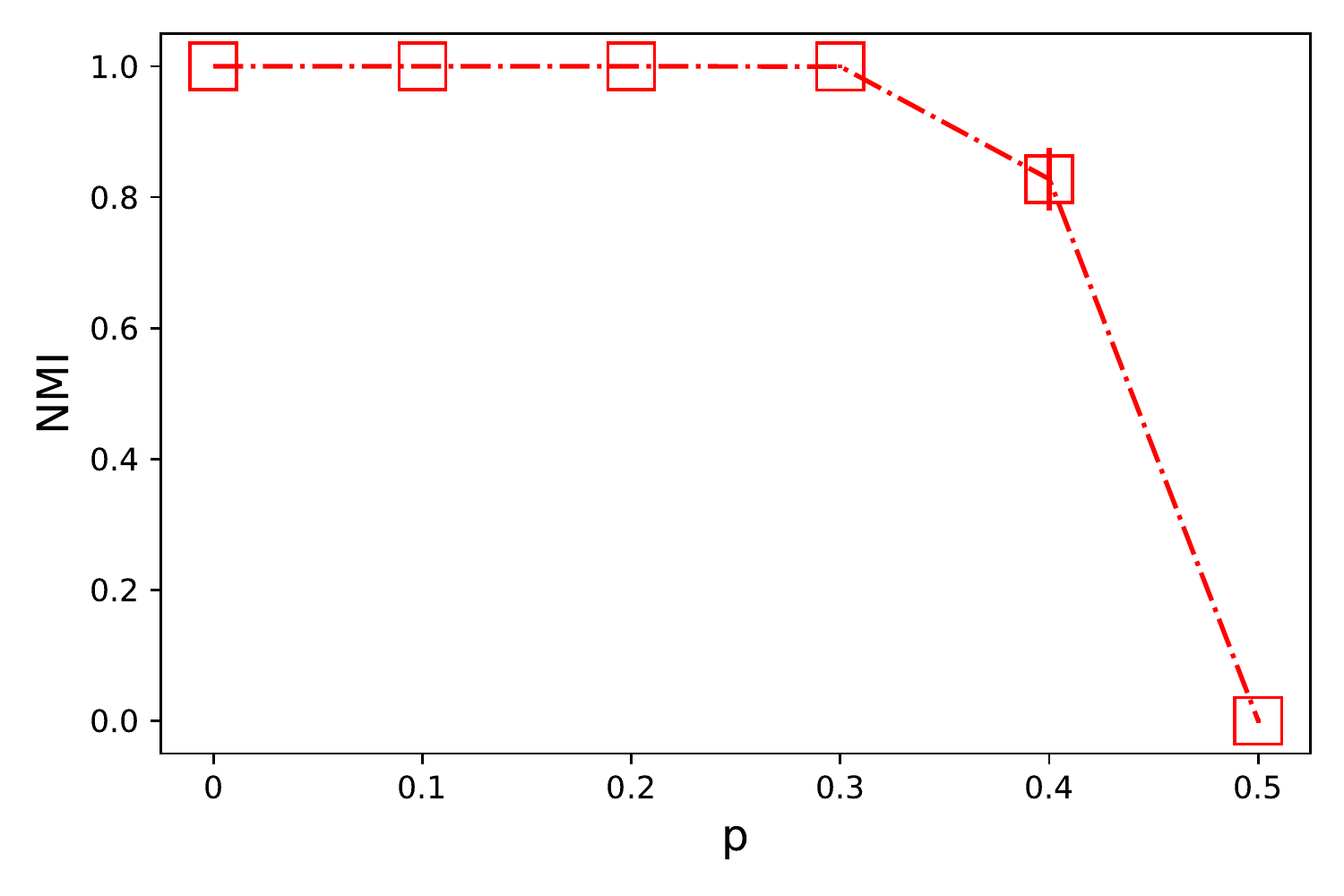}
\hspace{5mm}
\includegraphics[width=0.42\textwidth]{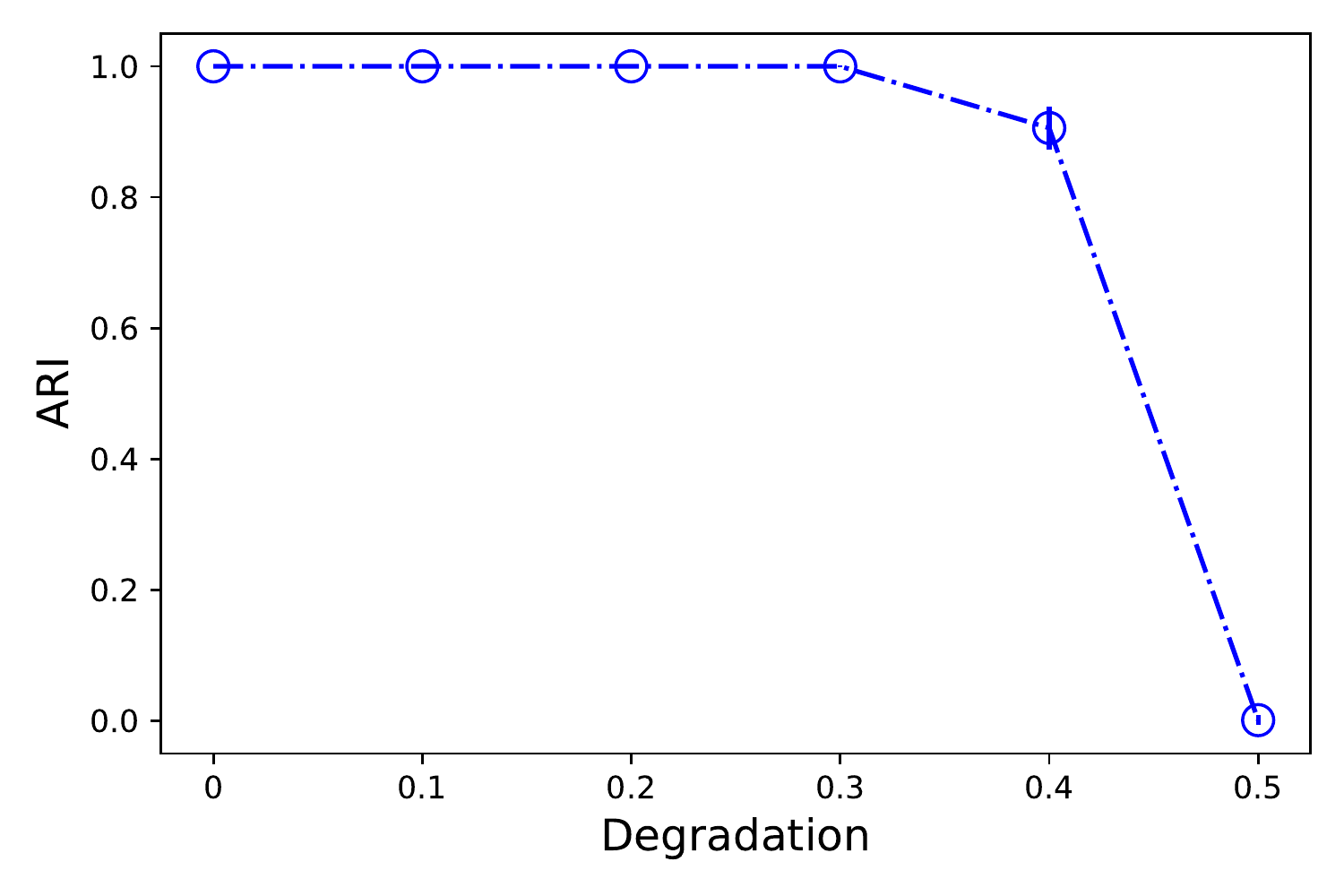}\\
\includegraphics[width=0.42\textwidth]{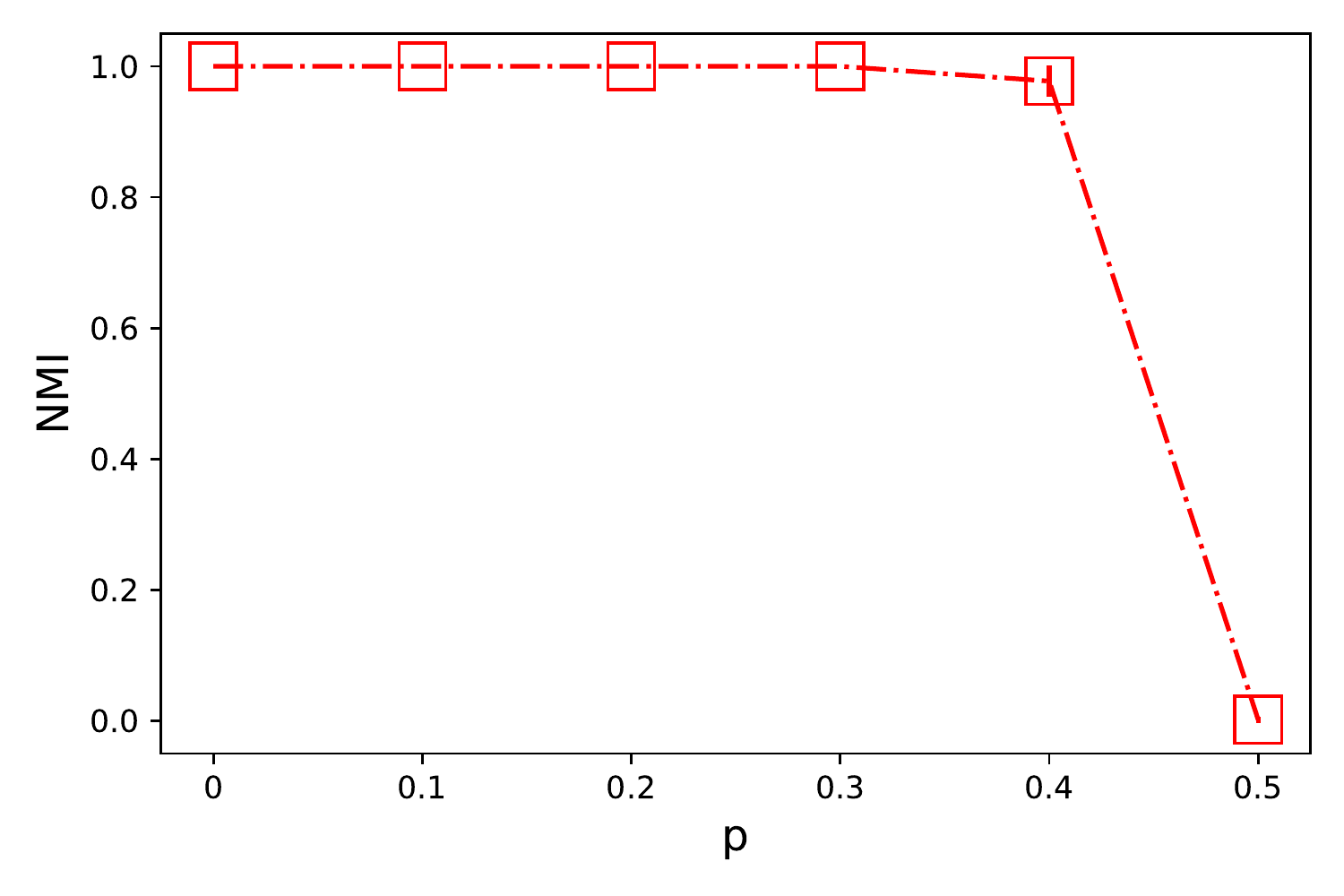}
\hspace{5mm}
\includegraphics[width=0.42\textwidth]{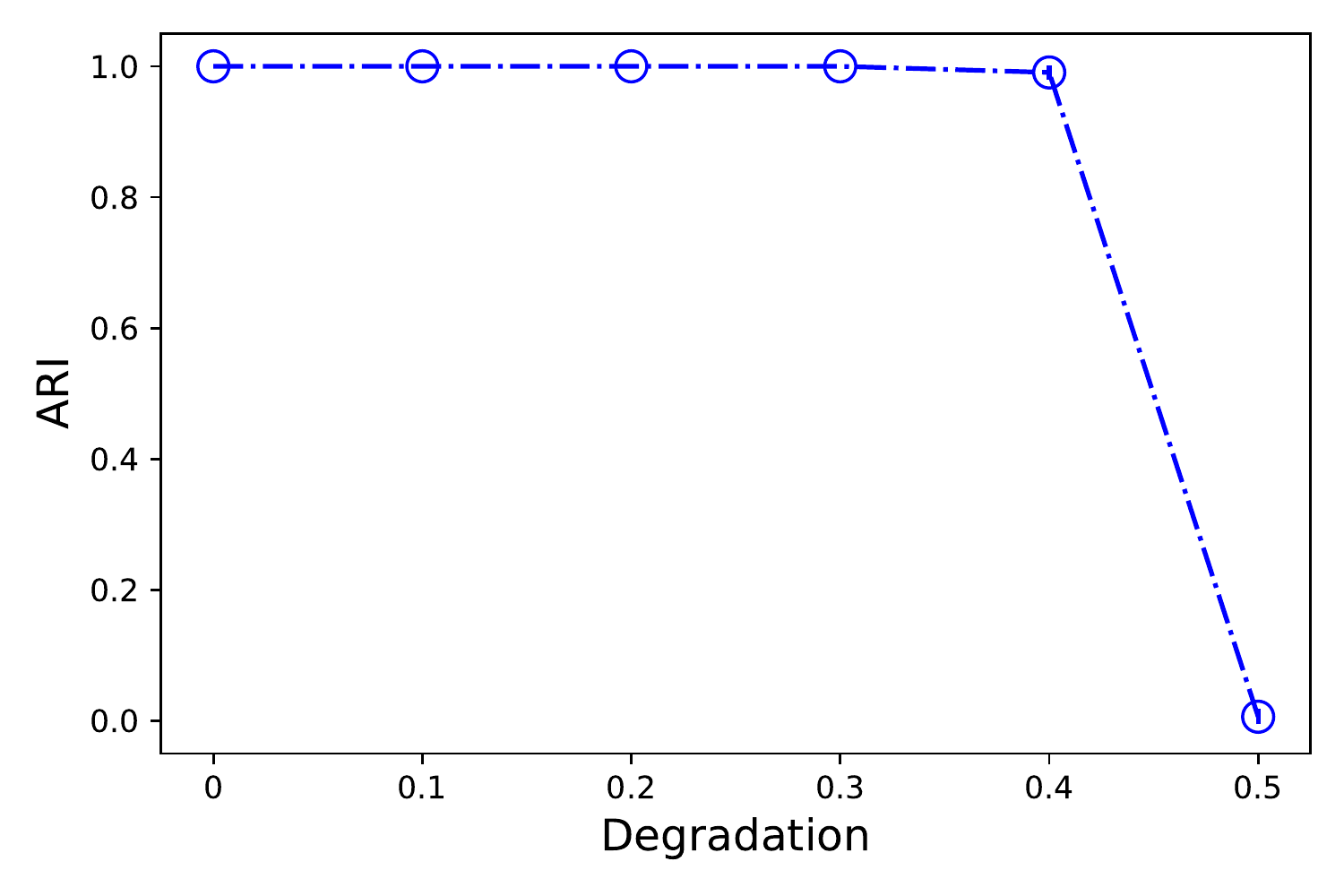}\\
\includegraphics[width=0.15\textwidth]{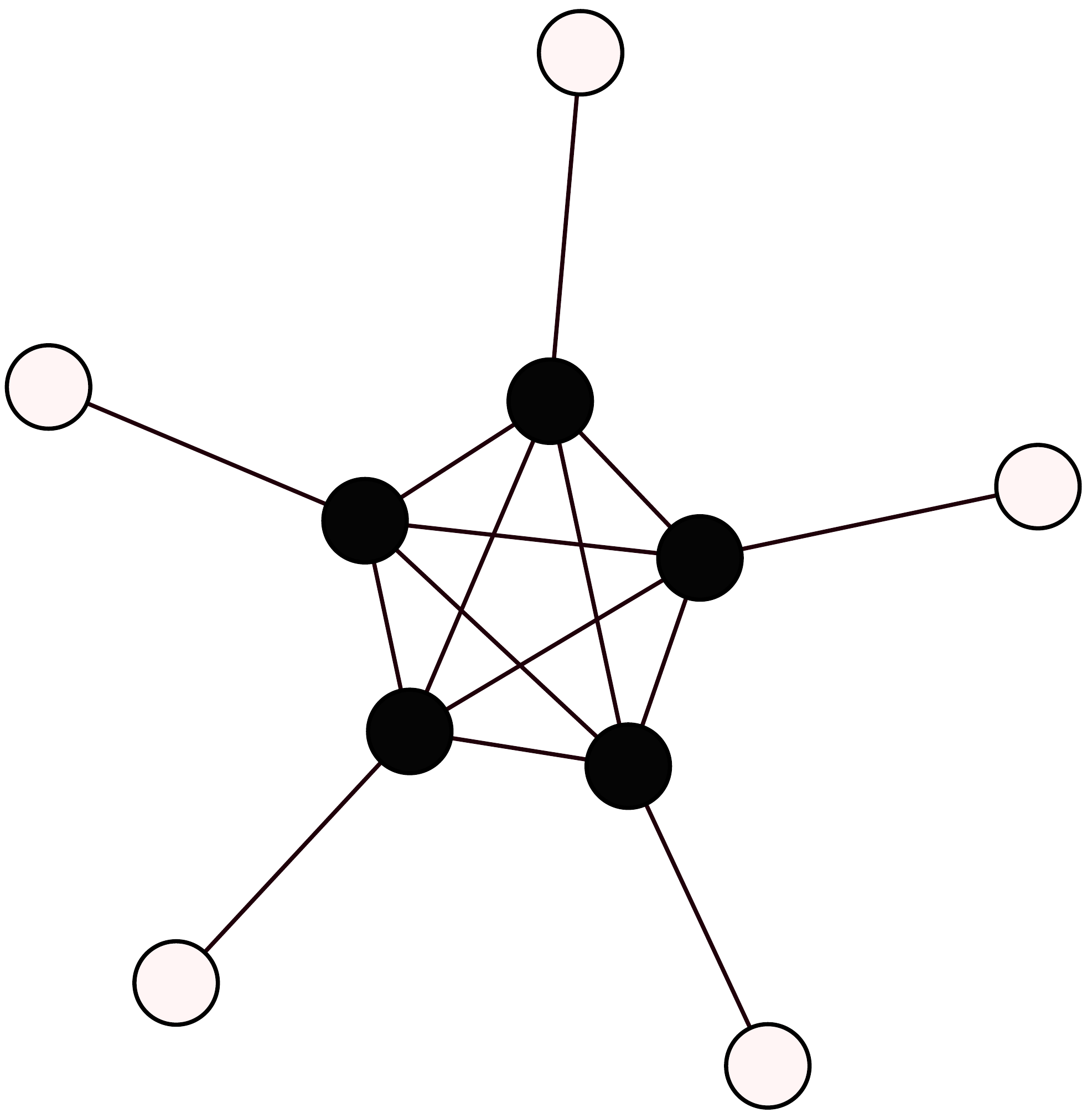}
\hspace{1mm}
\includegraphics[width=0.15\textwidth]{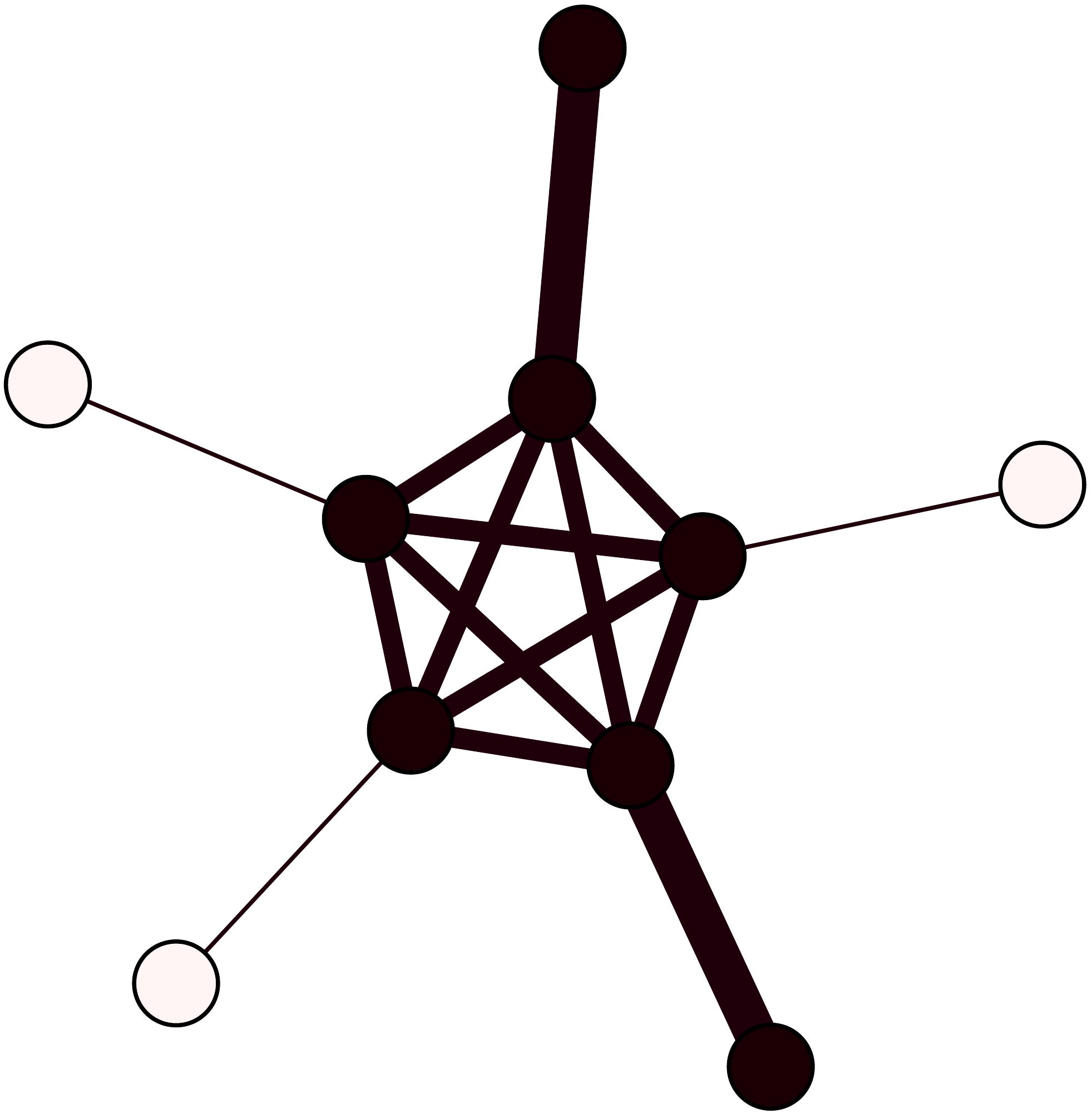}
\hspace{3mm}
\includegraphics[width=0.25\textwidth]{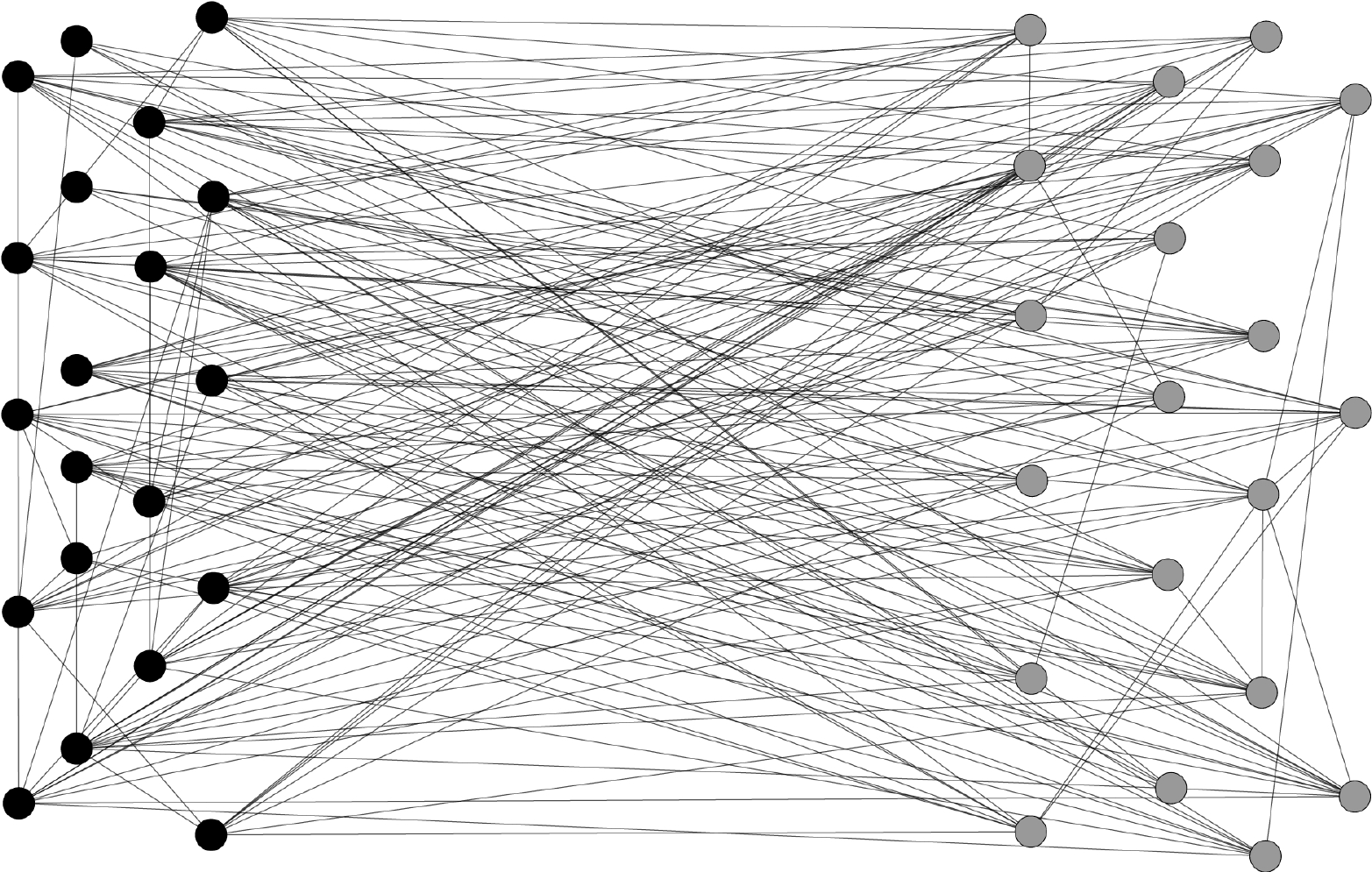}
\hspace{1mm}
\includegraphics[width=0.25\textwidth]{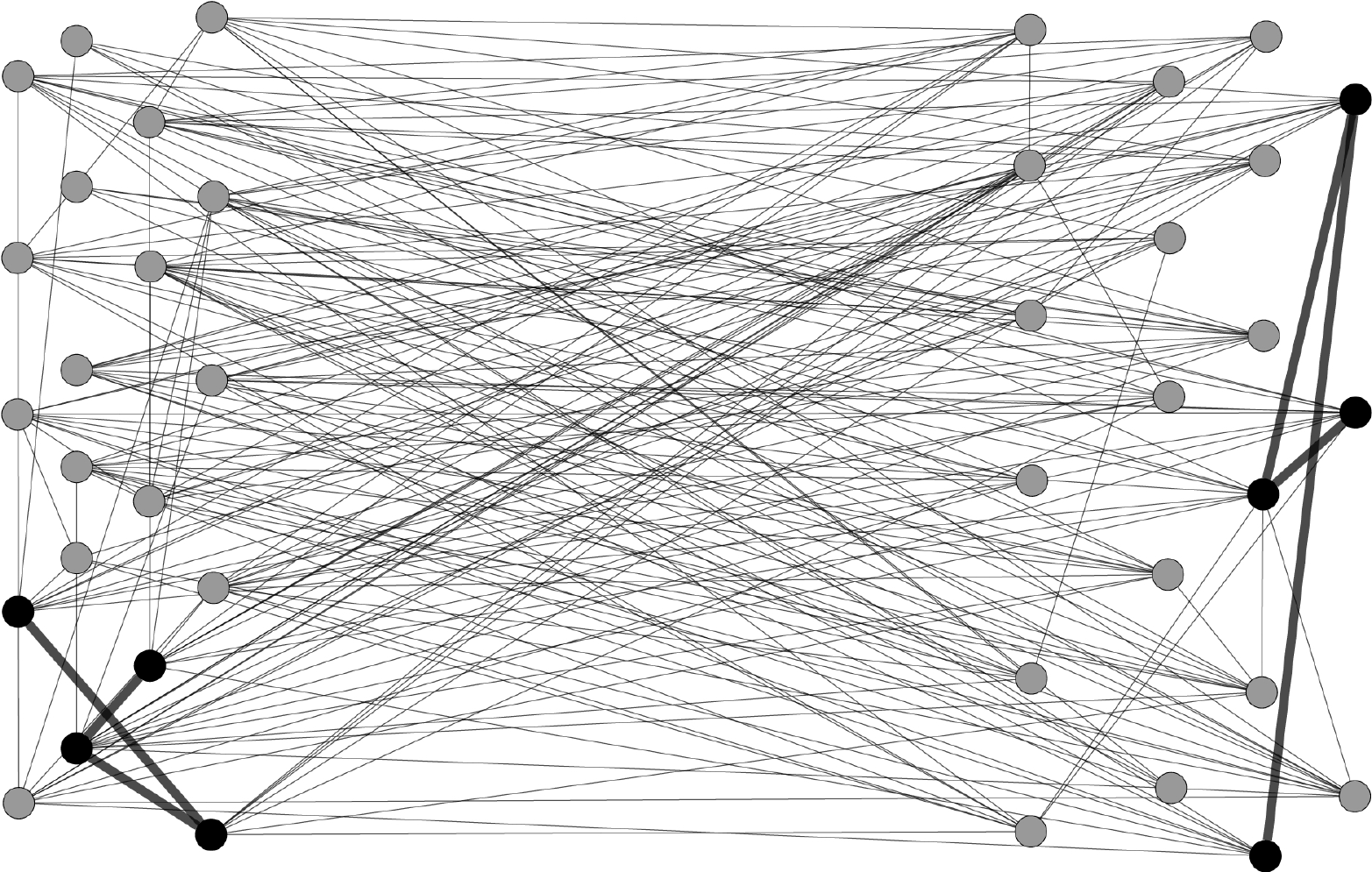}
\caption{\textcolor{black}{First, second and third row: benchmark for testing surprise on the recovery of core-periphery structures (left column: binary, undirected networks; right column: binary, directed networks). The one we have defined here progressively `degrades' an initial configuration, defined by 1) a completely connected core, 2) an empty periphery, 3) an intermediate part whose link density amounts at $p_{cp}=0.5$. Such a configuration is `degraded' by progressively filling the periphery and emptying the core. This is achieved by 1) considering all peripherical node pairs and link them with probability $q$; 2) considering all core node pairs and keep them linked with probability $1-q$: varying $q$ in the interval $[0,p_{cp}]$ allows us to span a range of configurations starting with Borgatti-Everett and ending with Erd\"os-R\'enyi. As expected, the performance of the surprise worsens as the `degradation' parameter becomes closer to $p_{cp}=0.5$; however, both the NMI and the ARI indices steadily remain very close to 1. Bottom row: the presence of weights affects the detection of `bimodular' mesoscale structures as well. In fact, rising the weight of any two links connecting the core with the periphery of a toy network allows the two nodes originally part of the periphery to be detected as belonging to the core; analogously, if a bipartite topology is modified by adding weights between some of the nodes belonging to the same layer, a core-periphery structure will now be detected as significant.}}
\label{fig8}
\end{figure*}

\noindent\textcolor{black}{{\bf Comparing mesoscale structures detection methods.} The previous subsections have been devoted to check the consistency of our surprise-based formalism and reveal the `statistical flaws' of other approaches. Let us now consider two popular algorithms for mesoscale structures detection, i.e. modularity maximization (now, $Q$ has been considered in its `full' definition, e.g. $\langle a_{ij}\rangle=p_{ij}=\frac{k_ik_j}{2L}$, $\forall\:i<j$ for binary, undirected configurations) and Infomap \cite{Rosvall2008}, and carry out more systematical comparisons between the former and the surprise. Upon doing so, we are able to compare one algorithm per class, i.e. modularity for the first class, surprise for the second class and Infomap for the third class.}

\textcolor{black}{To this aim, we have focused on different kinds of benchmarks, i.e. classes of synthetic networks with well-defined planted partitions, the aim being that of inspecting the goodness of a given algorithm in recovering the imposed partition. As an indicator of the goodness of the partition retrieved by each algorithm, we have followed \cite{Bongiorno2017} and employed three different indices. The first one is the \emph{normalized mutual information} (NMI), defined as}

\begin{equation}
\overline{I}(X,Y)=\frac{2I(X,Y)}{H(X)+H(Y)}
\end{equation}
\textcolor{black}{where partitions $X$ and $Y$ are compared, $H(X)=-\sum_xf_x\ln f_x$ and $f_x=\frac{n_x}{n}$ is the fraction of nodes assigned to the cluster labeled with $x$; analogously, for the partition $Y$. The term $I(X,Y)=\sum_x\sum_yf_{xy}\ln\left(\frac{f_{xy}}{f_xf_y}\right)$ is the `proper' \emph{mutual information} and $f_{xy}=\frac{n_{xy}}{n}$ is the fraction of nodes assigned to cluster $x$ in partition $X$ and to cluster $y$ in partition $Y$. Naturally, $\overline{I}(X,Y)$ equals 1 if the partitions are identical and 0 if the partitions are independent.}

\textcolor{black}{The second index we have considered is the \emph{adjusted Rand index} (ARI), defined as}

\begin{equation}
\text{ARI}=\frac{TP+TN-\langle TP+TN\rangle}{TP+FP+TN+FN-\langle TP+TN\rangle}
\end{equation}
\textcolor{black}{and representing a sort of accuracy\footnote{The number of true positives (TP) is the number of pairs of nodes being in the same community both in the considered and in the reference partition; the number of false positives (FP) is the number of pairs of nodes being in the same community in the considered partition but in different communities in the reference partition; the number of true negatives (TN) is the number of pairs of nodes being in the same community neither in the considered nor in the reference partition; the number of false negatives (FN) is the number of pairs of nodes being in the same community in the reference partition but not in the considered partition.} `corrected' by a term that quantifies the agreement between the reference partition and a random partition - the term `random' referring to the Permutation Model; equivalently, the closer the ARI to 0, the more `random' the provided partition.}

\begin{figure*}[t!]
\includegraphics[width=0.35\textwidth]{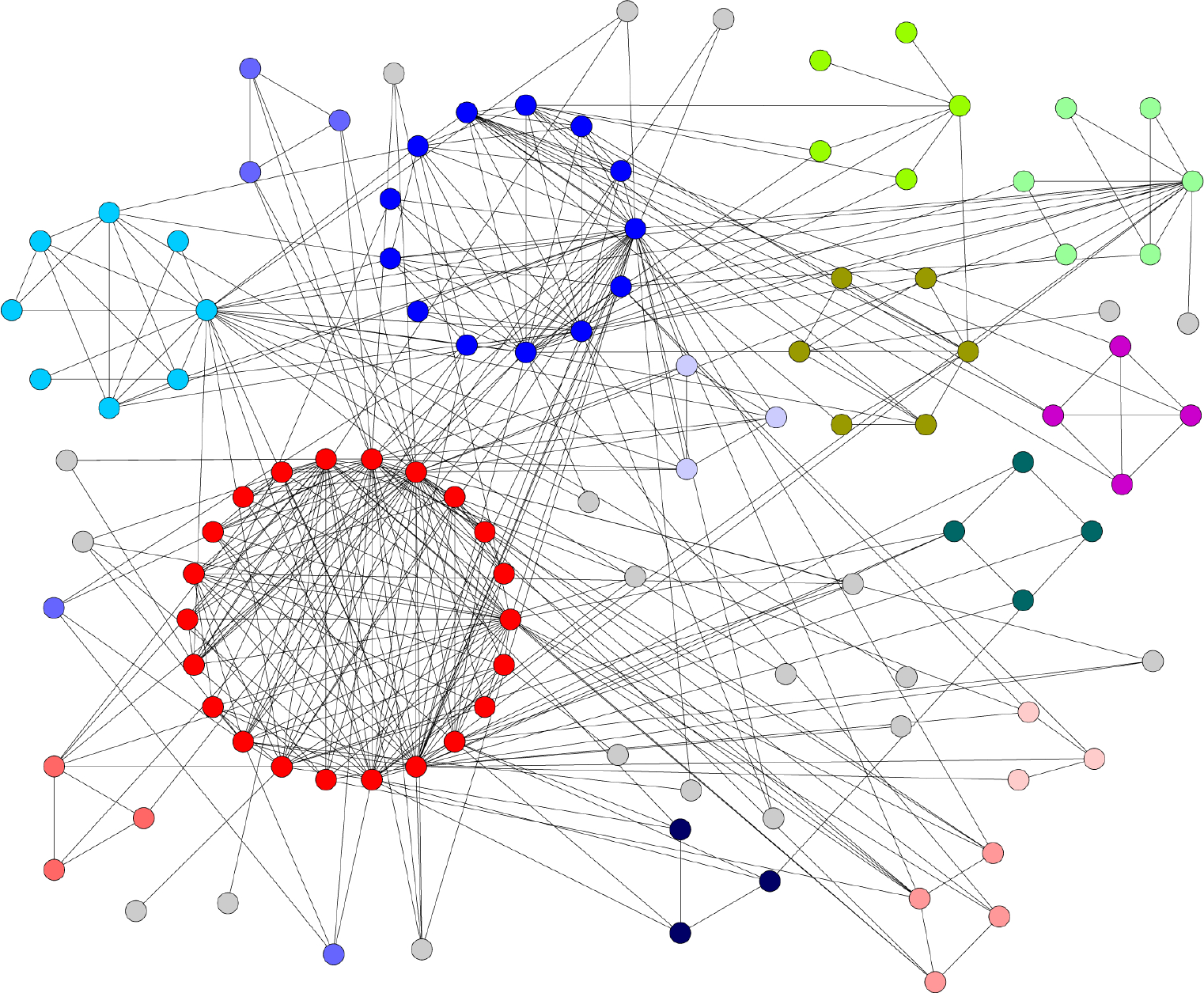}
\hspace{1cm}
\includegraphics[width=0.49\textwidth]{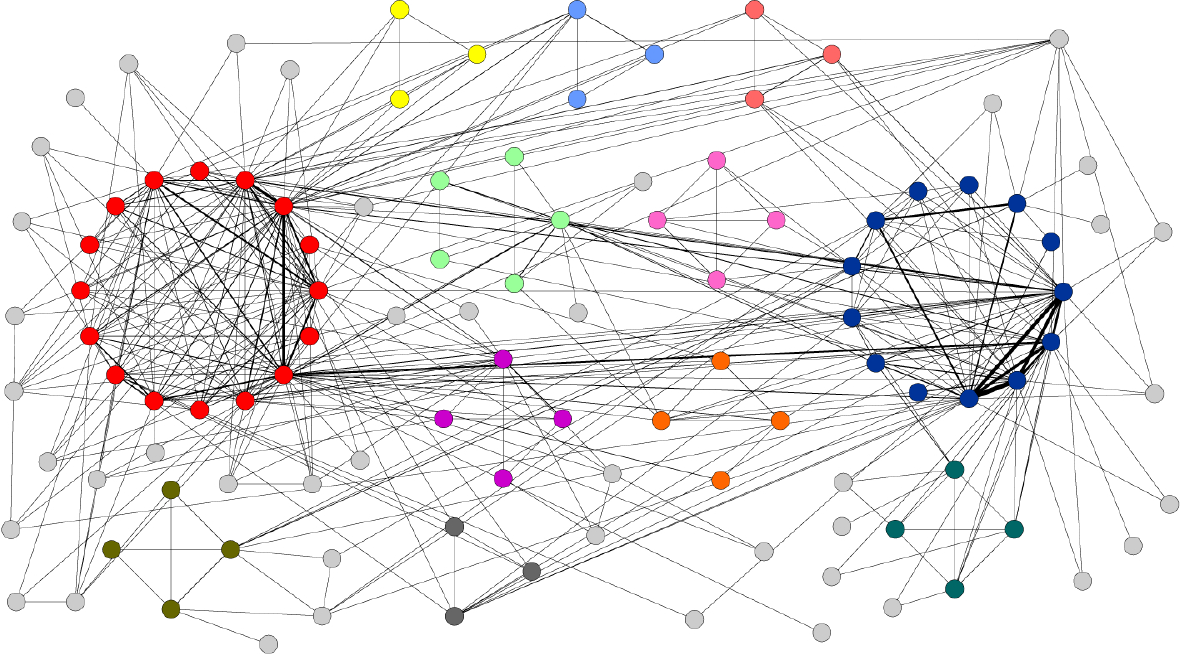}\\
\vspace{1cm}
\includegraphics[width=0.45\textwidth]{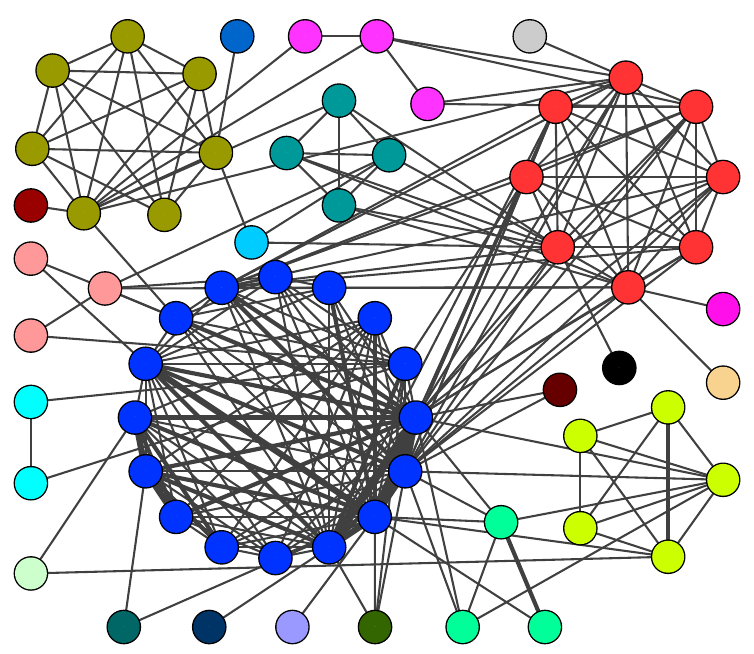}
\hspace{1cm}
\includegraphics[width=0.4\textwidth]{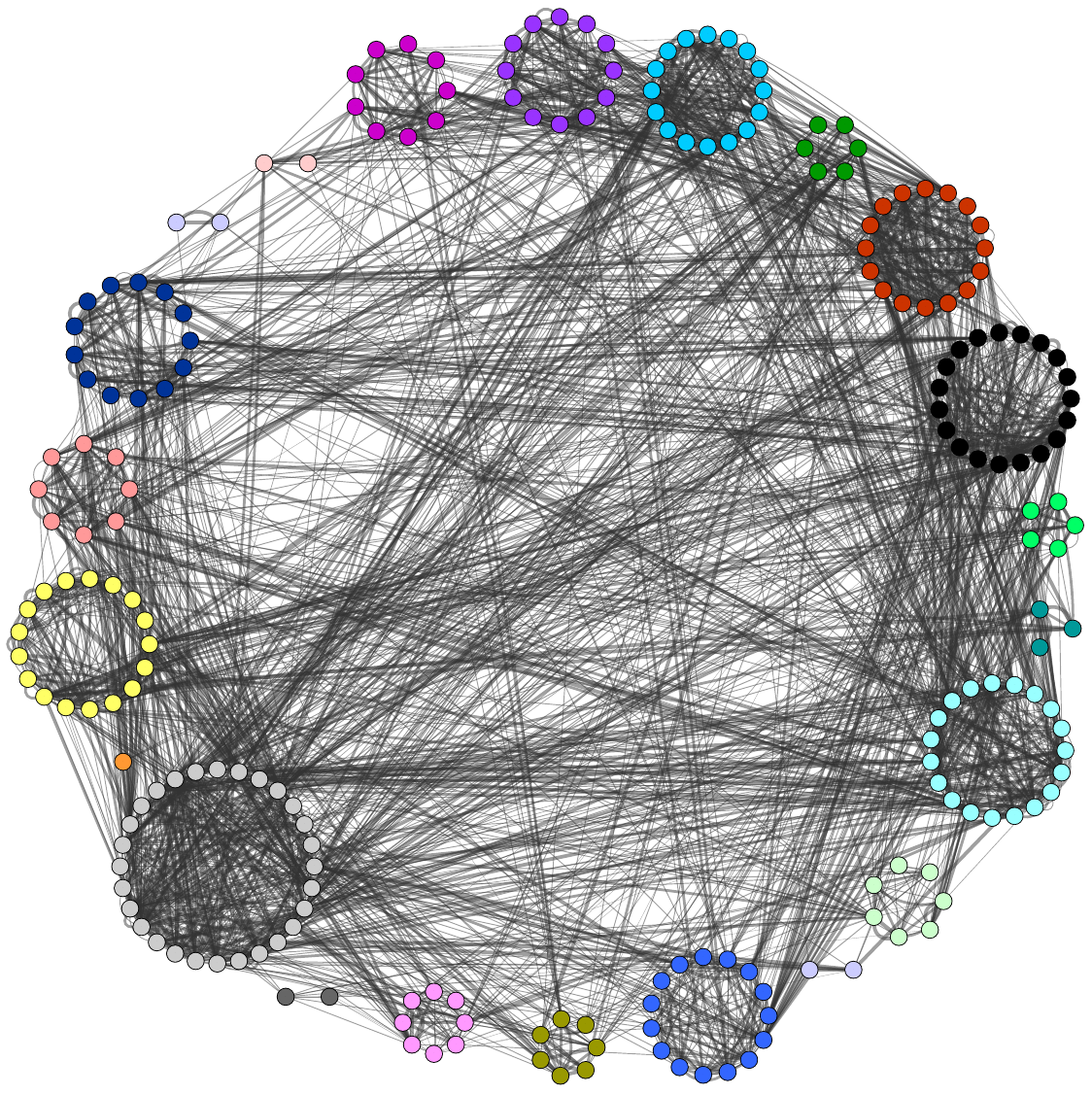}
\caption{Application of our framework for the detection of binary and weighted communities on a bunch of social networks: (top left) binary network of co-occurrences between Star Wars characters \cite{repository0}, (top right) weighted network of co-occurrences between Star Wars characters \cite{repository0}, (bottom left) weighted `friendship' network among the terrorists involved in the train bombing of Madrid in 2004 \cite{repository}, (bottom right) weighted `friendship' network among the residents living in an Australian University Campus \cite{repository}. Overall, link weights refine the picture provided by just considering the presence of links. A first example is provided by the two `friendship' networks where sparse subgraphs can be considered as communities whenever their connections are `heavy' enough; a second example is provided by the weighted communities of Star Wars characters: the three major clusters, in this case, are the one induced by the characters of Episodes I-III, the one induced by the characters of Episodes IV-IX and the one contaning the villains of Episodes VII-IX.}
\label{fig9}
\end{figure*}

\textcolor{black}{The third index we have considered is the \emph{adjusted Wallace index} (AWI), defined as}

\begin{equation}
\text{AWI}=\frac{TP-\langle TP\rangle}{TP+FP-\langle TP\rangle}
\end{equation}
\textcolor{black}{and representing a sort of `corrected' positive predicted value. Again, the closer the AWI to 0, the more `random' the provided partition.}\\

\textcolor{black}{First, let us inspect the performance of modularity, surprise and Infomap to detect cliques arranged in a ring. Specifically, we have considered 7, different ring-like configurations, each one linking 20 binary cliques (i.e. $K_3$, $K_4$, $K_5$, $K_8$, $K_{10}$, $K_{15}$, $K_{20}$). As fig. \ref{fig4} reveals, surprise always recovers the planted partition; on the other hand, modularity maximization leads to miss the partitions with $K_3$ and $K_4$ and Infomap misses the partition with $K_3$, a result that may be a consequence of the resolution limit, affecting both the aforementioned algorithms.}

Let us now ask us if the presence of weights affects the detection of mesoscale structures. Generally speaking, the answer is yes, as the comparison between the `ring of binary cliques' and the `ring of weighted cliques' cases shows. In particular, the result according to which surprise minimization is able to discriminate the inter-linked cliques changes once the weight of the links connecting any two cliques is risen: in fact, this leads the algorithm to reveal as `communities' two tightly-connected pairs of cliques, now. We explicitly notice that the results shown in fig. \ref{fig4} also depend on the relative magnitude of the weights of the intra-cliques and of the inter-cliques links: however, as long as the inter-cliques weight is up to two orders of magnitude larger than the intra-cliques one, it holds true.\\

\begin{figure}[t!]
\includegraphics[width=0.23\textwidth]{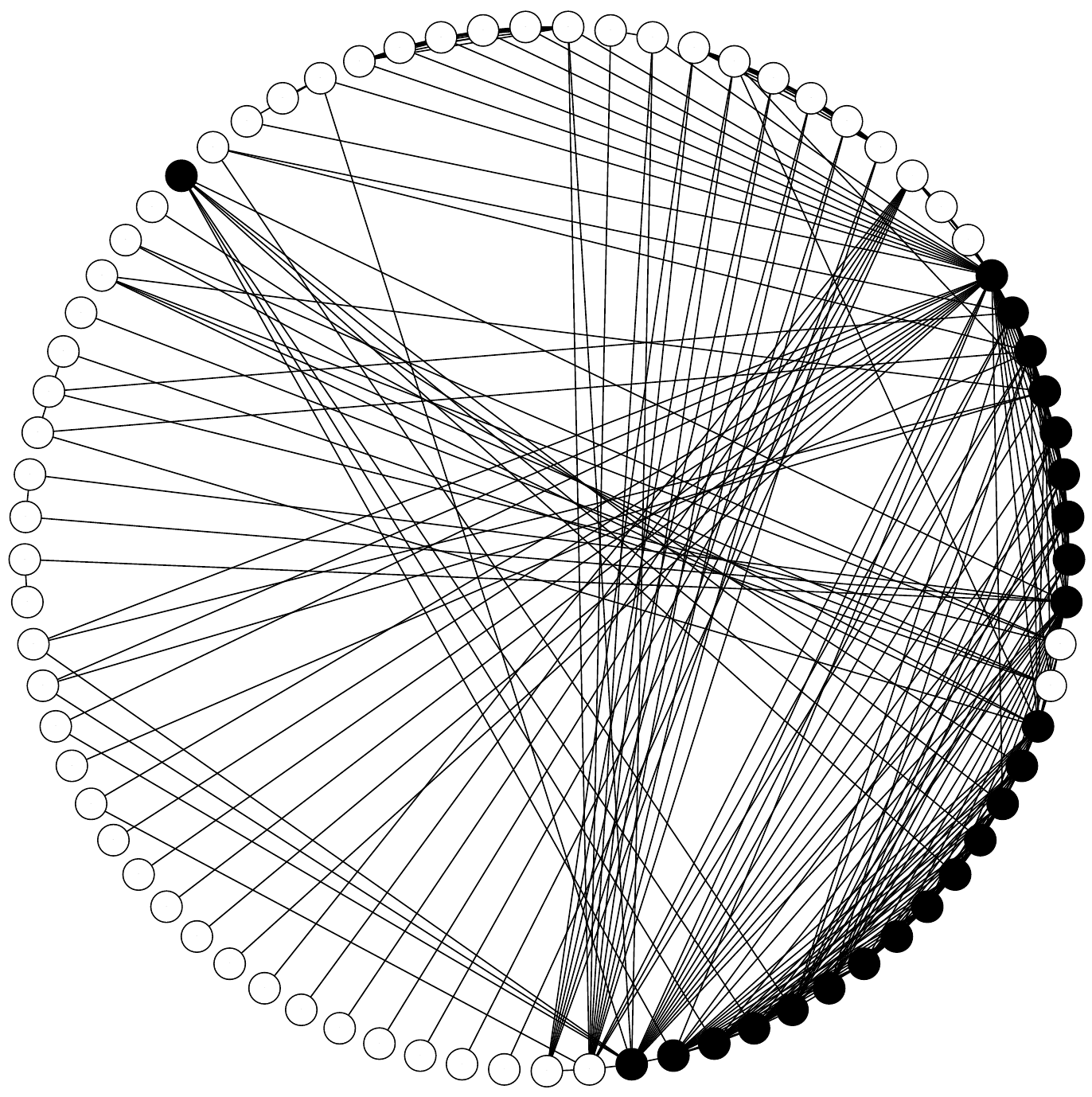}
\includegraphics[width=0.23\textwidth]{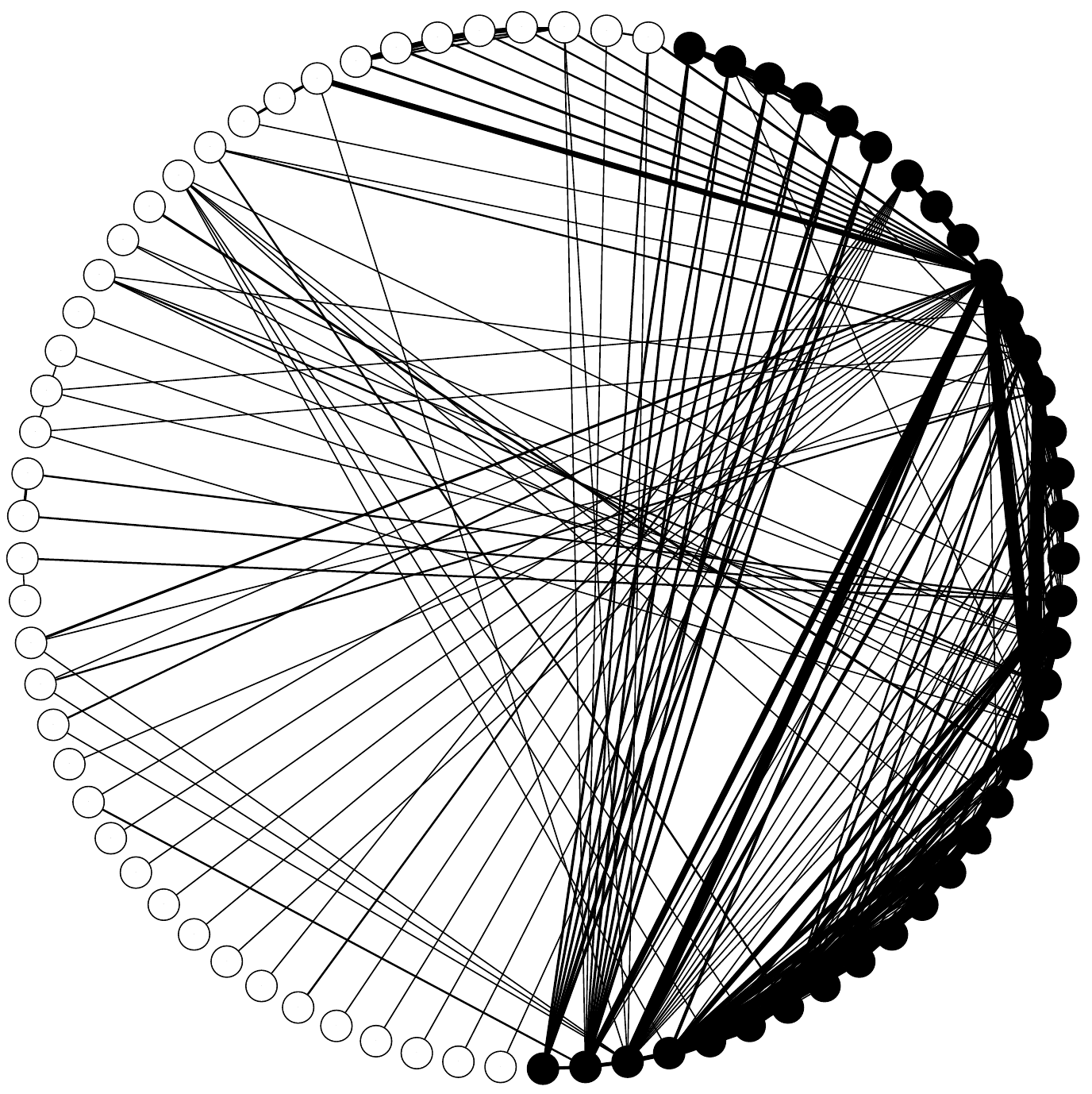}\\
\vspace{5mm}
\includegraphics[width=0.23\textwidth]{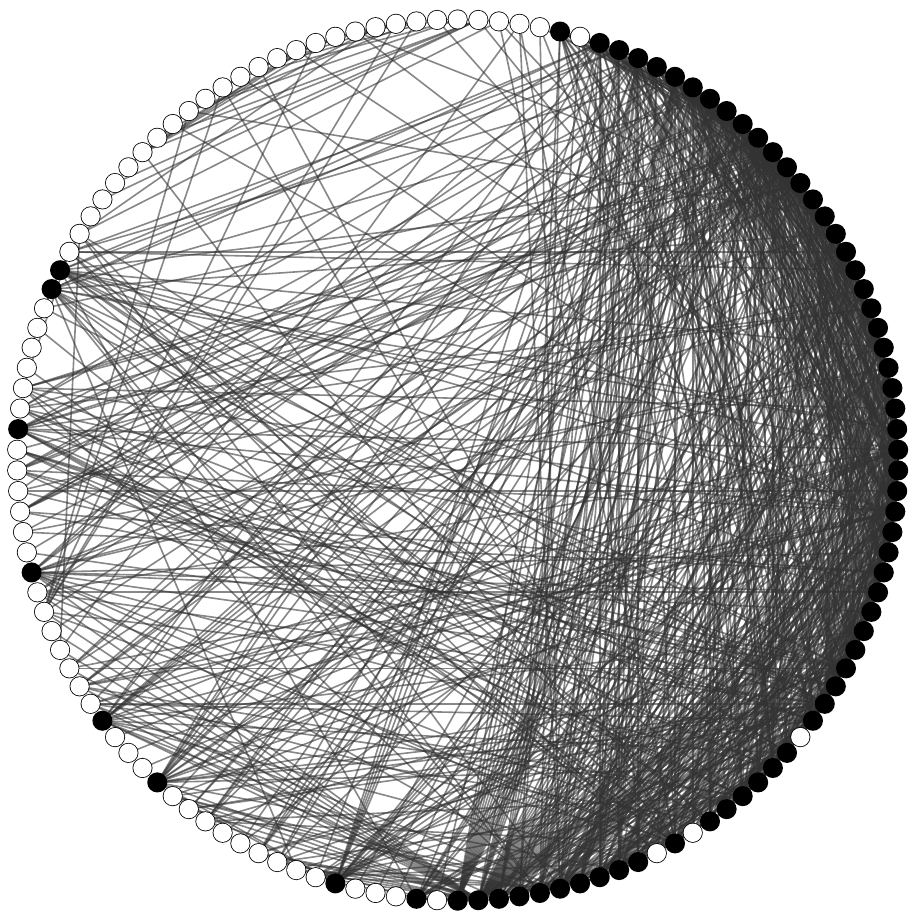}
\includegraphics[width=0.23\textwidth]{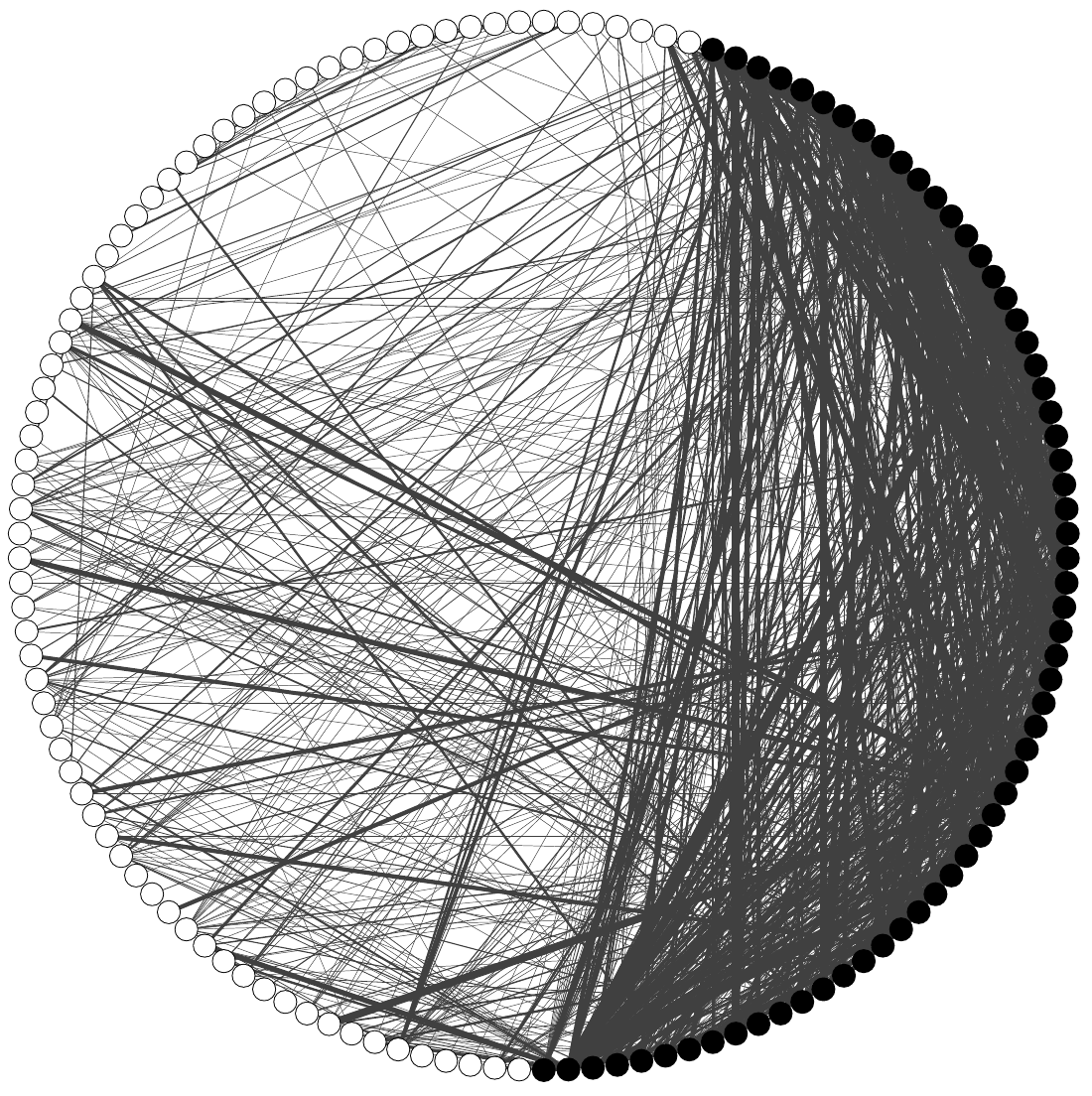}
\caption{Top panels: results of the application of our framework for the detection of `bimodular' structures on the network of co-occurrences of the `Les Miserables' characters \cite{repository} (first panel: binary version; second panel: weighted version). Link weights refine the picture provided by just considering the presence of links: the core (in black) is, in fact, constituted by the nodes connected by the `heavier' links, irrespectively from the link density of the former one. Bottom panels: results of the application of our framework for the detection of weighted `bimodular' structures on the electronic Italian Interbank Money Market (e-MID) \cite{Iori2006}, for the maintenance period approximately corresponding to May 2009 (third panel: binary version; fourth panel: weighted version). As the picture shows, the (purely binary) core links are also the `heavier' ones - an evidence confirming an ubiquitous tendency in economic and financial systems, i.e. binary and weighted quantities are closely related.}
\label{fig10}
\end{figure}

\textcolor{black}{In order to expand the set of comparisons, we have focused on two different kinds of well-established benchmarks, i.e. the Lancichinetti-Fortunato-Radicchi (LFR) one and the Aldecoa's `relaxed-caveman' (RC) one \cite{Aldecoa2021,Lancichinetti2009}.}

\textcolor{black}{The LFR benchmark is a special case of the planted-partition model, in which groups have different sizes and nodes have different degrees - hence constituting a refinement of the GN benchmark, where groups have equal size and nodes have the same expected degree \cite{Lancichinetti2009}. When binary, undirected configurations are considered, the degrees of nodes are distributed according to a power-law with exponent $\tau_1$ while the sizes of communities are distributed according to a power-law with exponent $\tau_2$. Once the sizes of communities have been drawn each node `receives' its own degree, say $k_i$: the percentage of these links connecting node $i$ with other internal nodes is, then, chosen to be $(1-\mu_t)k_i$, with $\mu_t$ playing the role of mixing parameter that controls for the sharpness of the planted partition. For binary, directed configurations, $\mu_t$ refers to in-degrees, which are distributed according to a power-law while the out-degrees are kept constant for all nodes; the other input parameters, instead, are the same used for undirected configurations.}

\textcolor{black}{Results on specific implementations of the LFR benchmark are shown in fig. \ref{fig6}. Infomap is, generally speaking, a strong performer; as evident upon looking at the first row, however, its performance decreases abruptly as the mixing parameter exceeds a threshold value that depends on the particular setting of the LFR benchmark. Modularity, instead, seems to be more robust (i.e. its performance `degrades' less rapidly as $\mu_t$ increases) although the resolution limit manifests itself when configurations with small communities are considered. Overall, the performance of surprise seems to constitute a good compromise between the robustness of modularity and the steadily high accuracy of Infomap. We would also like to stress that surprise competes with modularity although it employes much less information than the latter: in fact, while the benchmark employed by modularity coincides with the (sparse version of the) Configuration Model - hence, encodes the information on the entire degree sequence - surprise compares the RGM with the SBM, hence employing the information on the link density, both in a global and in a block-wise fashion.}

\textcolor{black}{Surprise becomes the best performer when binary, directed configurations are considered - see the second row of fig. \ref{fig6}: while the performance of modularity starts decreasing as soon as the value of $\mu_t$ is risen and $\text{NMI}_{Infomap}\simeq0$ when $\mu_t$ crosses the value of $0.6$, the performance of surprise `degrades' much more slowly - in fact, for some instances of the LFR benchmark, it achieves a large value of NMI even for values $\mu_t\geq 0.8$.}\\

\textcolor{black}{Let us now comment on the performance of our algorithms when weighted configurations are considered. The LFR benchmark, in fact, can be extended to account for weights as well. In this case, there are two mixing parameters: while the first one is the usual, topological mixing parameter, the second one, that will be indicated with $\mu_w$, accounts for the heterogeneity of weights - more precisely, controlling for the percentage of a node strength to be distributed on the links that connect it to the nodes outside its own community. The third parameter to be determined is the exponent of the strength distribution - that has been set to 1.5 for all realizations considered here. The results are, again, shown in fig. \ref{fig6} (to be noticed that we have kept one of the two parameters fixed and studied the dependence of NMI and ARI on the other: specifically, we have frozen the topological mixing parameter and studied the dependence of the results on $\mu_w$, thus inspecting the performance of our algorithms as the weights are redistributed on a fixed topology).}

\textcolor{black}{Infomap is, again, a strong performer although its performance keeps decreasing abruptly as $\mu_w$ exceeds a threshold value depending on the particular setting of the LFR benchmark; modularity, instead, performs worse than in the binary case although it is still more robust than Infomap. Although `degrading' less sharply than Infomap, the performance of the purely weighted surprise seems to be the worst, here; on the other hand, the enhanced surprise outperforms the competing algorithms for intermediate values of the topological mixing parameter, irrespectively from the size of the communities: in fact, $\text{NMI}_{surprise}=1$ even for $\mu_w=0.8$. Similar considerations hold true when weighted, directed configurations are considered (with the only difference that, now, modularity steadily performs worse than the other algorithms, except for the largest values of the mixing parameter). As for the binary cases, surprise competes with modularity although it employes much less information than the latter: in fact, while the benchmark employed by modularity now coincides with the (sparse version of the) Weighted Configuration Model - hence, encodes the information on the entire strength sequence - surprise compares the WRGM with the WSBM, hence employing the information on the magnitude of the total weight, both in a global and in a block-wise fashion.}\\

\begin{figure*}[t!]
\includegraphics[width=0.47\textwidth]{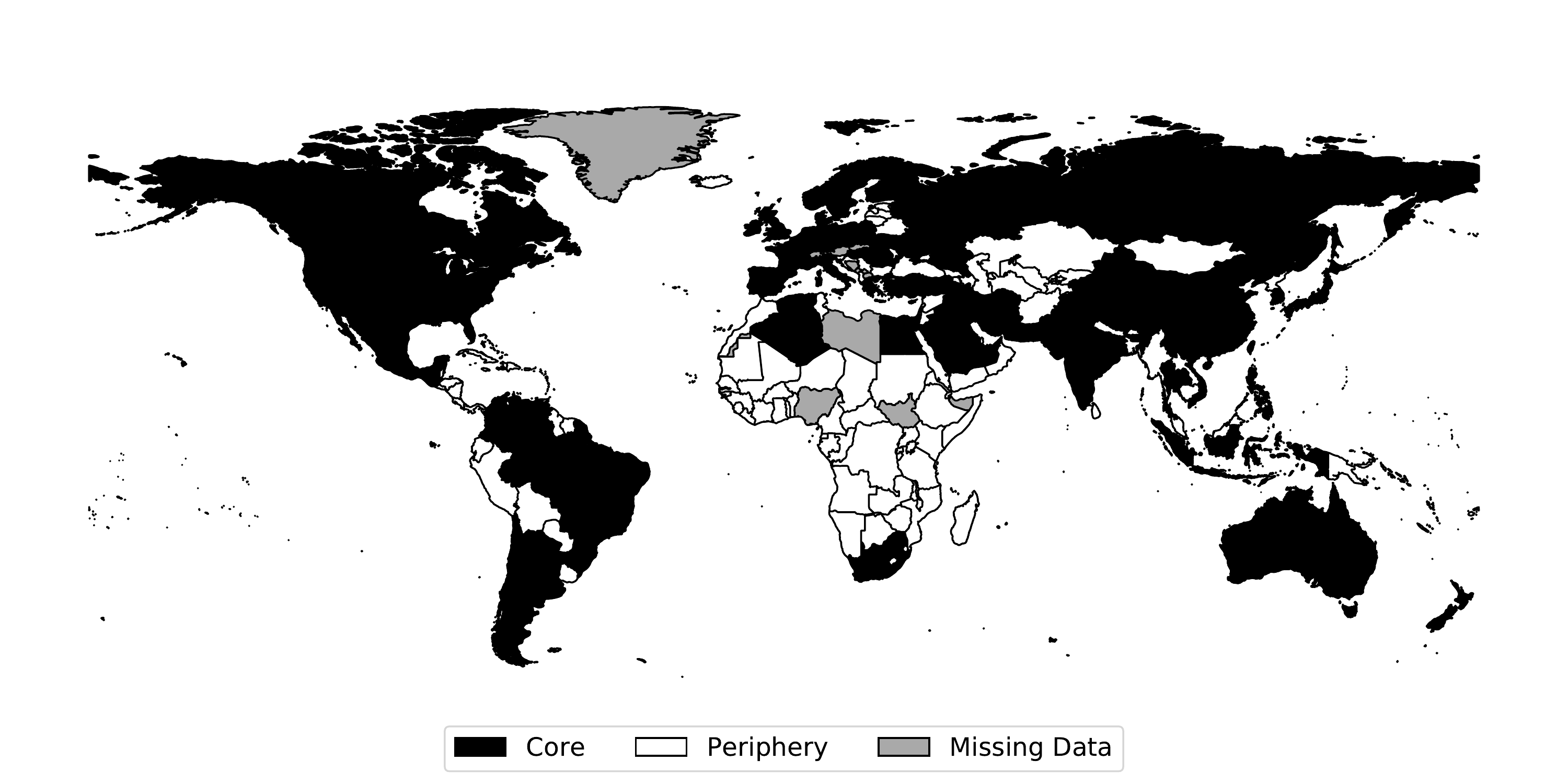}
\hspace{5mm}
\includegraphics[width=0.47\textwidth]{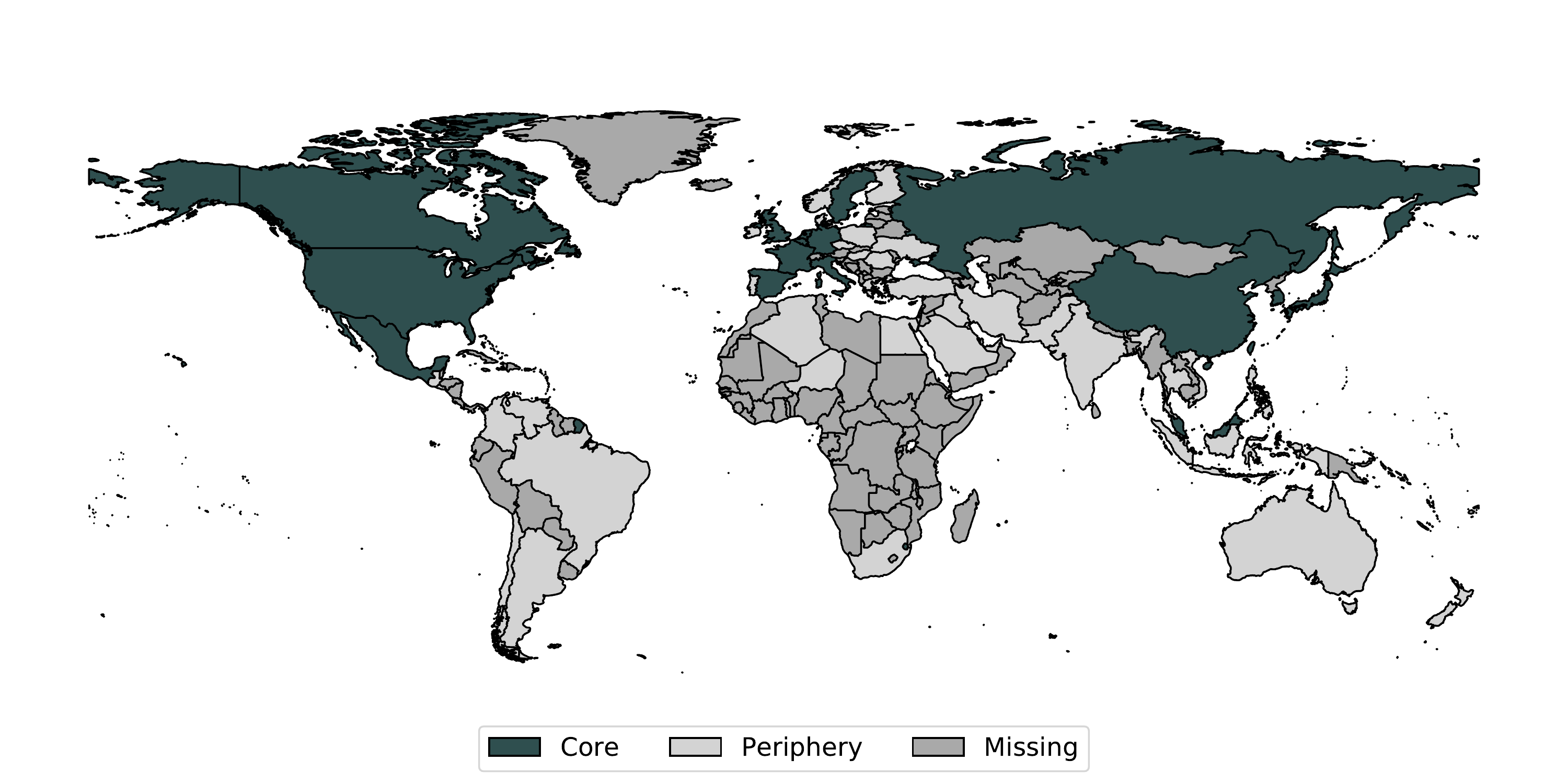}
\includegraphics[width=0.47\textwidth]{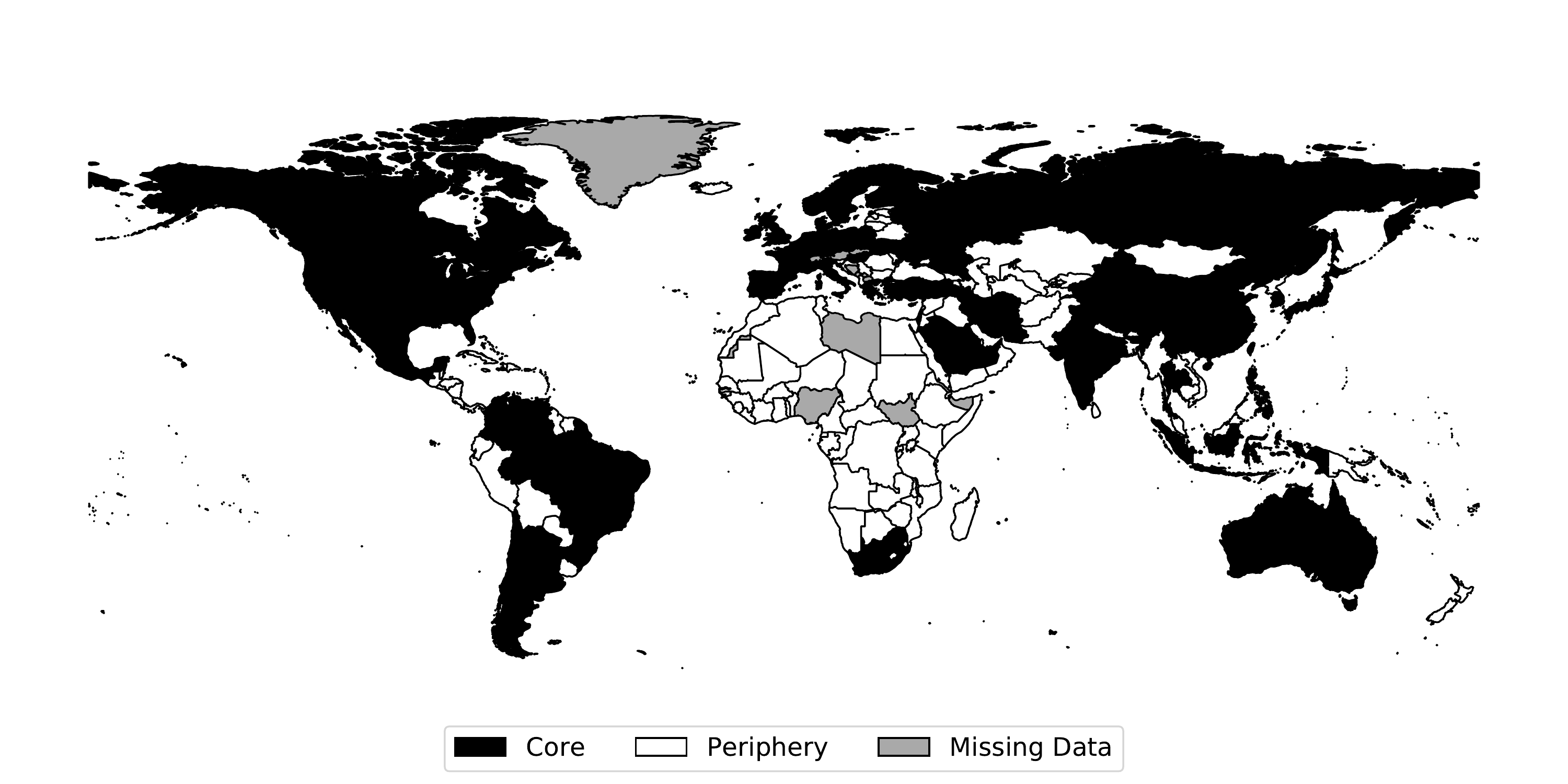}
\hspace{5mm}
\includegraphics[width=0.47\textwidth]{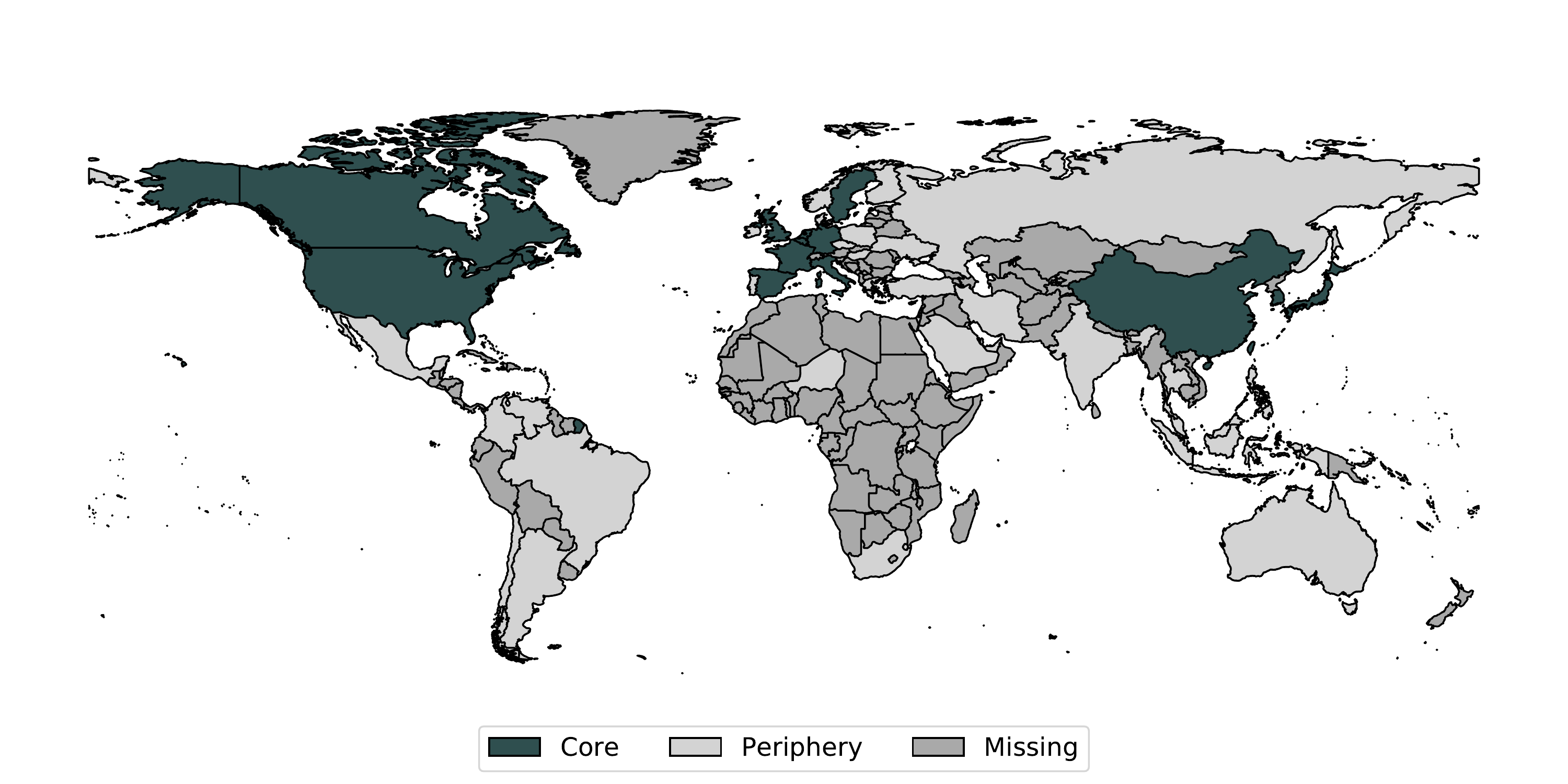}
\caption{Result of the application of our framework for the detection of weighted `bimodular' structures on the WTW in the year 2000 \cite{Gleditsch2002}. Top panels: upon running $\mathscr{W}_\sslash$ in a hierarchical fashion, a `core-within-the-core' is detected, revealing that the richest world countries (i.e. Canada, USA, the richest European countries, China and Russia - right panel, in dark green) constitute an even more tightly-linked cluster of nations among the ones with the largest trading activity (left panel, in black). Bottom panels: $\mathscr{E}_\sslash$ penalizes the countries with both a small degree and a small strength; in fact, the second run exlcudes the Russia, a result seemingly indicating that while its strength is `large enough' to be a member of the core, its degree is not.}
\label{fig11}
\end{figure*}

\textcolor{black}{Let us now consider the RC benchmark. It consists of 512 nodes, grouped in 16 communities arranged in a ring-like fashion and whose sizes obey a power-law whose exponent has been set to 1.8; the smallest community is composed by 3 nodes. Such a configuration is progressively `degraded' according to the following mechanism: first, a percentage $p$ of links is selected randomly and removed; afterwards, a percentage $p$ of links is selected randomly and rewired. In other words, a single `degradation' parameter $p$ drives the evolution of the initial ring-of-cliques towards a progressively less-defined clustered configuration.}

\textcolor{black}{Results on specific implementations of the RC benchmark are shown in fig. \ref{fig7}: surprise outperforms both competing algorithms across the entire domain of the `degradation' parameter $p$. More specifically, while modularity `degrades' slowly as the value of $p$ is risen, Infomap `degrades' abruptly as $p\geq0.4$. Hence, for small values of such a parameter, Infomap outperforms modularity; on the other hand, for large values of $p$, modularity outperforms Infomap (although both $\text{NMI}_{modularity}$ and $\text{ARI}_{modularity}$ achieve a value which is around $0.6$, i.e. already far from the maximum). Interestingly, for small values of $p$, the performance of Infomap and that of surprise overlap, both achieving NMI and ARI values which are very close to 1: as $p$ crosses the value of $0.4$, however, the two trends become increasingly different with Infomap being outperformed by modularity which is, in turn, outperformed by surprise. From a more general perspective, these results confirm what has been already observed elsewhere \cite{Aldecoa2013}, i.e. that the best-performing algorithms on the LFR benchmarks often perform poorly on the RC benchmarks and vice versa.}\\

\textcolor{black}{Let us now inspect the performance of surprise in recovering binary `bimodular' structures. To this aim, we have defined a novel benchmark mimicking the philosophy of the RC one, i.e. progressively `degrading' an initial, well-defined configuration:}

\begin{itemize}
\item \textcolor{black}{let us consider $N_c$ core nodes and $N_p$ periphery nodes. The core is completely connected (i.e. the link density of the $N_c\times N_c$ block is 1) and the periphery is empty (i.e. the link density of the $N_p\times N_p$ block is 0). So far, our benchmark is reminiscent of a core-periphery structure \emph{\`a la Borgatti-Everett};}
\item \textcolor{black}{let us now focus on the topology of the $N_c\times N_p$ bipartite network embodying the connections between the core and the periphery: in particular, let us consider each entry of such an adjacency matrix and pose $a_{cp}=1$ with probability $p_{cp}$. Upon doing so, such a subgraph will have a link density amounting precisely at $p_{cp}$;}
\item \textcolor{black}{let us now `degrade' such an initial configuration, by progressively filling the periphery and emptying the core. This can be achieved by 1) considering all peripherical node pairs and link them with probability $q$; 2) considering all core node pairs and keep them linked with probability $1-q$ (or, equivalently, disconnect them with probability $q$). Upon doing so, we end up with a core whose link density is precisely $1-q$ and with a periphery whose link density is precisely $q$. Now, varying $q$ in the interval $[0,p_{cp}]$ allows us to span a range of configurations starting with Borgatti-Everett and ending with Erd\"os-R\'enyi.}
\end{itemize}

\begin{center}
\begin{table*}[t!]
\begin{tabular}{l|c|c|c|c|c|c|c|c|c}
\hline
\hline
\multicolumn{1}{c}{\text{Binary networks}} & \multicolumn{3}{|c|}{$\#$ \text{communities}} & \multicolumn{6}{c}{} \\
\hline
 & \text{Modularity} & \text{Infomap} & \text{Surprise} & $\text{NMI}_\text{SM}$ & $\text{NMI}_\text{SI}$ & $\text{ARI}_\text{SM}$ & $\text{ARI}_\text{SI}$ & $\text{AWI}_\text{SM}$ & $\text{AWI}_\text{SI}$ \\
\hline
Madrid bombing terrorists & 6 & 5 & 26 & 0.50 & 0.44 & 0.33 & 0.27 & 0.83 & 0.73 \\
\hline
Star Wars characters & 9 & 5 & 35 & 0.37 & 0.53 & 0.15 & 0.28 & 0.39 & 0.94 \\
\hline
Australian University Campus & 13 & 6 & 30 & 0.47 & 0.72 & 0.20 & 0.46 & 0.76 & 0.93 \\
\hline
`Les Miserables' characters & 8 & 6 & 33 & 0.56 & 0.56 & 0.46 & 0.50 & 0.92 & 0.83 \\
\hline
\hline
\multicolumn{1}{c}{\text{Weighted networks}} & \multicolumn{3}{|c|}{$\#$ \text{communities}} & \multicolumn{6}{c}{} \\
\hline
 & \text{Modularity} & \text{Infomap} & \text{Surprise} & $\text{NMI}_\text{SM}$ & $\text{NMI}_\text{SI}$ & $\text{ARI}_\text{SM}$ & $\text{ARI}_\text{SI}$ & $\text{AWI}_\text{SM}$ & $\text{AWI}_\text{SI}$ \\
\hline
Madrid bombing terrorists & 8 & 5 & 22 & 0.50 & 0.67 & 0.37 & 0.62 & 0.65 & 0.94 \\
\hline
Star Wars characters & 10 & 6 & 52 & 0.36 & 0.43 & 0.18 & 0.24 & 0.79 & 0.79 \\
\hline
Australian University Campus & 14 & 6 & 25 & 0.47 & 0.74 & 0.22 & 0.46 & 0.72 & 0.91 \\
\hline
`Les Miserables' characters & 10 & 6 & 32 & 0.51 & 0.61 & 0.37 & 0.49 & 0.72 & 0.74 \\
\hline
\end{tabular}
\caption{\textcolor{black}{Table comparing the performance of modularity, Infomap and surprise on the bunch of real-world networks considered here. While Infomap is the algorithm producing the smallest number of clusters, surprise is the one producing the largest number of clusters; as a consequence, the three partitions output by our algorithms have an overall small overlap. Interestingly enough, the values of the AWI are quite large: since it just focuses on the percentage of true positives, a good performance under such an index indicates that the tested algorithms `agree' on the nodes to be clustered together: hence, the discrepancy between the ARI and the AWI may be explained by the presence of statistical `noise' (i.e. pairs of nodes on whose classification the algorithms `disagree') around the bulk of nodes to be put together - on which the algorithms `agree'.}}
\label{tab1}
\end{table*}
\end{center}

\textcolor{black}{Specifically, here we have considered $N_c=100$, $N_p=300$ and $p_{cp}=0.5$. The result of our exercise is shown in fig. \ref{fig8}: as expected, the performance of the surprise worsens as the `degradation' parameter becomes closer to $p_{cp}=0.5$; however, both the NMI and the ARI indices steadily remain very close to 1 - a result meaning that surprise optimization is not only able to correctly classify true positives (i.e. to keep the nodes originally in the same communities together) but also the other, possible kinds of node pairs. As with the exercise on community detection, let us now ask us if the presence of weights affects the detection of mesoscale structures. Generally speaking, the answer is, again, yes. Let us consider a toy core-periphery network: rising the weight of any two links connecting the core with the periphery allows the two nodes originally part of the periphery to be detected as belonging to the core.} \textcolor{black}{Analogously, if a bipartite topology is modified by adding weights between some of the nodes belonging to the same layer, $\mathscr{W}_\sslash$ will detect a core-periphery structure as significant, the core nodes being the ones linked by the `heaviest' connections.}\\

\noindent\textcolor{black}{{\bf Testing surprise on real-world networks.}} Let us now apply our formalism to the detection of mesoscale structures in real-world networks.

When coming to study real systems, particularly insightful examples are provided by social networks. \textcolor{black}{To this aim, let us consider the one induced by the co-occurrences of characters within the Star Wars saga (i.e. the three trilogies) \cite{repository0}. As shown in fig. \ref{fig9} we have both considered the binary and the weighted version of it. In both cases two major clusters are visible. For what concerns the binary version of such a network, the optimization of $\mathscr{S}$ reveals the presence of two major clusters: remarkably, those clusters are induced by the characters of Episodes I-III (e.g. Yoda, Qui-Gon, Obi-Wan, Anakin, Padme, the Emperor, Count Dooku, etc.) and by the characters of Episodes IV-IX (e.g. C-3PO, Leia, Han, Lando, Poe, Finn); a third cluster, instead, concerns the villains of Episodes VII-IX (i.e. Snoke, Kylo Ren, Phasma, Hux). Interestingly, Rey, BB-8, Maz-Kanata and other characters living on Jakku are clustered together. Quite remarkably, the interactions between the characters of Episodes IV-VI and those of Episodes VII-IX causes the former ones and the latter ones to be recovered within the same cluster. This picture is further refined once weights are taken into account: in fact, two of the aforementioned clusters are now merged, giving origin to the cluster of heroes of Episodes IV-IX (e.g. C-3PO, Leia, Han, Lando, Poe, Finn, Rey, Maz-Kanata).}\\

Let us now inspect the effectiveness of our framework in revealing weighted communities by considering the `friendship' networks among the terrorists involved in the train bombing of Madrid in 2004 \cite{repository} and the one among the residents living in an Australian University Campus \cite{repository} (see fig. \ref{fig9}). As the optimization of $\mathscr{W}$ reveals, while fully connected subsets of nodes are considered as communities in case links have unitary weights, sparser subgraphs can be considered as communities as well whenever their inner connections are `heavy' enough. On the other hand, both bottom panels of fig. \ref{fig9} seem to confirm that one of the main limitations of surprise-like functionals is that of recovering a large number of small cluster of nodes.

Let us now compare the performance of $\mathscr{S}_\sslash$ and $\mathscr{W}_\sslash$ in order to see if, and how, the presence of weights affects the `bimodular' mesoscale organization of networks. To this aim, let us focus on the network of co-occurrences of the characters of the novel `Les Miserables' \cite{repository}. As shown in fig. \ref{fig10}, link weights indeed modify the picture provided by just considering the simple presence of links (see also \cite{DeJeude2019}): the core of the weighted network is, in fact, constituted by the nodes connected by the `heavier' links, irrespectively from the link density of the former one.

After having applied our framework to the analysis of social networks, let us move to consider financial networks. One of the most popular examples of the kind is provided by the electronic Italian Interbank Money Market (e-MID) \cite{Iori2006}, depicted in fig. \ref{fig10}. Notice that, for such a network, the vast majority of core links are also the `heavier' ones, an evidence confirming a tendency that is ubiquitous in financial and economic systems, i.e. binary and weighted quantities - even at the mesoscale - are closely related.\\

Remarkably, the surprise-based formalism presented in this paper can be employed in a hierarchical fashion to highlight either nested communities or nested `bimodular' structures. To clarify this point, let us consider the World Trade Web (WTW) in the year 2000 as a case-study \cite{Gleditsch2002}. First, let us run $\mathscr{W}_\sslash$ to highlight the core portion of the weighted version of such a network; as fig. \ref{fig11} shows, the bipartition distinguishes countries with a large strength from those whose trade volume is low (basically, a bunch of African, Asian and South-American countries). Repeating our analysis \emph{within} the core portion of the network allows us to discover the presence of a (statistically-significant) nested core: in fact, the secound-round optimization reveals that the `core-inside-the-core' is composed by countries such as Canada, USA, the richest European countries, China and Russia.

\textcolor{black}{Let us now compare it with $\mathscr{E}_\sslash$, run in a hierarchical fashion as well. The results of our exercise are shown in fig. \ref{fig11}. As evident from looking at it, the enhanced surprise is more restrictive than the purely weighted one, as a consequence of constraining the degrees beside the strengths. Hence, while the first run excludes the countries with both a small degree and a small strength, the second run exlcudes the Russia, a result seemingly indicating that while its strength is `large enough' to allow it to be a member of the core, its degree is not.}

\textcolor{black}{In a sense, the optimization of $\mathscr{E}_\sslash$ corrects the picture provided by the optimization of $\mathscr{W}_\sslash$ as the core becomes less populated by `low degree' nodes - an effect which is likely to become more evident on systems that are neither financial nor economic in nature.}\\

\textcolor{black}{Let us now run, and compare, modularity, Infomap and surprise on the bunch of real-world networks above. Table \ref{tab1} sums up the results. A first observation concerns the number of detected communities: while Infomap is the algorithm producing the smallest number of clusters, surprise is the one producing the largest number of clusters - more precisely, surprise outputs \emph{more} and \emph{smaller} clusters than the other two methods. As a consequence, our three algorithms produce partitions with an overall small overlap, as indicated by the NMI; the ARI confirms such an observation - although indicating that the pictures provided by Infomap and surprise are (overall) more similar than those provided by modularity and surprise (and, as a consequence, by modularity and Infomap). Interestingly enough, the values of the AWI are quite large - and larger than the corresponding NMI and ARI values: since it just focuses on the percentage of true positives, a good performance under such an index indicates that the two tested algorithms `agree' on the nodes to be clustered together (although they may not - and, in general, will not - agree on the number of communities). Hence, the discrepancy between the ARI and the AWI may be explained by the presence of statistical `noise' (i.e. `misclassified' pairs of nodes, although the word may not be correct as the information about the `true' partition is not available) around the bulk of nodes to be put together.}

\textcolor{black}{Let us stress once more that, whenever real-world networks are considered, information about the existence of a `true' partition is rarely available; for this reason, exercises as the one we have carried out here may be useful to gain insight on the system under study: instead of trusting just one algorithm, combining pairs of them - e.g. by considering as communities the subsets of nodes output by both - may be the right solution to overcome the limitations affecting each single method.}\\

\noindent{\textcolor{black}{{\bf The `SurpriseMeMore' Python package.} As an additional result, we release a comprehensive package, coded in Python, that implements all aforementioned surprise-like variants. Its name is `SurpriseMeMore' - the name recalls the package released in 2014 \cite{Aldecoa2014} - and is freely downloadable at the following URL: \texttt{https://github.com/EmilianoMarchese/SurpriseMeMore}.}}

\section*{CONCLUSIONS}

\textcolor{black}{The hypergeometric distribution - together with its many variants - has recently revived the interest of researchers who have employed it to define novel network ensembles \cite{Casiraghi2018}, recipes for projecting bipartite networks \cite{Cimini2021b}, etc.}

\textcolor{black}{Remarkably, the distributions related to it allow for a wide variety of benchmarks to be defined, each one embodying a different set of constraints. In the present paper we have explored the `power' of the hypergeometric formalism to carry out the detection of mesoscale structures: remarkably, it allows proper statistical tests to be definable for revealing the presence of modular, core-periphery and bipartite structures on any kind of network, be it binary or weighted, undirected or directed.} \textcolor{black}{According to the classification proposed in the introductory section of the paper, we believe surprise to belong to the second class of algorithms - its asymptotic expression embodying a sort of LRT aimed at choosing between alternative hypotheses (the RGM and the SBM, the WRG and the WSBM, etc.).}

\textcolor{black}{More in general, our approach reveals the superiority of the algorithms for mesoscale structures detection belonging to the second and to the third class with respect to those belonging to the first one: still, the two classes of statistically-grounded approaches compete on some, specific benchmarks, as the comparisons carried out on the LFR and the RC ones clearly show (specifically, methods performing well on LFR benchmarks do not on RC benchmarks and vice versa). This also suggests a strategy to handle mesoscale structures detection on real-world networks: as the information on the presence of a possible, `true' partition is rarely available, a good strategy may be that of running different algorithms on the same empirical networks, check the consistency of their output and combine them, e.g. by taking the overlap - in a way that is reminiscent of multi-model inference.}

\textcolor{black}{Although the surprise-based approach is powerful and versatile, its} \textcolor{black}{downside is represented by the specific kinds of tests that are induced by the optimization of surprise-like score functions: comparisons between benchmarks `ignoring' the local structure of nodes (i.e. their degree) are, in fact, carried out. While this seems to be perfectly reasonable when considering core-periphery structures - see also the contribution \cite{Kojaku2018} whose authors claim that a core-periphery structure is always compatible with a network degree sequence - this is no longer true for the community detection task \cite{Karrer2010}} \textcolor{black}{- indeed, as it has been noticed elsewhere, ignoring degrees may be at the origin of the large number of singletons output by surprise as assigning nodes with few neighbors to larger clusters may be disfavored from a statistical point of view.}

\textcolor{black}{The observations above call for the `extension' of the hypergeometric formalism we have studied here to include more refined benchmarks as the ones constraining the entire degree sequence.}

\section*{DATA AVAILABILITY}

Network data that support the findings of this study are freely downloadable at the following URLs: \texttt{https://github.com/evelinag/StarWars-social-network}, \texttt{http://konect.cc/networks}, \texttt{http://ksgleditsch.com/exptradegdp.html}.

\section*{CODE AVAILABILITY}

The code employed for the analyses carried out in this study is freely downloadable at the following URL: \texttt{https://github.com/EmilianoMarchese/SurpriseMeMore}.

\section*{COMPETING INTERESTS}

\textcolor{black}{The authors declare no competing financial interests.}

\section*{ACKNOWLEDGEMENTS}

The authors acknowledge support from the EU project SoBigData-PlusPlus (grant no. 871042).

\clearpage
\newpage

\appendix
\widetext
\counterwithin{figure}{section}

\section{Detecting mesoscale structures in binary networks}

As stressed in the main text, community detection on binary networks can be implemented by carrying out an exact statistical test whose function $f$ must be identified with a binomial hypergeometric distribution. Hence, we can compactly write

\begin{equation}
f(l_\bullet)\equiv\text{H}(l_\bullet|V,V_\bullet,L)=\frac{\prod_{i=\bullet,\circ}\binom{V_i}{l_i}}{\binom{V}{L}}=\frac{\binom{V_\bullet}{l_\bullet}\binom{V_\circ}{l_\circ}}{\binom{V}{L}}=\frac{\binom{V_\bullet}{l_\bullet}\binom{V-V_\bullet}{L-l_\bullet}}{\binom{V}{L}}
\end{equation}
with a clear meaning of the symbols. The correctedness of such the definition of can be explicitly checked upon calculating its normalization via the Vandermonde's identity $\sum_{k=0}^r\binom{m}{k}\binom{n}{r-k}=\binom{m+n}{r}$. While the binomial coefficient $\binom{V_\bullet}{l_\bullet}$ enumerates the number of ways $l_\bullet$ links can be redistributed within communities, i.e. over the available $V_\bullet$ node pairs, the binomial coefficient $\binom{V-V_\bullet}{L-l_\bullet}$ enumerates the number of ways the remaining $L-l_\bullet$ links can be redistributed between communities, i.e. over the remaining $V-V_\bullet$ node pairs. The position above, in turn, induces the definition of the binary surprise, i.e. $\mathscr{S}\equiv\sum_{l_\bullet\geq l_\bullet^*}f(l_\bullet)$.\\

The generalization of the formalism above to detect `bimodular' structure on binary networks is straightforward. It is enough to consider the multivariate (or multinomial) version of the hypergeometric distribution, i.e.

\begin{equation}
f(l_\bullet,l_\circ)\equiv\text{MH}(l_\bullet,l_\circ|V,V_\bullet,V_\circ,L)=\frac{\prod_{i=\bullet,\circ,\top}\binom{V_i}{l_i}}{\binom{V}{L}}=\frac{\binom{V_\bullet}{l_\bullet}\binom{V_\circ}{l_\circ}\binom{V_\top}{l_\top}}{\binom{V}{L}}=\frac{\binom{V_\bullet}{l_\bullet}\binom{V_\circ}{l_\circ}\binom{V-(V_\bullet+V_\circ)}{L-(l_\bullet+l_\circ)}}{\binom{V}{L}};
\end{equation}
the correctedness of such a definition of can be checked upon calculating its normalization by applying the identity $\sum_{k=0}^r\binom{m}{k}\binom{n}{r-k}=\binom{m+n}{r}$ twice. The position above, in turn, induces the definition of the binary bimodular surprise, i.e. $\mathscr{S}_\sslash\equiv\sum_{l_\bullet\geq l^*_\bullet}\sum_{l_\circ\geq l^*_\circ}f(l_\bullet,l_\circ)$.\\

The asymptotic expression of $\mathscr{S}$ and $\mathscr{S}_\sslash$ can be derived by applying the Stirling approximation $n!\simeq\sqrt{2\pi n}\left(\frac{n}{e}\right)^n$ to the binomial coefficients appearing into the definition of the versions of the surprise described above, leading to expressions like 

\begin{eqnarray}
\binom{V}{L}&=&\frac{V!}{L!(V-L)!}\simeq\frac{\sqrt{2\pi V}V^Ve^{-V}}{\sqrt{2\pi L}L^Le^{-L}\sqrt{2\pi(V-L)}(V-L)^{V-L}e^{-(V-L)}}\nonumber\\
&=&\frac{1}{\sqrt{2\pi V\frac{L}{V}\left(1-\frac{L}{V}\right)}}\cdot\frac{V^V}{L^L(V-L)^{V-L}}\nonumber\\
&=&\frac{1}{\sqrt{2\pi Vp(1-p)}}\cdot\frac{1}{\left(\frac{L}{V}\right)^L\left(1-\frac{L}{V}\right)^{V-L}}\nonumber\\
&=&\frac{1}{\sqrt{2\pi Vp(1-p)}}\cdot\frac{1}{p^L(1-p)^{V-L}}=\frac{1}{\sqrt{2\pi\sigma^2}\cdot\text{Ber}(V,L,p)}
\end{eqnarray}
where $\text{Ber}(x,y,z)=z^y(1-z)^{x-y}$ is a Bernoulli probability mass function, $z=\frac{y}{x}$ and $p=\frac{L}{V}$.\\

The time, in seconds, to find the corresponding optimal partitions is reported in tables \ref{tab1A} and \ref{tab2A}: overall, denser networks require more time.

\begin{table}[b!]
\begin{tabular}{@{}lllllll@{}}
\toprule
\textbf{Network} & \multicolumn{1}{l}{\textbf{\# nodes} $(N)$} & \multicolumn{1}{l}{\textbf{\# links} $(L)$} & \multicolumn{1}{l}{\textbf{Connectance}} & \multicolumn{1}{l}{\textbf{Binary} (s)} & \multicolumn{1}{l}{\textbf{Weighted} (s)} & \multicolumn{1}{l}{\textbf{Enhanced} (s)}\\
\midrule
\hline
Madrid terrorists & 64 & 486 & 0.24 & 0.05 & 0.05 & 0.07\\
\hline
Star Wars & 110 & 398 & 0.07 & 0.13 & 0.14 & 0.24\\
\hline
Australian college & 217 & 2672 & 0.11 & 0.34 & 0.30 & 0.38\\
\hline
\bottomrule
\end{tabular}
\caption{Time, in seconds, employed by $\mathscr{S}$, $\mathscr{W}$ and $\mathscr{E}$ to find the corresponding optimal partition.}
\label{tab1A}
\end{table}

\section{Detecting mesoscale structures in weighted networks}

When analyzing weighted networks, one needs to move from the binomial hypergeometric distribution to its negative counterpart, i.e. to

\begin{equation}
f(w_\bullet)=\text{NH}(w_\bullet|V+W,W,V_\bullet)=\frac{\prod_{i=\bullet,\circ}\binom{V_i+w_i-1}{w_i}}{\binom{V+W-1}{W}}=
\frac{\binom{V_\bullet+w_\bullet-1}{w_\bullet}\binom{V_\circ+w_\circ-1}{w_\circ}}{\binom{V+W-1}{W}}=\frac{\binom{V_\bullet+w_\bullet-1}{w_\bullet}\binom{V-V_\bullet+(W-w_\bullet)-1}{W-w_\bullet}}{\binom{V+W-1}{W}}
\end{equation}
with a clear meaning of the symbols. The soundness of the definition above can be cheked upon calculating the normalization of the distribution $\text{NH}(w_\bullet|V+W,W,V_\bullet)$ via the identity $\sum_{k=0}^r\binom{k+m}{k}\binom{n-m-k}{r-k}=\binom{n+1}{r}$: in fact

\begin{eqnarray}
\sum_{w_\bullet=0}^W\frac{\binom{w_\bullet+V_\bullet-1}{w_\bullet}\binom{w_\circ+V_\circ-1}{w_\circ}}{\binom{V+W-1}{W}}=\sum_{w_\bullet=0}^W\frac{\binom{w_\bullet+V_\bullet-1}{w_\bullet}\binom{W-w_\bullet+V-V_\bullet-1}{W-w\bullet}}{\binom{V+W-1}{W}}=\sum_{w_\bullet=0}^W\frac{\binom{w_\bullet+V_\bullet-1}{w_\bullet}\binom{(W+V-2)-(V_\bullet-1)-w_\bullet}{W-w_\bullet}}{\binom{V+W-1}{W}}=1.
\end{eqnarray}

The position above induces the definition of the weighted surprise as $\mathscr{W}=\sum_{w_\bullet\geq w_\bullet^*}f(w_\bullet)$.\\

The generalization of the formalism above to detect `bimodular' structure on weighted networks is straightforward. It is enough to consider the multivariate (or multinomial) version of the negative hypergeometric distribution, i.e.

\begin{eqnarray}
f(w_\bullet,w_\circ)=\text{MNH}(w_\bullet,w_\circ|V+W,W,V_\bullet,V_\circ)&=&\frac{\prod_{i=\bullet,\circ,\top}\binom{V_i+w_i-1}{w_i}}{\binom{V+W-1}{W}}=
\frac{\binom{V_\bullet+w_\bullet-1}{w_\bullet}\binom{V_\circ+w_\circ-1}{w_\circ}\binom{V_\top+w_\top-1}{w_\top}}{\binom{V+W-1}{W}}\nonumber\\
&=&\frac{\binom{V_\bullet+w_\bullet-1}{w_\bullet}\binom{V_\circ+w_\circ-1}{w_\circ}\binom{V-(V_\bullet+V_\circ)+W-(w_\bullet+w_\circ)-1}{W-(w_\bullet+w_\circ)}}{\binom{V+W-1}{W}}
\end{eqnarray}
whose correctedness can be checked upon calculating its normalization, by applying the identity $\sum_{k=0}^r\binom{k+m}{k}\binom{n-m-k}{r-k}=\binom{n+1}{r}$ twice. The position above, in turn, induces the definition of the weighted bimodular surprise, as $\mathscr{W}_\sslash=\sum_{w_\bullet\geq w^*_\bullet}\sum_{w_\circ\geq w^*_\circ}f(w_\bullet,w_\circ)$.\\

\begin{table}[b!]
\begin{tabular}{@{}lllllll@{}}
\toprule
\textbf{Network} & \multicolumn{1}{l}{\textbf{\# nodes} $(N)$} & \multicolumn{1}{l}{\textbf{\# links} $(L)$} & \multicolumn{1}{l}{\textbf{Connectance}} & \multicolumn{1}{l}{\textbf{Binary} (s)} & \multicolumn{1}{l}{\textbf{Weighted} (s)} & \multicolumn{1}{l}{\textbf{Enhanced} (s)}\\
\midrule
\hline
`Les Miserables' & 77 & 254 & 0.09 & 0.22 & 0.31 & 0.48\\
\hline
e-MID 2009 & 134 & 4609 & 0.26 & 1.42 & 0.80 & 4.12\\
\hline
WTW 2000 & 187 & 20105 & 0.58 & 12.16 & 12.05 & 21.23\\
\hline
\bottomrule
\end{tabular}
\caption{Time, in seconds, employed by $\mathscr{S}_\sslash$, $\mathscr{W}_\sslash$ and $\mathscr{E}_\sslash$ to find the corresponding optimal partition.}
\label{tab2A}
\end{table}

The asymptotic expression of $\mathscr{W}$ and $\mathscr{W}_\sslash$ can be derived by applying the Stirling approximation $n!\simeq\sqrt{2\pi n}\left(\frac{n}{e}\right)^n$ to the binomial coefficients appearing into the definition of the versions of the surprise described above, leading to expressions like 

\begin{eqnarray}
\binom{V+W-1}{W}&=&\frac{(V+W-1)!}{W!(V-1)!}\simeq\frac{\sqrt{2\pi(V+W-1)}(V+W-1)^{V+W-1}e^{-(V+W-1)}}{\sqrt{2\pi W}W^We^{-W}\sqrt{2\pi(V-1)}(V-1)^{V-1}e^{-(V-1)}}\nonumber\\
&=&\frac{1}{\sqrt{2\pi (V-1)\left(\frac{W}{V+W-1}\right)}}\cdot\frac{(V+W-1)^{V+W-1}}{W^W(V-1)^{V-1}}\nonumber\\
&\simeq&\frac{1}{\sqrt{2\pi Vq}}\cdot\frac{1}{\left(\frac{W}{V+W-1}\right)^W\left(\frac{V-1}{V+W-1}\right)^{V-1}}\nonumber\\
&\simeq&\frac{1}{\sqrt{2\pi Vq}}\cdot\frac{1}{q^W(1-q)^{V-1}}=\frac{1}{\sqrt{2\pi\mu}\cdot\text{Geo}(V,W,q)}
\end{eqnarray}
where $\text{Geo}(x,y,z)=z^y(1-z)^x$ is a geometric probability mass function, $z=\frac{y}{x+y}$ and $q\simeq\frac{W}{V+W}$.\\

The time, in seconds, to find the corresponding optimal partitions is reported in tables \ref{tab1A} and \ref{tab2A}: overall, denser networks require more time.

\section{Enhanced detection of mesoscale structures}

The enhanced hypergeometric distribution is a negative hypergeometric distribution `conditional' to the existence of connections within communities, i.e. $l_\bullet>0$. It reads

\begin{equation}
\text{EH}(l_\bullet,w_\bullet)=\left\{\begin{array}{lr}
\frac{\binom{V_\bullet}{l_\bullet}\binom{V_\circ}{l_\circ}}{\binom{V}{L}}\cdot\frac{\binom{w_\bullet-1}{w_\bullet-l_\bullet}\binom{w_\circ-1}{w_\circ-l_\circ}}{\binom{W-1}{L-1}}, & 0<l_\bullet<L\\
\frac{\binom{V_\bullet}{L}}{\binom{V}{L}}\cdot\frac{\binom{w_\bullet-1}{w_\bullet-L}}{\binom{W}{L}}, & l_\bullet=L
\end{array}
\right.
\end{equation}
the first expression being valid when $l_\bullet<L$ and the second expression being valid in the `extreme' case $l_\bullet=L$ (i.e. whenever the total weight has to be redistributed on top of links which are all put inside the communities). Normalization is ensured by the fact that

\begin{eqnarray}
&&+\sum_{l_\bullet=1}^{L-1}\frac{\binom{V_\bullet}{l_\bullet}\binom{V-V_\bullet}{L-l_\bullet}}{\binom{V}{L}}\cdot\sum_{w_\bullet-l_\bullet=0}^{W-L}\frac{\binom{(w_\bullet-l_\bullet)+(l_\bullet-1)}{w_\bullet-l_\bullet}\binom{(W-2)-(l_\bullet-1)-(w_\bullet-l_\bullet)}{(W-L)-(w_\bullet-l_\bullet)}}{\binom{W-1}{L-1}}\nonumber\\
&&+\frac{\binom{V_\bullet}{L}}{\binom{V}{L}}\cdot\sum_{w_\bullet=L}^W\frac{\binom{w_\bullet-1}{w_\bullet-L}}{\binom{W}{L}}=\sum_{l_\bullet=1}^L\frac{\binom{V_\bullet}{l_\bullet}\binom{V-V_\bullet}{L-l_\bullet}}{\binom{V}{L}}=1-\frac{\binom{V_\circ}{L}}{\binom{V}{L}}
\end{eqnarray}
where we have used the relationships $\sum_{k=0}^r\binom{k+m}{k}\binom{n-m-k}{r-k}=\binom{n+1}{r}$, $\sum_{k=r}^n\binom{k}{r}=\sum_{k=r}^n\binom{k}{k-r}=\binom{n+1}{r+1}$ and $\sum_{k=0}^r\binom{m}{k}\binom{n-m}{r-k}=\binom{n}{r}$ and assumed that $L<V_\bullet$. For what concerns the expected value, we have that

\begin{eqnarray}
\langle l_\bullet\rangle&=&\sum_{l_\bullet=1}^{L-1}l_\bullet\cdot\frac{\binom{V_\bullet}{l_\bullet}\binom{V-V_\bullet}{L-l_\bullet}}{\binom{V}{L}}\cdot\sum_{w_\bullet-l_\bullet=0}^{W-L}\frac{\binom{(w_\bullet-l_\bullet)+(l_\bullet-1)}{w_\bullet-l_\bullet}\binom{(W-2)-(l_\bullet-1)-(w_\bullet-l_\bullet)}{(W-L)-(w_\bullet-l_\bullet)}}{\binom{W-1}{L-1}}\nonumber\\
&&+L\cdot\frac{\binom{V_\bullet}{L}}{\binom{V}{L}}\cdot\sum_{w_\bullet=L}^W\frac{\binom{w_\bullet-1}{w_\bullet-L}}{\binom{W}{L}}=\sum_{l_\bullet=1}^Ll_\bullet\cdot\frac{\binom{V_\bullet}{l_\bullet}\binom{V-V_\bullet}{L-l_\bullet}}{\binom{V}{L}}=L\frac{V_\bullet}{V}
\end{eqnarray}
as it should be: in fact, the expected value of (the number of) connections should not depend on the weights put on the connections. Moreover,

\begin{eqnarray}
\langle w_\bullet-l_\bullet\rangle&=&\sum_{l_\bullet=1}^{L-1}\sum_{w_\bullet-l_\bullet=0}^{W-L}(w_\bullet-l_\bullet)\cdot\frac{\binom{V_\bullet}{l_\bullet}\binom{V-V_\bullet}{L-l_\bullet}}{\binom{V}{L}}\cdot\frac{\binom{(w_\bullet-l_\bullet)+(l_\bullet-1)}{w_\bullet-l_\bullet}\binom{(W-2)-(l_\bullet-1)-(w_\bullet-l_\bullet)}{(W-L)-(w_\bullet-l_\bullet)}}{\binom{W-1}{L-1}}\nonumber\\
&&+(W-L)\cdot\frac{\binom{V_\bullet}{L}}{\binom{V}{L}}\cdot\sum_{w_\bullet=L}^W\frac{\binom{w_\bullet-1}{w_\bullet-L}}{\binom{W}{L}}\nonumber\\
&=&\sum_{l_\bullet=1}^{L-1}\frac{\binom{V_\bullet}{l_\bullet}\binom{V-V_\bullet}{L-l_\bullet}}{\binom{V}{L}}\cdot\sum_{w_\bullet-l_\bullet=0}^{W-L}(w_\bullet-l_\bullet)\cdot\frac{\binom{(w_\bullet-l_\bullet)+(l_\bullet-1)}{w_\bullet-l_\bullet}\binom{(W-2)-(l_\bullet-1)-(w_\bullet-l_\bullet)}{(W-L)-(w_\bullet-l_\bullet)}}{\binom{W-1}{L-1}}+(W-L)\cdot\frac{\binom{V_\bullet}{L}}{\binom{V}{L}}
\nonumber\\
&=&\sum_{l_\bullet=1}^{L-1}\frac{\binom{V_\bullet}{l_\bullet}\binom{V-V_\bullet}{L-l_\bullet}}{\binom{V}{L}}\cdot l_\bullet\cdot\frac{W-L}{L}+(W-L)\cdot\frac{\binom{V_\bullet}{L}}{\binom{V}{L}}
\nonumber\\
&=&L\left(\frac{V_\bullet}{V}-\frac{\binom{V_\bullet}{L}}{\binom{V}{L}}\right)\left(\frac{W-L}{L}\right)+(W-L)\cdot\frac{\binom{V_\bullet}{L}}{\binom{V}{L}}=(W-L)\frac{V_\bullet}{V}.
\end{eqnarray}

\begin{figure*}[t!]
\includegraphics[width=0.45\textwidth]{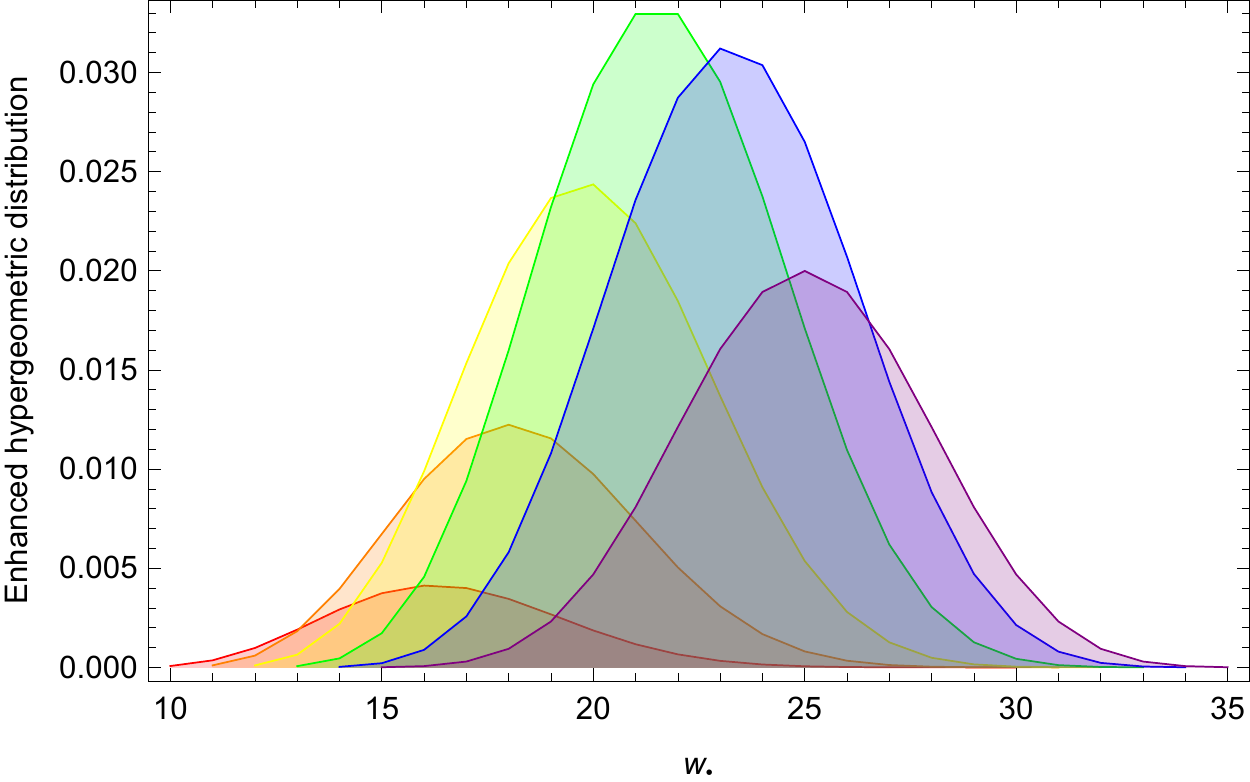}
\hspace{5mm}
\includegraphics[width=0.45\textwidth]{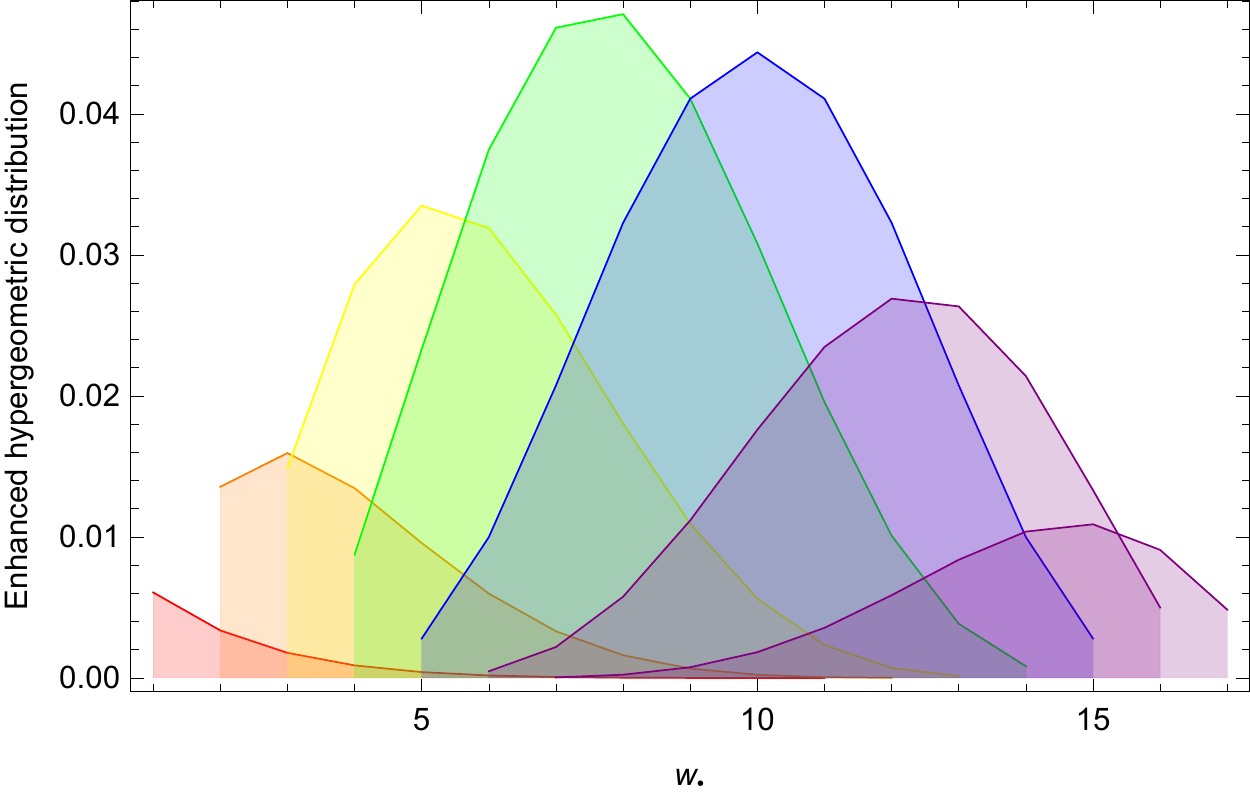}\\
\includegraphics[width=0.45\textwidth]{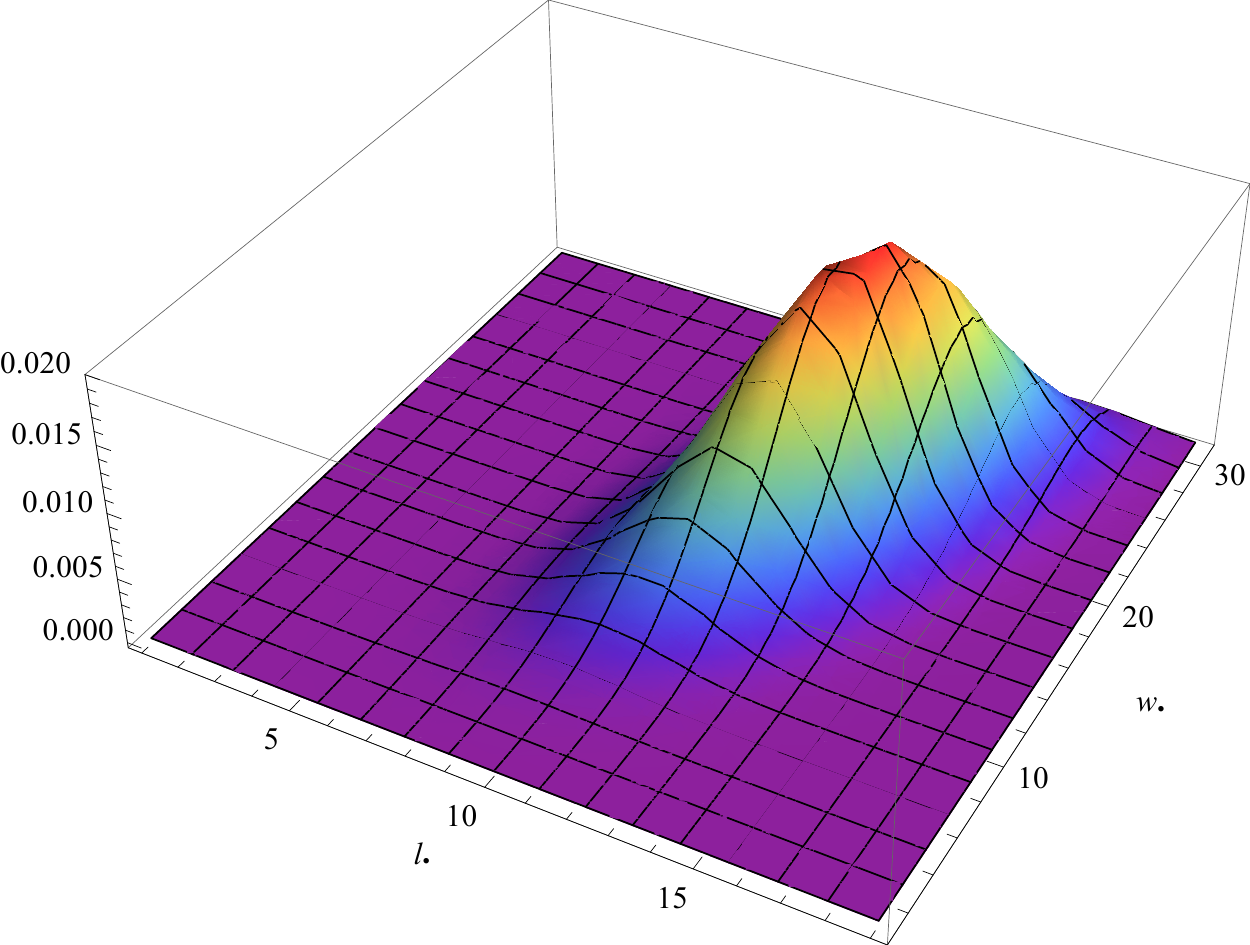}
\caption{Examples of the enhanced hypergeometric distribution $\text{EH}(l_\bullet,w_\bullet|V,V_\bullet,L,W)$. Top-left panel: instances characterized by the values $V=45$, $V_\bullet=20$, $L=30$, $W=50$ and $l_\bullet=10,11,12,13,14,15$. Top-right panel: instances characterized by the values $V=45$, $V_\bullet=20$, $L=10$, $W=20$ and $l_\bullet=1,2,3,4,5,6,7$. Bottom panel: full view of the enhanced hypergeometric distribution with parameters $V=80$, $V_\bullet=50$, $L=20$, $W=50$ and $0<l_\bullet<L$.}
\label{fig1A}
\end{figure*}

The asymptotic expression of $\mathscr{E}$ can be derived by applying the Stirling approximation $n!\simeq\sqrt{2\pi n}\left(\frac{n}{e}\right)^n$ to the binomial coefficients appearing into the definition of the enhanced surprise, leading to expressions like 

\begin{eqnarray}
\binom{V}{L}\binom{W-1}{W-L}&=&\frac{V!}{L!(V-L)!}\cdot\frac{(W-1)!}{(W-L)!(L-1)!}\nonumber\\
&\propto&\frac{V^V}{L^L(V-L)^{V-L}}\cdot\frac{(W-1)^{W-1}}{(W-L)^{W-L}(L-1)^{L-1}}\nonumber\\
&\simeq&\frac{V^V}{L^L(V-L)^{V-L}}\cdot\frac{W^{W}}{(W-L)^{W-L}L^{L}}\nonumber\\
&=&\frac{1}{\left(\frac{L}{V-L}\right)^L\left(1-\frac{L}{V}\right)^V}\cdot\frac{1}{\left(1-\frac{L}{W}\right)^W\left(\frac{L}{W-L}\right)^L}\nonumber\\
&=&\frac{1}{\left(\frac{L}{V-L}\right)^L\left(1-\frac{L}{V}\right)^V}\cdot\frac{1}{\left(1-\frac{L}{W}\right)^{W-L}\left(\frac{L}{W}\right)^L}\nonumber\\
&=&\frac{1}{p^L(1-p)^{V-L}}\cdot\frac{1}{q^{W-L}(1-q)^L}\nonumber\\
&=&\text{BF}[V,W,L,p,q]=\text{Ber}(V,L,p)\cdot\text{Geo}(L,W-L,q)
\end{eqnarray}
with a clear meaning of the symbols. A graphical representation of the enhanced hypergeometric distribution is provided in fig. \ref{fig1A}.\\

Let us now briefly consider the distribution inducing the definition of the enhanced bimodular surprise, i.e. the `multivariate' enhanced hypergeometric distribution. As before, it is `conditional' to the existence of connections within either modules one is looking at (i.e. either $l_\bullet>0$ or $l_\circ>0$) and reads

\begin{equation}
\text{MEH}(l_\bullet,l_\circ,w_\bullet,w_\circ)=\left\{\begin{array}{lr}
\frac{\binom{V_\bullet}{l_\bullet}\binom{V_\circ}{l_\circ}\binom{V_\top}{l_\top}}{\binom{V}{L}}\cdot\frac{\binom{w_\bullet-1}{w_\bullet-l_\bullet}\binom{w_\circ-1}{w_\circ-l_\circ}\binom{w_\top-1}{w_\top-l_\top}}{\binom{W-1}{W-L}}, & 0<l_\bullet, l_\circ<L\\
\frac{\binom{V_\bullet}{L}}{\binom{V}{L}}\cdot\frac{\binom{w_\bullet-1}{w_\bullet-L}}{\binom{W}{L}}, & l_\bullet=L,\:l_\circ=0\\
\frac{\binom{V_\circ}{l_\circ}}{\binom{V}{L}}\cdot\frac{\binom{w_\circ-1}{w_\circ-L}}{\binom{W}{L}}, & l_\bullet=0,\:l_\circ=L\\
\end{array}
\right.
\end{equation}
with a clear meaning of the symbols. Notice that the cases $l_\bullet=0$, $0<l_\circ<L$ and $l_\circ=0$, $0<l_\bullet<L$ can be easily recovered from the first row of the definition above.

\section{Pseudocodes for surprise optimization}

Let us now provide a description of the algorithms implemented in the `SurpriseMeMore' Python package (freely downloadable at the URL \texttt{https://github.com/EmilianoMarchese/SurpriseMeMore}) for surprise optimization: to this aim, we will also show the corresponding pseudocodes. Since an exhaustive exploration of the space of all possible partitions of a network is not feasible (especially when dealing with large graphs), one often proceeds heuristically, in either an \emph{agglomerative} fashion or fixing the number of clusters to be detected \emph{a priori}.

\vspace{5mm}

\noindent\textit{Community detection on binary and weighted networks.} The algorithm for community detection on binary networks is called PACO (PArtitioning Cost Optimization) and its pseudocode can be found in \cite{Nicolini2016}. The algorithm implemented there is agglomerative in nature: in the initial configuration of the algorithm, in fact, nodes are treated as singletons (hence, there are as many communities as nodes). Then, edges are sorted according to the value of their Jaccard index, in a decreasing order: as the Jaccard index of an edge is proportional to the number of common neighbours of the two connected nodes, a large value of it likely points out that the two nodes must be assigned to the same module. For further details about PACO, see \cite{Nicolini2016}.

Here we adopt a similar strategy to the one implemented by PACO in the binary case, with two main differences. The first one concerns the edge-sorting step: edges, in fact, are no longer sorted according to the Jaccard index but randomly ordered; this strategy avoids getting stuck in local minima more effectively. The second one concerns the possibility of letting entire communities merge (instead of moving single nodes from one to another) with a certain probability $p_{mix}$.

\vspace{5mm}

\begin{algorithmic}[1]
\Function{CalculateAndUpdateSurprise}{$C,C'$}
	\State $S\gets calculateSurprise(C)$
	\State $S'\gets calculateSurprise(C')$
	\If{$S'<S$}
		\State $C\gets C'$
		\State $S\gets S'$
	\EndIf \\
\Return $C$
\EndFunction
\State
\State $C \gets $ array of length $N$, randomly initialized with $N$ different integers, i.e. $1,2\dots N$;
\State $E \gets $ randomly sorted edges;
\For{edge $(u,v)\in E$}
	\If{$C[u] \neq C[v]$}
		\State $C'\gets C$
	\If{$0\leq uniformDistribution(0,1)\leq p_{mix}$}
		\For{node $t \in C[u]$}
			\State $C'[t] \gets C[v] $
		\EndFor
		\State $C\gets$\Call{CalculateAndUpdateSurprise}{$C,C'$}
	\EndIf
	\If{$p_{mix}\leq uniformDistribution(0,1)\leq p_{single}$}
		\State $C'[u]\gets C[v]$
		\State $C\gets$\Call{CalculateAndUpdateSurprise}{$C,C'$}
	\EndIf
	\If{$p_{single}\leq uniformDistribution(0,1)\leq 1$}
		\State $C'[v]\gets C[u]$
		\State $C\gets$\Call{CalculateAndUpdateSurprise}{$C,C'$}
	\EndIf
\EndIf
\State $\Rightarrow$ consider each node and swap its membership with the one of each first neighbour of it; accept the move if $\mathscr{S}$, $\mathscr{W}$ or $\mathscr{E}$ decreases;
\EndFor
\State{$\Rightarrow$ repeat the for-loop to improve the chance of finding the optimal partition}
\end{algorithmic}

\vspace{5mm}

Remarkably, it is also possible to define a version of the surprise-based community-detection algorithm, where the number of clusters to be detected is fixed a priori:

\vspace{5mm}

\begin{algorithmic}[1]
\Function{CalculateAndUpdateSurprise}{$C,C'$}
	\State $S\gets calculateSurprise(C)$
	\State $S'\gets calculateSurprise(C')$
	\If{$S'<S$}
		\State $C\gets C'$
		\State $S\gets S'$
	\EndIf \\
\Return $C$
\EndFunction
\State
\State $C \gets $ array of length $N$, randomly initialized with $n_{clust}$ different integers, i.e. $1,2\dots n_{clust}$;
\State $E \gets $ randomly sorted edges;
\For{edge $(u,v)\in E$}
	\State $C'\gets C$
	\If{$C'[u] \neq C[v]$}
		\State $C'[u]\gets C[v]$
		\State $C[v]\gets C'[u]$
		\State $C\gets$\Call{CalculateAndUpdateSurprise}{$C,C'$}
	\Else
		\State $C'[u]\gets randomInteger(1,n_{clust})$
		\State $C\gets$\Call{CalculateAndUpdateSurprise}{$C,C'$}
	    \State $C''[v]\gets randomInteger(1,n_{clust})$
	    \State $C\gets$\Call{CalculateAndUpdateSurprise}{$C,C''$}
	\EndIf
\State $\Rightarrow$ consider each node and swap its membership with the one of each first neighbour of it; accept the move if $\mathscr{S}$, $\mathscr{W}$ or $\mathscr{E}$ decreases;
\EndFor
\State{$\Rightarrow$ repeat the for-loop to improve the chance of finding the optimal partition}
\end{algorithmic}

\vspace{5mm}

\noindent\textit{`Bimodular' structures detection on binary and weighted networks.} In this case, the algorithm we implement fixes the number of clusters to be detected a priori, since each node of the network can belong to only one \emph{out of two} possible subsets (e.g. the core and the periphery) - a circumstance that is reflected into the initialization of the algorithm, i.e. a vector whose entries are randomly chosen to be either 0 or 1. A second difference with respect to the community detection case concerns the criterion according to which edges are sorted (for purely bipartite graphs, in fact, the Jaccard index is zero for all edges, as nodes connected by an edge always lie on different layers): in our modified version of the PACO algorithm, when binary graphs are considered, we sort links according to their eigenvector centrality, which has been shown to proxy `coreness' to quite a good extent. As a final step, we consider a number of random re-assignments of nodes with the aim of preventing the possibility of getting stuck in a local minimum (a random move consists of selecting 3 nodes belonging to the same group and evaluating if assigning them to the other group would further minimize surprise).

The edge-sorting strategy described above is not the best one in the weighted case: in this case, in fact, randomly-ordered edges are to be preferred, as they avoid getting stuck in local minima more effectively.

\vspace{5mm}

\begin{algorithmic}[1]
\Function{CalculateAndUpdateSurprise}{$C,C'$}
	\State $S\gets calculateSurprise(C)$
	\State $S'\gets calculateSurprise(C')$
	\If{$S'<S$}
		\State $C\gets C'$
		\State $S\gets S'$
	\EndIf \\
\Return $C$
\EndFunction
\State
\State $C \gets $ array of length $N$, randomly initialized with binary entries, i.e. $0,1$;
\State $E \gets $ sorted edges (either via the eigenvector centrality, in case of binary networks, or randomly, in case of weighted networks);
\For{edge $(u,v)\in E$}
	\State $C'\gets C$
	\If{$C'[u] \neq C[v]$}
		\State $C'[u]\gets C[v]$
		\State $C\gets$\Call{CalculateAndUpdateSurprise}{$C,C'$}
	\Else
		\State $C'[u]\gets 1-C[v]$
		\State $C\gets$\Call{CalculateAndUpdateSurprise}{$C,C'$}
	\EndIf
\State $\Rightarrow$ randomly switch the membership of 3 nodes in the same group and accept the move if $\mathscr{S}_\sslash$, $\mathscr{W}_\sslash$ or $\mathscr{E}_\sslash$ decreases;
\EndFor
\State{$\Rightarrow$ repeat the for-loop to improve the chance of finding the optimal partition}
\end{algorithmic}

\end{document}